\documentclass[a4paper,fleqn]{cas-sc}

\usepackage[numbers,sort&compress]{natbib}

\def\tsc#1{\csdef{#1}{\textsc{\lowercase{#1}}\xspace}}
\tsc{WGM}
\tsc{QE}
\tsc{EP}
\tsc{PMS}
\tsc{BEC}
\tsc{DE}

\usepackage{float}
\usepackage{graphicx}
\usepackage[utf8]{inputenc}
\usepackage[english]{babel}
\usepackage{amssymb,amsthm,amsmath,amstext,amsbsy,amsopn}
\usepackage[cal=boondoxo]{mathalpha}
\usepackage{subfigure}
\usepackage{mathrsfs}
\usepackage{bbm}
\usepackage{nicefrac}
\usepackage{slashed}
\usepackage{pstricks}
\usepackage{hyperref}
\usepackage{leftidx}
\usepackage{environ}
\usepackage{mathtools}
\usepackage{xspace}
\usepackage{overpic}
\usepackage{epstopdf}
\usepackage{calligra}
\usepackage{soul}
\usepackage{cancel}
\usepackage[commandnameprefix=ifneeded]{changes}

\newcommand{\ur}[1]{\textcolor{red}{#1}}

\begin{document}
\let\WriteBookmarks\relax
\def\floatpagepagefraction{1}
\def\textpagefraction{.001}
\shorttitle{Universal characterization of Efimovian $D^0 nn$ System {\it via} Faddeev Techniques}
\shortauthors{G.Meher et~al.}

\title [mode = title]{Universal characterization of Efimovian $D^0 nn$ System {\it via} Faddeev Techniques}

\author[1,2,3]{Ghanashyam Meher}

\ead{ghanashyam@phys.ntu.edu.tw}


\author[1]{Sourav Mondal}
\ead{sm206121110@iitg.ac.in}

\author[1]{Udit Raha}
\ead{udit.raha@iitg.ac.in}

\cormark[1]

\affiliation[1]{organization={Department of Physics, Indian Institute of Technology Guwahati},
                city={Guwahati},
                postcode={781039},
                country={India}}

\affiliation[2]{organization={Department of Physics and Center for Theoretical Physics},
                institution={National Taiwan University},
                city={Taipei},
                postcode={10617},
                country={Taiwan}}
\affiliation[3]{organization=   {Physics Division, National Center for Theoretical Sciences}, city={Taipei}, postcode={10617}, country={Taiwan}}

\cortext[cor1]{Corresponding author}

\begin{abstract}
We demonstrate remnant structural universality in a putative S-wave $2n$-halo-bound $D^0nn$ system in the $J=0, T=3/2$
channel by invoking the zero-coupling limit (ZCL), which eliminates sub-threshold decay channels. Within this framework,
we evaluate the one- and two-body matter density form factors, their associated root mean-square radii, and 
the $n$–$D^0$–$n$ opening angle. Our analysis is carried out at leading order using a quantum mechanical Faddeev 
technique in the momentum representation. Employing Jacobi momenta, we construct a complete partial-wave 
basis to expand the full three-body $D^0nn$ wave function across distinct rearrangement channels. Projection onto this
basis yields a coupled set of Faddeev integral equations that govern the multiple-scattering dynamics of the 
constituent coupled spin–isospin sub-systems. By introducing short-range separable interactions and expressing the 
two-body scattering amplitudes {\it via} spectator functions, we establish a direct correspondence with the 
familiar Skornyakov–Ter-Martirosyan equations from halo-EFT approach at leading order. A 
regulator-dependent analysis highlights the Efimov-like character of the three-body observables, with ground state 
properties exhibiting marked sensitivity to cutoff variations. However, the inclusion of a three-body force suppresses 
this dependence, as expected from renormalization-group invariance. We thereby conclude that, for sufficiently 
shallow three-body binding, the $D^0nn$ system in the ZCL exhibits a universal halo-bound structure. The subtle 
implications of range-like corrections at LO are addressed at a qualitative level in this analysis. 
\end{abstract}


\maketitle
\section{Introduction}
\label{sec:Intro}
Few-body universality broadly refers to the insensitivity of long-distance/low-energy characteristics of few-body systems 
to short-distance/high-energy details of interactions. The universality of different few-body systems can manifest in 
markedly distinct ways. A common example of universality is what is manifested in the two-body sector, accompanied by the 
occurrence of shallow bound {\it dimers} whenever the S-wave two-body scattering length $a_0$ becomes positive and, 
in essence, much larger than the corresponding interaction range $r_0$. A renormalization group (RG) analysis of 
two-body contact interactions reveals the proximity of a {\it non-trivial infrared fixed-point, corresponding to} the 
{\it unitary limit} of the two-body interactions. However, in the pioneering works of Efimov and 
Danilov~\cite{Efimov:1970zz,Efimov:1971zz,Efimov:1973awb,Danilov:1961}, a rather exceptional universal behavior was 
demonstrated in the three-boson system interacting {\it via} short-distant two-body $\sim 1/r^2$ singular potential, with 
the scattering lengths tuned to the unitary limit. Solving the Faddeev equations~\cite{Faddeev:1961} in the 
{\it hyperspherical} co-ordinate representation led to the emergence of a sequence of geometrically spaced arbitrarily 
shallow S-wave {\it trimers} accumulating to zero-energy threshold, as $|a_0|\to \infty$. Such an emergent phenomenon, 
termed as the {\it Efimov effect}~\cite{Efimov:1970zz,Efimov:1971zz,Efimov:1973awb,Danilov:1961,Braaten:2004rn,Naidon:2016dpf}, 
heralds the partial breakdown of the exact scale invariance of the system's observables into a {\it discrete scaling} behavior.
A study of the scale variation of three-body interactions shows the onset of an RG {\it limit cycle}~\cite{Braaten:2004rn}. 

Over the past two decades, Efimov universality has been extensively explored within {\it pionless} effective field theory 
(${}^{\pi\!\!\!\!/}$EFT)~\cite{Braaten:2004rn,Kaplan:1996xu,Kaplan:1998tg,Kaplan:1998we,Kaplan:1998sz,vanKolck:1998bw} 
using the momentum-space formulation of the Faddeev scattering equations, the so-called {\it Skornyakov-Ter-Martirosyan} 
(STM) equations~\cite{STM1,STM2}, across a broad range of nuclear systems. These include ordinary shallow-bound nuclear
isotopes (e.g., ${}^{\,\,\,3}{\rm H}$,${}^{\,\,\,3}{\rm He}$,${}^{\,\,\,5}{\rm He}$,${}^{\,\,\,6}{\rm He}$,${}^{\,\,\,11}{\rm Li}$,
${}^{\,\,\,12}{\rm Be}$,${}^{\,\,\,14}{\rm Be}$,${}^{\,\,\,18}{\rm C}$,${}^{\,\,\,20}{\rm  C}$, 
etc.)~\cite{Bedaque:1998mb,Bedaque:1998kg,Bedaque:1999ve,Bertulani:2002sz,Bedaque:2002yg,Canham:2008jd,DL-Canham:2009,Rotureau:2012yu,Ji:2014wta,Vanasse:2016hgn,Ryberg:2017tpv,Gobel:2019jba}, 
exotic nuclear clusters, such as hypernuclei (e.g., ${}^3_\Lambda$H, $nn\Lambda$, 
${}^{\,\,\,\,\,4}_{\Lambda\Lambda}{\rm He}$, ${}^{\,\,\,\,\,5}_{\Lambda\Lambda}{\rm H}$, 
${}^{\,\,\,\,\,5}_{\Lambda\Lambda}{\rm He}$, ${}^{\,\,\,\,\,6}_{\Lambda\Lambda}{\rm He}$, $\Xi^-nn$, etc.)~\cite{Hammer:2001ng,Ando:2015fsa,Hildenbrand:2019sgp,Ando:2013kba,Ando:2014mqa,Contessi:2018qnz,Contessi:2019csf,Meher:2020yjr,Meher:2020ark,Schafer:2020vzl,Schafer:2020rba,Schafer:2021qbt}, as well as pure mesonic clusters that include heavy charm and bottom 
mesons~\cite{Braaten:2003he,Raha:2017ahu,Raha:2020sse}. 
In addition, we refer the reader to Ref.~\cite{Hammer:2019poc} for a recent detailed review on the applications of 
${}^{\pi\!\!\!\!/}$EFT in the search and analyses of exotic Efimov-bound states. In particular, some of these putative
bound states constitute a universal class of exotic nuclear few-body systems termed as {\it halo nuclei}. Their quantum 
descriptions are defined in terms of a few physical scales independent of range and other details of the short-distance 
pairwise interactions among the constituents. Such halo systems have been extensively exploited to explore 
three-body universality~\cite{Riisager:1994zz,Zhukov:1993aw,Hansen:1995pu,Tanihata:1996,Jensen:2004zz}. Often {\it neutron rich} 
three-body hadronic clusters are identified as good halo-bound candidates owing to their typical diffuse structure and large 
radius with one or more loosely held ``valence'' neutrons orbiting around their compact cores. The core itself may 
consist of either a single structureless particle (as in our case) or a hadronic cluster with excitation energies much larger
than the outer halo-neutron separation energies. This characteristic scale separation makes them amenable to a low-energy 
EFT description. In order to tackle this multi-scale threshold dynamics of halo nuclei at scales typically lower than in 
applicability of standard ${}^{{\pi}\!\!\!/}$EFT, an important variant, the so-called 
{\it halo}-EFT~\cite{Bertulani:2002sz,Rotureau:2012yu,Vanasse:2016hgn,Ryberg:2017tpv}, needs to be deployed (for a recent 
review, see Ref.~\cite{Hammer:2017tjm}). This framework provides a natural starting point for exploring the possible 
existence and low-energy properties of a putative $2n$-halo-bound $D^0nn$ $(J=0, T=3/2)$ system, as previously 
initiated in Refs.~\cite{Raha:2017ahu,Raha:2020sse}.

Concomitantly, meson–nucleon interactions have attracted considerable attention for their role in strangeness nuclear physics,
particularly in elucidating hadronic {\it clustering} phenomena. This has led to the identification of numerous exotic bound 
systems involving mesons, nucleons, and hyperons. Especially, the study of the interaction of (anti-) kaons with nucleons or 
nuclei gained considerable impetus since the seminal works of 
Dalitz~\cite{Dalitz:1959dn,Dalitz:1959dq,Dalitz:1960du,Dalitz:1961dx,Dalitz:1961dv}, leading to the discovery of the well-known
resonance state $\Lambda(1405)$, just below the $\bar{K}N$ threshold, thereby establishing a strongly attractive nature of the
$\bar{K}N$ interaction in the $T=0$ isospin channel. This subsequently prompted a flurry of theoretical investigations on the 
inherent nature of the resonance state and its possible influence on the formation of exotic kaonic nuclei in the context of 
various potential and chiral EFT-based model analyses~\cite{Oller:2000fj,Jido:2002zk,Mai:2012dt,Cieply:2015pwa,Mai:2014xna,Cieply:2016jby,Feijoo:2018den,Sadasivan:2018jig,Nogami:1963xqa,Akaishi:2002bg,Yamazaki:2007cs,Shevchenko:2006xy,Shevchenko:2007zz,Ikeda:2007nz,Ikeda:2008ub,Dote:2008in,Dote:2008hw,Wycech:2008wf,Ikeda:2010tk,Hyodo:2011ur,Barnea:2012qa,Ohnishi:2017uni} 
(also see Ref.~\cite{Gal:2016boi} for a comprehensive review). However, it is a well-known fact that the dynamics of decay 
into other open coupled channels tend to ``wash out'' emergent features of three-body universality, diminishing the 
likelihood of an Efimov-bound state formation~\cite{Braaten:2003he,Raha:2017ahu,Raha:2020sse,Hyodo:2013zxa}. For example, 
given that $\Lambda (1405)$ has a large decay width, $\Gamma\sim 40-70$~MeV, compared to its binding energy $\sim 20$~MeV, 
it is expected that a bound Efimov-like $\bar{K}NN\, (J=0, T=1/2)$ state is practically unfeasible. In fact, by 
performing a momentum cutoff ($\Lambda_{\rm reg}$)-dependent analysis of the STM integral equations, the
${}^{\pi\!\!\!\!/}$EFT study of Ref.~\cite{Raha:2017ahu} demonstrated that even the comparatively particle-stable
$\bar{K}NN (J=0, T=3/2)$ system remains predominantly unbound despite Efimov attraction. 

An analogous situation arises in the charm sector in the presence of the resonance state $\Lambda^+_c(2595)$ just below the
$DN$ threshold ($T=0$ isospin channel)~\cite{Hofmann:2005sw,Mizutani:2006vq}, having an identical set of quantum numbers as
in $\Lambda(1405)$. However, unlike $\Lambda(1405)$, the $\Lambda^+_c(2595)$ has a narrow width $\Gamma\sim 2.6$~MeV, 
which makes it more particle-stable. Furthermore, the potential model analysis of Ref.~\cite{Bayar:2012dd} posits that the 
$DNN\, (J=0, T=1/2)$ system is likely a deeply-bound trimer with binding energy $\sim 250$~MeV. However, assessing its 
Efimov signature has been difficult, complicated by resonance and Coulomb effects. In contrast, the $DN$ ($T=1$) channel is 
comparatively less attractive with a near-threshold resonance or {\it quasi-bound} state, the 
$\Sigma_c(2800)$~\cite{ParticleDataGroup:2016lqr}. Being manifestly free of complications due to Coulomb effects, the 
latter may prove to be more favorable in exhibiting three-body universal signatures in the $DNN\, (J=0, T=3/2)$ channel, as
hinted in Ref.~\cite{Yasui:2010zz}. Indeed, in Refs.~\cite{Raha:2017ahu,Raha:2020sse}, it was demonstrated that the $DN$ 
($T=1$) channel manifests two-body universality upon invoking a {\it zero-coupling limit} (ZCL) idealization that excluded 
decay and coupled-channel effects. Furthermore, by invoking the ZCL ansatz in the three-body sector, the neutron separation 
($S_n$) energy arguably becomes much smaller than the excitation energy of the $D^0$-meson to the $D^{*0}(2007)$ 
resonance state. This distinct separation of scales allows for a halo-EFT~\cite{Hammer:2017tjm} description of the S-wave 
$D^0 nn\, (J=0, T=3/2)$ system. By solving the STM integral equations for the three-body scattering amplitudes, the 
analysis in Ref.~\cite{Raha:2017ahu} demonstrated that an Efimov-like $D^0nn$ threshold bound state becomes feasible at a 
substantially small value of the {\it critical cutoff} (i.e., the minimal regulator scale at which the first Efimov-bound 
state appears above the $n+(D^0n)$ {\it particle-dimer break-up} threshold), namely 
$\Lambda^{(0)}_{\rm crit}\approx 48$~MeV, compared to the ${}^{{\pi}\!\!\!/}$EFT hard/breakdown scale 
$\lambda_H\sim m_\pi$ (where $m_\pi$ is the pion mass). This indicates an encouraging prospect of an S-wave $D^0nn$ system
manifesting itself as an shalow bound Efimov state.

This work extends the previous analyses of Refs.~\cite{Raha:2017ahu,Raha:2020sse} by employing leading order (LO)
Faddeev techniques in momentum representation, as developed in 
Refs.~\cite{Canham:2008jd,DL-Canham:2009,Yamashita:2004pv,Platter:2004he,Platter:2004ns,Platter:2004zs}. Within this 
framework, we formulate an effective quantum mechanical approach to investigate the {\it geometrical} features of a
plausible S-wave $D^0nn$ Efimov-bound system. Such an EFT-inspired description provides a systematic and model-independent
framework for correlating the low-energy structure of a halo-bound system with the characteristic features expected of a  
universal Efimov state. The structural predictions can not only shed light on the underlying nature of the 
hadronic clustering mechanisms, but also provide valuable input for other {\it ab initio} approaches, including lattice QCD
simulations. 

The paper is organized as follows: In Sec.~\ref{sec:Faddeev-Dnn},
we elaborate on our theoretical halo-EFT framework needed for the Faddeev-based analyses. We begin with the mathematical 
formulation of the LO Faddeev equations~\cite{Faddeev:1961,Glockle_1983} needed to describe the three-body dynamics of a 
$2n$-halo-bound $D^0nn$ system. For this purpose, we employ Jacobi momentum representation to build a complete set of basis 
states using partial-wave decomposition needed to describe the component wave functions representing all possible 
re-arrangements of the binary sub-systems. By projecting the operator form of the Faddeev equations onto the chosen basis, we 
obtain a set of coupled integral equations that describe the LO dynamics of the coupled spin-isospin channels for the 
$D^0nn$ system. Subsequently, we establish a one-to-one correspondence of our halo-EFT results with the analogous STM 
integral equations derived in Ref.~\cite{Raha:2017ahu} using LO ${}^{\pi\!\!\!\!/}$EFT, whose solutions determine the 
low-energy $D^0nn$ observables. We further reconstruct the full three-body $D^0nn$ wave function in terms of different 
fragmentation component wave functions (cf. Fig.~\ref{fig:channels}) and show how their solutions could be used to extract 
various geometrical features of a plausible bound $D^0nn$ system. Our numerical results within the ZCL model framework are 
presented in Sec.~\ref{sec:results}, followed by a discussion of their implications for universal physics. A concise summary 
and concluding remarks are given in Sec.~\ref{sec:summary}. Technical details of the formalism and numerical 
implementation are deferred to the appendices.

\section{Theory and Methodology: Faddeev analysis of $D^0nn$ universality}
\label{sec:Faddeev-Dnn}
The LO ${}^{\pi\!\!\!\!/}$EFT analysis presented in Refs.~\cite{Raha:2017ahu,Raha:2020sse} indicate that the S-wave 
$D^0nn\,(J=0, T=3/2)$ system is likely to support Efimov states when the $DN$ ($J=1/2,\,T=1$) interactions tuned close to 
the unitary limit, {with couplings} to sub-threshold hadronic decay channels switched off.\footnote{The eliminated coupled
decay channels include hadronic product states, such as $\pi^- \Lambda_c^+$, $\pi^0\Sigma_c^0$, and $\pi^-\Sigma_c^+$, which lie at
least 200~MeV below the $D^0n$ channel threshold. Within the kinematic regime of applicability of \textit{pionless} EFT, these 
channels can therefore be regarded as effectively closed and do not explicitly contribute to the near-threshold dynamics. A 
consistent inclusion of such decay channels would require extending the framework beyond pionless EFT, for instance to a 
\textit{pionful} chiral EFT with explicit pion degrees of freedom and dynamical charmed baryons, such as $\Lambda_{c},\, 
\Sigma_{c}^{0}$ and $\Sigma_{c}^{+}$, significantly increasing the theoretical complexity. Moreover, once open channels are included,
the problem acquires a non-Hermitian character, and assessing their impact on Efimov physics would necessitate dedicated treatments
(e.g., three-body complex scaling and coupled-channel approaches), which lie beyond the scope of the present work.} With such a ZCL
idealization, the otherwise {\it quasi-bound} $DN$ state with a small imaginary part of the {\it complex} $n$-$D^0$ S-wave scattering
length turns into a {\it real-bound} state with a large positive scattering length (see Ref.~\cite{Raha:2017ahu} for details of the 
extraction of the scattering lengths in the framework of a dynamical coupled-channel model). By invoking the ZCL model scenario, the 
$D^0nn$ system is pictured as a $2n$-halo nucleus with three two-body fragmentation channels: two $n+(D^0n)$ channels, each
representing a bound $D^0n$ sub-system with a spectator neutron, and one $D^0+(nn)$ channel, representing a {\it virtual} bound $nn$
sub-system with a spectator $D^0$-meson. Consequently, the $D^0nn$ system can be described using the three Faddeev components. The 
first two components are represented by the states $|\psi_{n_1}\rangle$ and $|\psi_{n_2}\rangle$, in which one of the two 
halo-neutrons acts as the spectator, while the other combines with the $D^0$-meson core to form the corresponding binary 
sub-system in each case. The third component is represented by the state $|\psi_{D}\rangle$ in which the core $D^0$-meson acts as the
spectator, while the two halo-neutrons form the binary sub-system. However, due to the fermionic symmetry between identical
configurations of the states $|\psi_{n_1}\rangle$ and $|\psi_{n_2}\rangle$ upon transposition of the two neutrons, the number of 
independent components reduces to two. If $\mathcal P$ denotes the operator permuting the two neutrons, then by anti-symmetry of
the full three-body wave function $\Psi$, as well as its components, we have the following:
\begin{eqnarray}
{\mathcal P}|\Psi\rangle \!\!\!&=&\!\!\!  {\mathcal P} \big(|\psi_{n_1}\rangle + |\psi_{n_2}\rangle + |\psi_D\rangle\big) 
= - |\psi_{n_2}\rangle - |\psi_{n_1}\rangle - |\psi_D\rangle  = -|\Psi\rangle\,.
\label{eq:antisymmetry}
\end{eqnarray}
Thus, with the component $|\psi_{n_2}\rangle$ expressed in terms of $|\psi_{n_1}\rangle$, we effectively have two Faddeev 
partitions, namely $|\psi_{n}\rangle$ and $|\psi_{D}\rangle$, describing all possible three-body re-arrangements. The complete 
three-body state is then expressed as $|\Psi\rangle = (1-{\mathcal P})|\psi_n\rangle + |\psi_D\rangle$. Assuming that the binary 
sub-systems $D^0n$ and $nn$ are associated with two-body T-matrices $t_n$ and $t_D$, respectively ($t_i$ is the standard 
{\it spectator} notation for the interaction between the particles $j$ and $k$, with $i$ being the spectator), the two-component 
coupled homogeneous Faddeev equations for the $D^0nn$ system can be expressed in matrix form:
\begin{equation}
\begin{aligned}
\begin{bmatrix}|\,\psi_n\rangle\, \\  \,|\psi_D\rangle\, \end{bmatrix}
= G_0(z)
\begin{bmatrix} -t_n(z){\mathcal P}                & t_n(z)\\ 
                 t_D(z) (1-{\mathcal P})           & 0\\
\end{bmatrix}
\begin{bmatrix}\,|\psi_n\rangle\, \\  \,|\psi_D\rangle\, \end{bmatrix}
\end{aligned}\,.
\label{eq:matrixform}
\end{equation}
Note that for simplicity's sake, no genuine three-body interactions are being assumed to be present in the system at this
stage. For the case of negative total three-body energy $E<0$, analytically continued to complex energies $z=E+i0_+$, the 
above matrix equation has nontrivial solutions when the eigenvalue of the kernel matrix is unity. However, before 
attempting to solve the aforementioned system of equations to extract the three-body observables, it is 
necessary to project them onto a complete set of basis states. This is conveniently achieved using the Jacobi 
representation in momentum-space.

\subsection{S-wave projected Faddeev equations with LO contact interactions}
\label{sec:2.1}
To construct the Faddeev equations in Jacobi momentum representation, we follow the methodology outlined in 
Refs.~\cite{Canham:2008jd,DL-Canham:2009} that employs an effective quantum mechanical technique where only LO effective
{\it separable} potentials are used. It is an essential adoption of the universality-based approach, originally developed
in Refs.~\cite{Platter:2004he,Platter:2004ns,Platter:2004zs} for studying resonant systems of three and four bosons, and
subsequently extended to the study of $2n$-halo nuclei, such as ${}^{6}$He, ${}^{20}$C, 
etc.~\cite{Canham:2008jd,DL-Canham:2009,Ji:2014wta,Gobel:2019jba}.\footnote{For instance, the isotope ${}^{20}$C is 
ostensibly identified as an exited $n-{}^{18}$C\,$-n$ Efimov 
cluster~\cite{Canham:2008jd,DL-Canham:2009,Amorim:1997mq,Mazumdar:2000dg,Bhasin:2020eus}.} 

In the three-body center-of-mass frame (barycenter), there are two independent Jacobi momentum variables, namely $\bf p$ 
and ${\bf q}$, where ${\bf p}$ is the relative three-momentum of a given two-body sub-system and ${\bf q}$ is the 
three-momentum of the third spectator particle relative to the center-of-mass of the two-body sub-system. Depending upon 
the choice of the spectator particle, three equivalent sets of Jacobi momenta are used to describe the same three-body 
system. Using the spectator indices ($ijk$) to denote the momentum variables, the different sets of Jacobi momenta can be
related by cyclic permutations of the indices. For completeness, a brief description of the quantum mechanical construction
of a complete set of three-particle partial-wave Jacobi basis states is provided in Appendix~\ref{sec:Appendix-A}. 
Based on that framework, we can define our set of Faddeev equations for the putative $2n$-halo-bound $D^0nn$ system. A 
generic partial-wave projected Jacobi basis constructed from plane wave states is represented as 
$|pq\,\mathcal{Q}_i\rangle_i \equiv |p_iq_i\,\mathcal{Q}_i\rangle$~\cite{Glockle_1983}, where $\mathcal{Q}_i$ is a 
collective index that specifies the spin-isospin state of the three-body system with particle $i$ as the spectator and 
$jk$ as the binary sub-system. Thus, the different partial-wave basis states for the $D^0nn$ system are denoted 
as\footnote{As discussed in the Appendix, we introduce the labels $n_1$ and $n_2$ solely as a {\it naive} bookkeeping 
device to distinguish configurations in which one neutron acts as the spectator while the other forms the $D^0 n$ 
sub-system. Nevertheless, the two neutrons remain indistinguishable fermions.} 
$|pq\,\mathcal{Q}_D\rangle_D, \,|pq\,\mathcal{Q}_n\rangle_{n_1}$ and $|pq\,\mathcal{Q}_n\rangle_{n_2}$, with the 
$D^0$-meson, $n_1$-neutron and $n_2$-neutron as the spectator particle, respectively. Since the two identical neutrons 
have opposite spins in the $nn$ sub-system, the $|pq\,\mathcal{Q}_D\rangle_D$ basis state is anti-symmetric with respect to
the two neutrons. Thus, the permutation operator $\mathcal P$ yields the eigenvalue -1 when operated on 
$|pq\,\mathcal{Q}_D\rangle_D$. As for the basis states $|pq\,\mathcal{Q}_n\rangle_{n_1}$ and 
$|pq\,\mathcal{Q}_n\rangle_{n_2}$, the binary sub-systems are symmetric with respect to the non-identical particles. The 
action of ${\mathcal P}$ on these states merely interchanges the position of the two neutrons with a negative sign to 
account for the overall parity of the three-body system [see Eq.~\eqref{eq:antisymmetry}]. Consequently, for an S-wave 
$D^0nn$ system, we must have 
\begin{eqnarray}
{\mathcal P} |pq\,\mathcal{Q}_D\rangle_D = -|pq\,\mathcal{Q}_D\rangle_D\,, \quad \text{and} \quad
{\mathcal P}|pq\,\mathcal{Q}_n\rangle_{n_1(n_2)} = -|pq\,\mathcal{Q}_n\rangle_{n_2(n_1)} \,.
\end{eqnarray}

In Appendix~\ref{sec:Appendix-B}, we present a derivation of the general form of the two-body T-matrix\footnote{The
two-particle T-matrices in our case arise from operators $t_n(z)$ and $t_D(z)$, which act only in the binary sub-systems. 
Hence, these T-matrices are understood as operator embeddings in the three-body space. They enter into the Faddeev 
equations off-shell at the ``shifted'' three-body energies, $z_n=z-\frac{q^{\prime 2}_n}{2\mu_{n(nD)}}$ and 
$z_D=z-\frac{q^{\prime 2}_D}{2\mu_{D(nn)}}$, respectively, because the evaluation of the operator products $G_0(z)t_n(z)$
and $G_0(z)t_D(z)$ imply an integration over all possible intermediate states. These shifted energies are those of the 
binary sub-systems obtained by subtracting the kinetic energy of the spectator particle from the three-body energy $z$. 
Here, $\mu_{n(nD)}=M_n(M_D+M_n)/(M_D+2M_n)$ and $\mu_{D(nn)}=2M_D M_n/(M_D+2M_n)$ denote the three-body reduced 
masses corresponding to the two distinct re-arrangement channels.} [see Eq.~\ref{eq:two-body_T-matrix}] starting 
from an S-wave non-local separable potential in momentum representation, characterized by the LO contact interaction 
$C_0$, given by
\begin{equation}
\langle {\mathbf p}| V_{\rm eff}|{\mathbf p}^\prime\rangle \,\stackrel{\rm S-wave}{\longrightarrow}\, \chi(p)\, C_0 \,\chi(p^\prime)\,,
\label{eq:V_eff}
\end{equation}
where ${\bf p}$ (${\bf p^\prime}$) is the two-body incoming (outgoing) relative three-momentum. The $\chi$'s denote the 
so-called {\it form factor} functions, which are related to the factorized residues at the poles of the corresponding 
T-matrix or propagator. The formalism is pertinent to the EFT treatment of halo systems, where the binary sub-systems 
are also likely to exhibit universality with distinct separation of scales. Moreover, owing to the separable nature of the 
T-matrix, the Lippmann-Schwinger equation is analytically solvable, allowing us to directly fix the coupling constant $C_0$ 
from the pole of the two-body bound/anti-bound state of a given sub-system, see Appendix~\ref{sec:Appendix-B}. In the absence
of sub-threshold hadronic decay channels,\footnote{Recall that we invoke ZCL model idealization in the treatment of the 
$D^0n$ binary sub-system. For details of the ZCL model approach, we refer to Ref.~\cite{Raha:2017ahu}.} the form factors are
real and may be used as regulators to suppress high-momentum modes beyond a certain hard cutoff scale, i.e., for 
$p,p^\prime \geq \Lambda_H$, where the effective potential breaks down. In this way, one can directly relate this to a 
low-energy halo-EFT description with zero-range contact interactions. For numerical implementation of renormalization, 
it is customary to introduce a finite sharp momentum cutoff regulator $\Lambda_{\rm reg}$ to rectify the ill-defined 
ultraviolet (UV) behavior of the Faddeev integral equations. The latter is equivalently implemented {\it via} a simple 
{\it Heaviside} step-function, $\chi(p)=\Theta(\Lambda_{\rm reg}-p)$, for the S-wave form factors in the integration kernels,
with the UV limits subsequently taken to infinity. However, it should be emphasized that, for consistency with RG 
invariance, low-energy observables are expected to become independent of the regulator in the limit 
$\Lambda_{\rm reg}\to \infty$. Since we wish to relate  our halo-EFT results of the effective potential approach to 
those obtained earlier in Ref.~\cite{Raha:2017ahu} using ${}^{\pi\!\!\!\!/}$EFT, the natural choice of the UV 
regulator cutoff is $\Lambda_{\rm reg}\gtrsim \Lambda_H\sim m_\pi$. 

By projecting the Faddeev matrix equation, Eq.~\eqref{eq:matrixform}, onto the appropriate Jacobi basis for the $D^0nn$ 
system, we obtain the following set of homogeneous S-wave integral 
equations~\cite{Canham:2008jd,DL-Canham:2009,Platter:2004he,Platter:2004ns,Platter:2004zs}:
\begin{eqnarray}
\psi_n (p,q) \!\!\!&=&\!\!\! \frac{1}{2} \mathcal{G}^{(n)}_0(p,q;B_3)\,\chi(p)\,\tau_n(q;B_3)
\int_0^{\infty}{\rm d}{q^{\prime\prime }} q^{\prime\prime 2}\int_{-1}^1 {\rm d}x
\bigg[\chi\left(\pi_{nn}(q,q^{\prime\prime};x)\right)\psi_n\left(\pi_{nn}(q^{\prime\prime},q;x),q^{\prime\prime}\right)
\nonumber\\
&&\hspace{5.5cm}\,+\, \chi\left(\pi_{nD}(q,q^{\prime\prime};x)\right)\psi_D\left(\pi^\prime_{nD}(q,q^{\prime\prime};x),q^{\prime\prime}\right)\bigg]\,, 
\quad \text{and}
\nonumber\\
\psi_D (p,q) \!\!\!&=&\!\!\! \mathcal{G}^{(D)}_0(p,q;B_3)\,\chi(p)\,\tau_D(q;B_3) 
\int_0^{\infty}{\rm d}{q^{\prime\prime}} q^{\prime\prime 2}\int_{-1}^1 {\rm d}x\,
\chi\left(\pi^\prime_{nD}(q^{\prime\prime},q;x)\right)
  \psi_n\left(\pi_{nD}(q^{\prime\prime},q;x),q^{\prime\prime}\right),
\label{eq:psi_D_Faddeev}
\end{eqnarray}
where $\psi_i(p,q) \equiv {}_i\langle pq\,\mathcal{Q}_i |\psi_i\rangle$ denotes the Faddeev component wave function 
in Jacobi momentum partial-wave basis. Here, the spectator index $i=n,D$ specifies the choice of spectator particle, 
corresponding to either a neutron ($i=n$) or the $D^0$ meson ($i=D$). Also $z=E=-B_3<0$ represents the three-particle 
binding energy of the system, $p$ is the relative three-momentum between the two particles $j$ and $k$ for the binary 
sub-system $jk$, and $q$ is the relative three-momentum of the third spectator particle $i$ relative to the center-of-mass
of the $jk$ sub-system. The functions $\mathcal{G}^{(i)}_0(p,q;B_3)$ in the above equations are proportional to the 
matrix element of the three-particle free Green's function $G_0(z)$, stemming from the kernel elements of the Faddeev 
equations evaluated on the Jacobi basis:
\begin{eqnarray}
{}_i\langle pq\,\mathcal{Q}_i|\,G_0(z)t_i(z)\,|p^{\prime}q^{\prime}\,\mathcal{Q}_i\rangle_i 
\!\!\!&=&\!\!\! -\mathcal{G}_0^{(i)}(p,q;B_3)\,\, {}_i\langle pq\,\mathcal{Q}_i|\,t_i(z)\,|p^{\prime}q^{\prime}\,\mathcal{Q}_i\rangle_i
\nonumber\\
&=&\!\!\! -\,4\pi \mathcal{G}_0^{(i)}(p,q;B_3)\frac{\delta(q-q^{\prime})}{qq^{\prime}} T_i(p,p^\prime;z_i)\,,
\end{eqnarray}
where
\begin{eqnarray}
\mathcal{G}_0^{(n)} (p,q;B_3) \!\!\!&=&\!\!\! \bigg[B_3+\frac{p^2}{2\mu_{nD}}+\frac{q^2}{2\mu_{n(nD)}}\bigg]^{-1}
= \bigg[B_3+\frac{y+1}{2yM_n}p^2+\frac{y+2}{2(y+1)M_n}q^2\bigg]^{-1}\,, \quad \text{and}
\nonumber\\
\mathcal{G}_0^{(D)} (p,q;B_3) \!\!\!&=&\!\!\! \bigg[B_3+\frac{p^2}{M_n}+\frac{q^2}{2\mu_{D(nn)}}\bigg]^{-1}
= \bigg[B_3+\frac{p^2}{M_n}+\frac{y+2}{4yM_n}q^2\bigg]^{-1}\,. 
\label{eq:G0_n_D_matrix_elements}
\end{eqnarray}
Here, $M_n$ and $M_D = y M_n$ denote the neutron and $D^0$-meson masses, respectively, while $\mu_{nD} = yM_n/(y+1)$ is 
the corresponding reduced mass. The Green's function $\mathcal{G}_0^{(n)}$ corresponds to the exchange of the $D^0$-meson 
between the two neutrons, whereas the Green's function $\mathcal{G}_0^{(D)}$ corresponds to the exchange of a neutron 
between the other neutron and the $D^0$-meson. Furthermore, the kernel elements after the S-wave projection can be related 
to the LO S-wave two-body T-matrices ($T_{n,D}$), evaluated at the two-body energies, $z_i=\mathscr{E}_i+i0_+$, and expressed
in the following separable form (see Appendix~\ref{sec:Appendix-B}):
\begin{eqnarray}
T_i(p,p^\prime;\mathscr{E}_i) = -\frac{1}{4\pi}\chi(p)\,\chi(p^\prime)\,\tau_i(q^\prime;B_3)\,.
\label{eq:t_separable}
\end{eqnarray}
Here, the terms containing information on the short-range LO S-wave effective interactions are given by
\begin{eqnarray}
\tau_n(q;B_3) \!\!\!&=&\!\!\! \frac{y+1}{\pi y M_n} \bigg[-\frac{1}{a_{nD}}\,\frac{2}{\pi}\arctan\big(|a_{nD}|{\Lambda_{\rm reg}}\big)
+\gamma_n(q;B_3)\,\frac{2}{\pi}\arctan\bigg(\frac{\Lambda_{\rm reg}}{\gamma_n(q;B_3)}\bigg)\bigg]^{-1} \,,\quad \text{and}
\nonumber\\
\tau_D(q;B_3) \!\!\!&=&\!\!\! \frac{2}{\pi M_n} \bigg[-\frac{1}{a_{nn}}\,\frac{2}{\pi}\arctan\big(|a_{nn}|{\Lambda_{\rm reg}}\big)
+\gamma_D(q;B_3)\,\frac{2}{\pi}\arctan\bigg(\frac{\Lambda_{\rm reg}}{\gamma_D(q;B_3)}\bigg)\bigg]^{-1}\,,
\label{eq:tn_tD_S-wave}
\end{eqnarray}
where $a_{nn}$ and $a_{nD}$ are the S-wave $n$-$n$ and $n$-$D^0$ scattering lengths, respectively. Since we aim to assess the 
plausibility of a halo-bound $D^0nn$ system, allowing for the possibility that its two-body sub-systems are either bound or virtual 
(anti-bound), it is useful to evaluate the two-body $T$-matrices at energies expressed in terms of the three-body binding energy 
$B_3$. Thus, for the binary sub-systems, we define the following momentum functions: 
\begin{eqnarray}
\gamma_n(q;B_3) \!\!\!& \equiv &\!\!\! \sqrt{-2\mu_{nD} \mathscr{E}_n(q)}\, 
=\, \sqrt{\frac{y}{y+1}\bigg(2M_nB_3 + \frac{y+2}{y+1}q^{2}\bigg)}\,, \quad \text{and}
\nonumber\\
\gamma_D(q;B_3) \!\!\!& \equiv &\!\!\! -\sqrt{-2\mu_{nn} \mathscr{E}_D(q)}\, 
=\, -\sqrt{M_nB_3 + \frac{y+2}{4y} q^{2}}\,.
\label{eq:gamma_n_d}
\end{eqnarray}
In particular, for $q=0$, the two-body bound (anti-bound) states correspond to the respective poles in the above T-matrices 
which at LO correspond to real (virtual) binding momenta, i.e.,\footnote{Here we re-emphasize that the di-neutron ($nn$) forms a
virtual or anti-bound sub-system with negative imaginary momentum $-i\sqrt{2\mu_{nn}B_{nn}}$ and S-wave scattering length 
$a_{nn}=-18.63$~fm~\cite{Chen:2008zzj}, whereas the $D^0n$ sub-system in the ZCL limit forms a real-bound sub-system with 
positive imaginary binding momentum $i\sqrt{2\mu_{nD}B_{nD}}$ and S-wave scattering length $a_{nD}= 4.141$~fm. The latter 
value was extracted in the dynamical coupled-channel model analysis of Ref.~\cite{Raha:2017ahu}, where the idealized limit 
of vanishing couplings to sub-threshold $DN$ ($T=1$) decay channels was considered.}
\begin{eqnarray}
\text{{\it real}-bound:} \quad \gamma_n(q=0;B_3=B_{nD}) \!\!\!&=&\!\!\! \sqrt{2\mu_{nD}B_{nD}}=\frac{1}{a_{nD}}>0\,, \quad \text{and} 
\nonumber\\
\text{{\it virtual}-bound:} \quad \gamma_D(q=0;B_3=B_{nn}) \!\!\!&=&\!\!\! -\sqrt{2\mu_{nn}B_{nn}}=\frac{1}{a_{nn}}<0\,, 
\end{eqnarray}
with the three-body binding energy $B_3$ coinciding with the two-body binding energy $B_2=B_{nn}$ and $B_2=B_{nD}$, respectively. 
We note that analogous expressions for the T-matrices have been derived in earlier 
works~\cite{Canham:2008jd,DL-Canham:2009,Platter:2004he,Platter:2004ns,Platter:2004zs}, \ur {both in} universality-based studies 
of bosonic systems and in the investigation of ${}^{20}$C as a $2n$-halo nucleus. 

The ``shifted'' relative momenta, $\boldsymbol{\pi}_{ij}$ or $\boldsymbol{\pi^\prime}_{ij}$, appearing in the coupled
Faddeev equations~\eqref{eq:psi_D_Faddeev}, arise from the evaluation of the overlap matrix elements of the form 
${}_i\langle pq\,\mathcal{Q}_i|p^\prime q^\prime\,\mathcal{Q}_j\rangle_j$ ($i,j=n,D$). The latter represents the transition 
amplitudes between different three-body sub-system re-arrangements, which are obtained by evaluating the re-coupling coefficient 
between different spectator Jacobi basis states (see Ref.~\cite{Glockle_1983} for details). Here we simply quote the results for
the S-wave projected overlap matrix elements relevant to the $D^0nn$ system. The generic results are expressed as
\begin{equation}
{}_i\langle pq\,\mathcal{Q}_i|p^\prime q^\prime\,\mathcal{Q}_j\rangle_j 
\equiv {}_i\langle p_iq_i\,\mathcal{Q}_i|p^\prime_i q^\prime_i\,\mathcal{Q}_j\rangle_j
= \int_{-1}^1 {\rm d}x\, \frac{\delta(p-\pi_{ij})}{p\pi_{ij}}\frac{\delta(p^\prime-\pi^\prime_{ij})}{p^\prime \pi^\prime_{ij}}\, 
\mathscr{G}^{(ij)}\,,
\label{eq:ij_overlap_matrix}
\end{equation}
where $x\equiv \hat{\bf q}_i\cdot \hat{\bf q}^\prime_i$ is the cosine of the angle between the initial and final spectator
momenta relative to the binary sub-system $jk$, such that
\begin{eqnarray}
\pi_{ij}(q,q^\prime;x) \!\!\!& = &\!\!\! \sqrt{\left(\frac{\mu_{jk}}{m_k}\right)^2q^2+q^{\prime 2} 
+ 2\left(\frac{\mu_{jk}}{m_k}\right)qq^\prime x}\,, \quad \text{and}
\nonumber\\
\pi^\prime_{ij}(q,q^\prime;x) \!\!\!& = &\!\!\! \sqrt{q^2+\left(\frac{\mu_{ki}}{m_k}\right)^2q^{\prime 2} 
+ 2\left(\frac{\mu_{ki}}{m_k}\right)qq^\prime x}\,\,,
\label{eq:pi_ij}
\end{eqnarray}
denote the ``shifted" three-momenta due to the re-couplings between the different Jacobi spectator bases. The 
quantities represented by
\begin{eqnarray}
\mathscr{G}^{(ij)}=\mathscr{G}^{(ji)} \equiv {}_i\langle \mathcal{Q}_i|\mathcal{Q}_j\rangle_j \!\!\!&=&\!\!\! \frac{1}{2} 
\sqrt{\hat{s}\,\hat{j}\,\hat{t}\,\hat{\mathcal{J}}\,\hat{s^\prime}\,
\hat{j^\prime}\,\hat{t^\prime}\,\hat{\mathcal{J}^\prime}}\,
(-1)^{t_i+t_j+t_k+T}\,(-1)^{s_i+s_j+s_k+S}
\nonumber\\
\!\!\!&&\!\!\!\times\,
\begin{Bmatrix}t_j & t_k & t\\
               t_i & T   & t^\prime\\
\end{Bmatrix}
\begin{Bmatrix}s_j & s_k & s\\
               s_i & S   & s^\prime\\
\end{Bmatrix}
\sum_S \hat{S} 
\begin{Bmatrix}  0 & s      & j\\
                 0 & \sigma & \mathcal{J}\\
                 0 & S      & J \\
\end{Bmatrix}
\begin{Bmatrix} 0 & s^\prime        & j^\prime\\
                0 & \sigma^{\prime} & \mathcal{J}^{\prime}\\
                0 & S               & J\\
\end{Bmatrix}\,,
\label{eq:g_ij}
\end{eqnarray}
are simple geometrical constants arising from the re-coupling of angular momenta and isospins between different spectator 
bases with indices $i$ and $j$ employing Wigner's $6j$ and $9j$ symbols. The hatted quantum numbers denote their 
multiplicities; for example, $\hat{S}\equiv 2S+1$. Each overlap matrix has four momenta, of which any two could be 
eliminated in favor of the others after integrating the delta-functions in Eq.~\eqref{eq:ij_overlap_matrix}. Here, we 
prefer to eliminate $p$ and $p^\prime$ in favor of $q$ and $q^\prime$.

\begin{figure}[tbp]
\centering
\includegraphics[scale=0.53]{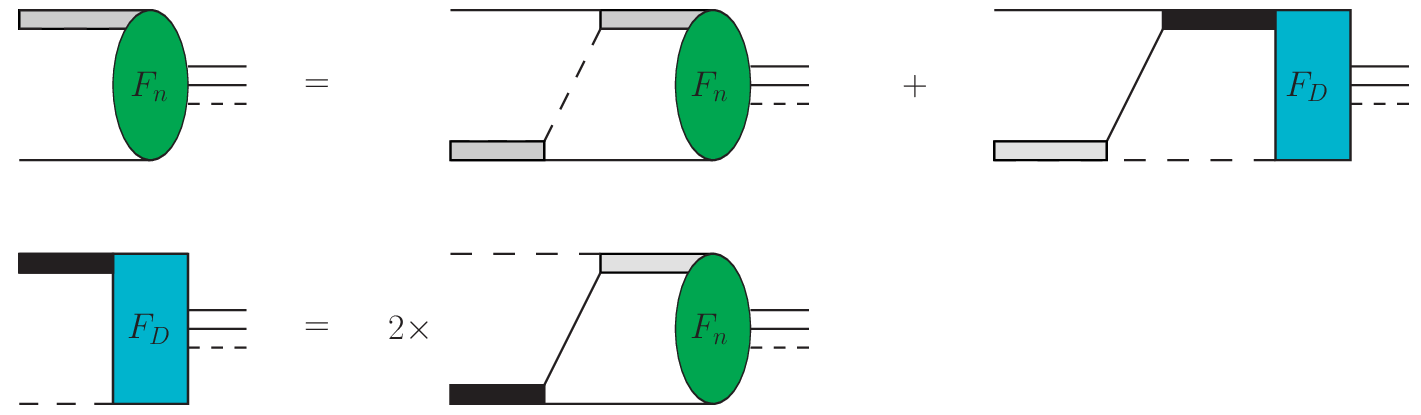}
    \caption{Feynman diagrams for the leading order coupled-channel homogeneous integral equations for the spectator 
             functions, $F_n$ and $F_D$, of an S-wave $2n$-halo $D^0nn$ system. The solid (dashed) lines represent the
             neutron ($D^0$-meson) field. The gray (black) shaded thick lines represent the iterated $n$-$D^0$ 
             ($n$-$n$) two-body S-wave T-matrix $\tau_n$ ($\tau_D$), Eq.~\eqref{eq:tn_tD_S-wave}, which in halo-EFT 
             corresponds to the renormalized dressed auxiliary field (or dimeron) propagator [see later in 
             Eqs.~\eqref{eq:dimeron_prop1} and ~\eqref{eq:dimeron_prop2}]. The elliptical (rectangular) blobs represent 
             the elastic (inelastic) channel $n+d_{(nD)}\to n+d_{(nD)}$ ($D^0+s_{(nn)}\to n+d_{(nD)}$) transition 
             amplitude which is related to the spectator function $F_n$ ($F_D$) in the vicinity of trimer pole energy 
             [see later in Eq.~\eqref{eq:T_F_relation}]. }
\label{Fig:Faddeev_eq} 
\end{figure}

For the sake of simplicity of numerical evaluations, it is advantageous to introduce the {\it spectator functions} 
$F_i(q)$~\cite{Mitra:1969} to define the distinct Faddeev components, namely
\begin{eqnarray}
\psi_i(p,q) = \mathcal{G}^{(i)}_0(p,q;B_3)\,\chi(p)\,\tau_i(q;B_3)\,F_i(q)\,; \quad i=n,D\,.
\label{eq:spectator_intro}
\end{eqnarray}
Thus, we obtain the following set of coupled integral equations in a single variable, which is considerably more
tractable for numerical solution than the preceding two-variable formulation,
Eq.~\eqref{eq:psi_D_Faddeev}~\cite{Canham:2008jd,DL-Canham:2009,Platter:2004he,Platter:2004ns,Platter:2004zs}, namely
\begin{eqnarray}
\label{eq:F_nD}
F_n(q) \!\!\!&=&\!\!\! \frac{1}{2} \int_0^{\infty}{\rm d}{q^\prime} q^{\prime 2}\int_{-1}^1 {\rm d}x 
\Big[\chi\left(\pi_{nn}(q,q^\prime;x)\right)\,\chi\left(\pi_{nn}(q^\prime,q;x)\right) 
\mathcal{G}^{(n)}_0(\pi_{nn}(q^\prime,q;x),q^\prime;B_3)\,\tau_n(q^\prime;B_3)\,F_n(q^\prime)
\nonumber\\
&&\hspace{2cm} +\,\chi\left(\pi_{nD}(q,q^\prime;x)\right)\,\chi\left(\pi^\prime_{nD}(q,q^\prime;x)\right)
\mathcal{G}^{(D)}_0(\pi^\prime_{nD}(q,q^\prime;x),q^\prime;B_3)\,\tau_D(q^\prime;B_3)\,F_D(q^\prime)\Big]\,, \,\,\,\, \text{and}
\nonumber\\
F_D(q) \!\!\!&=&\!\!\! \int_0^{\infty}{\rm d}{q^\prime} {q^\prime}^2\int_{-1}^1 {\rm d}x 
\Big[\chi\left(\pi_{nD}(q^\prime,q;x)\right)\,\chi\left(\pi^\prime_{nD}(q^\prime,q;x)\right)
\mathcal{G}^{(n)}_0(\pi_{nD}(q^\prime,q;x),q^\prime;B_3)\,\tau_n(q^\prime;B_3)F_n(q^\prime)\Big]\,.
\end{eqnarray}
The above coupled integral equations describe the LO three-body dynamics of multiple scattering between the core $D^0$-meson
and the two halo-neutrons, as diagrammatically illustrated in Fig.~\ref{Fig:Faddeev_eq}. As we shall next demonstrate, 
these equations are completely analogous to the homogeneous parts of the STM integral equations for the $D^0nn$ system 
derived in Ref.~\cite{Raha:2017ahu} in the context of LO ${}^{\pi\!\!\!\!/}$EFT employing a sharp momentum cutoff 
regularization.\footnote{In the analogous ${}^{\pi\!\!\!\!/}$EFT scenario, the integral equations are described by 
{\it half-off-shell} transition amplitudes instead of the spectator functions $F_i$, which connect the spectator $i$ and the 
interacting pair $jk$ to form the three-body bound states. Furthermore, to describe the two-body bound state dynamics, the
iterated T-matrices $\tau_i$ are replaced in ${}^{\pi\!\!\!\!/}$EFT by the renormalized dressed auxiliary field propagators 
in the STM equations~\cite{Braaten:2004rn,Bedaque:1998mb,Bedaque:1998kg,Bedaque:1999ve,Bedaque:1998km,Birse:1998dk,Beane:2000fi,Ando:2004mm,Kaplan:1996nv}.}
Analogously, they yield the three-body binding energy $B_3$ for which they have a nontrivial solution. Written as a matrix 
eigenvalue equation of the form 
\begin{equation}
\begin{bmatrix} F_n\\F_D \end{bmatrix} = \begin{bmatrix}\mathcal{K}_{nn}&\mathcal{K}_{nD}\\
\mathcal{K}_{Dn}&0\end{bmatrix} \begin{bmatrix} F_n\\F_D \end{bmatrix}\,,
\label{eq:Matrix_eq_Fn_FD}
\end{equation}
with discretized incoming $q$ and outgoing $q^\prime$ relative momenta, the binding energy $B_3$ corresponds to the eigenvalue
of 1 for the kernel matrix $\mathcal{K}_{ij}$ [constructed from Eq.~\eqref{eq:F_nD}]. For completeness, we include a brief
description of the numerical methodology for determining the eigenvectors of the integral equations in 
Appendix~\ref{sec:Appendix-C}. Next, we demonstrate how these Faddeev components ($F_{n,D}$) could be used to determine the 
structural characteristic of a plausible halo-bound $D^0nn$ system. In what follows, we discuss a simple correspondence 
between the effective quantum mechanical halo-EFT framework and the standard LO ${}^{\pi\!\!\!\!/}$EFT approach in the
analysis of the $D^0nn$ system~\cite{Raha:2017ahu}. However, for simplicity's sake, we only demonstrate this correspondence, 
excluding possible genuine three-body interactions in the Faddeev equations. Establishing a more rigorous EFT correspondence with
three-body potentials and wave functions included in the effective quantum mechanical halo-EFT formalism is rather 
intricate and, to our current knowledge, has only been formally established for a system of 
three-bosons~\cite{Platter:2004he,Meier:1983gsw}. In this work, we do not intend to pursue the rigorous correspondence 
between these frameworks for the $D^0nn$ system. Nevertheless, as we demonstrate subsequently, the inclusion of genuine 
three-body contact interactions (needed at LO in our case) is straightforwardly achieved in the ${}^{\pi\!\!\!\!/}$EFT 
framework in momentum-space, which we utilize to obtain our numerical results.

\subsection{Faddeev equations with sharp momentum cutoff: A LO ${}^{\pi\!\!\!\!/}$EFT connection}
\label{sec:2.2}
\begin{figure}[tbp]
\centering
\includegraphics[scale=0.5]{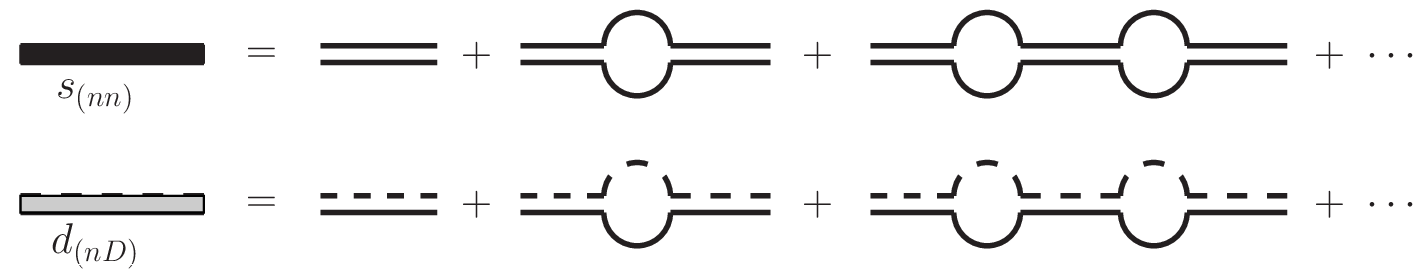}
    \caption{Renormalized dressed dimeron propagators associated with the $nn$ and $D^0n$ sub-systems. The solid
             (dashed) lines represent the neutron ($D^0$-meson) field propagator.}
\label{fig:dimerons} 
\end{figure}

In Ref.~\cite{Raha:2017ahu}, the S-wave $D^0nn$ system was investigated for the manifestation of the Efimov effect under the
idealized zero coupling limit ansatz that eliminated sub-threshold $DN$ ($T=1$) decay channels in the charm sector, e.g., 
$\pi^{-}\Lambda_{c},\, \pi^{0}\Sigma_{c}^{0}$ and $\pi^{-}\Sigma_{c}^{+}$. In this scenario, the $D^0nn$ system is pictured as 
a three-particle $2n$-halo cluster, with the so-called {\it Samba} configuration~\cite{Yamashita:2004pv}, having two 
real-bound $D^0n$ sub-systems and a virtual-bound $nn$ sub-system. The fact that the $D^0n$ sub-system in ZCL tends to develop
a large scattering length makes the $D^0nn$ system Efimovian in nature. Using a mathematical framework similar to the well-known 
${}^{{\pi}\!\!\!/}$EFT~\cite{Braaten:2004rn,Kaplan:1996xu,Kaplan:1998tg,Kaplan:1998we,Kaplan:1998sz,vanKolck:1998bw} at LO, 
qualitative features of three-body universality can be explored. The non-relativistic ${}^{\pi\!\!\!\!/}$EFT Lagrangian at
LO, consistent with low-energy symmetries (P, C, T, and Galilean invariances), is constructed as the sum of one-, two-, and 
three-body parts~\cite{Raha:2017ahu,Raha:2020sse}:
\begin{equation}
{\mathcal L} = \left[{\mathcal L}_n + {\mathcal L}_D\right]_{\rm 1-body} + 
\left[{\mathcal L}_{s(nn)} + {\mathcal L}_{d(nD)}\right]_{\rm 2-body} + {\mathcal L}_{\rm 3-body}\,, \quad \text{with}
\end{equation}
\begin{eqnarray}
{\mathcal L}_n(x) \!\!\!&=&\!\!\! n^\dagger(x) \left[i\partial_t + \frac{\nabla^2}{2M_n}\right]n(x) + \cdots\,,
\qquad
{\mathcal L}_D(x) = D^\dagger(x) \left[i\partial_t + \frac{\nabla^2}{2yM_n}\right] D(x) + \cdots\,,
\nonumber\\
{\mathcal L}_{s(nn)}(x) \!\!\!&=&\!\!\! -\, s_{(nn)}^\dagger(x) \left[i\partial_t + \frac{\nabla^2}{4M_n} \right] 
s_{(nn)}(x) - g_{2s}\left[s_{(nn)}^\dagger(x) \left(n^T(x)\, 
\hat{P}_{(nn)}^{(^1S_0)}\,n(x)\right)+ \mbox{\rm h.c.} \right] + \cdots\,,
\nonumber\\
{\mathcal L}_{d(nD)}(x) \!\!\!&=&\!\!\! -\, d_{(nD)}^\dagger(x) \left[i\partial_t + \frac{\nabla^2}{2M_n(1+y)} \right] 
d_{(nD)}(x) - g_{2d}\left[d_{(nD)}^\dagger(x)\, n(x)\, D(x) + \mbox{\rm h.c.} \right] + \cdots\,, 
\label{eq:L_12}
\end{eqnarray}
where the mass ratio, $y=M_D/M_n$, is introduced for brevity of notation, and
\begin{eqnarray}
\mathcal{L}_{\rm 3-body}(x) \!\!\!&=&\!\!\! -\, \frac{M_ng_3(\Lambda_{\rm reg})}{\Lambda_{\rm reg}^2} 
\Bigg[yg_{2d}^2\left\{d^T_{(nD)}(x)\, \hat{P}^{(^1S_0)}_{(nd)}\,n(x)\right\}^{\dagger} \left\{d^T_{(nD)}(x)\,\hat{P}^{(^1S_0)}_{(nd)}\,n(x)\right\}
\nonumber\\
&&\hspace{2.0cm}+\, 2g_{2s}g_{2d}\left\{d^T_{(nD)}(x)\, \hat{P}^{(^1S_0)}_{(nd)}\,n(x)\right\}^{\dagger} \left\{s_{(nn)}(x)\,D(x)\right\} 
+ {\rm h.c.}\Bigg] + \cdots \,,\quad\,
\label{eq:L_3BF}
\end{eqnarray}
where the ellipses are used to denote two- and three-body derivative interactions, which by the standard power-counting of 
${}^{{\pi}\!\!\!/}$EFT~\cite{Braaten:2004rn,Kaplan:1996xu,Kaplan:1998tg,Kaplan:1998we,Kaplan:1998sz,vanKolck:1998bw} 
contribute at subleading orders. It must also be emphasized here that by the same power-counting argument, the kinetic terms
corresponding to the dimeron fields are of higher order, and hence omitted in our expression for the three-body Lagrangian, 
Eq.~\eqref{eq:L_3BF}. As explicit degrees of freedom, we include the neutron $n(x)$ and $D^0$-meson $D(x)$ as the 
elementary fields in the theory whose kinetic parts constitute the one-body terms of the effective Lagrangian. In view of 
the possible existence of bound or anti-bound binary sub-systems, it is customary to introduce the auxiliary composite fields
called 
dimerons~\cite{Braaten:2004rn,Bedaque:1998mb,Bedaque:1998kg,Bedaque:1999ve,Bedaque:1998km,Birse:1998dk,Beane:2000fi,Ando:2004mm,Kaplan:1996nv}, 
namely the spin singlet $nn$ di-neutron field $s_{(nn)}(x)$, and the spin-doublet $D^0n$-dimer or dihadron field 
$d_{(nD)}(x)$.\footnote{The construction of the {\it genuine} three-body interactions in the ${}^{\pi\!\!\!\!/}$EFT
Lagrangian is quite straightforward using auxiliary fields. In contrast, the general methodology of including three-body 
interaction in the quantum mechanical halo-EFT framework is rather involved, as demonstrated in 
Refs.~\cite{Platter:2004he,Meier:1983gsw}.} The corresponding renormalized dressed propagators (cf. Fig.~\ref{fig:dimerons}) 
are given as
\begin{eqnarray}
iS^R_d\left(l_0,{\bf l}\right) \!\!\!&=&\!\!\! i\left[\frac{1}{a_{nD}}
-\sqrt{\frac{y}{y+1}\left(-2M_nl_0+\frac{{\bf l}^2}{y+1}-i0_+\right)}-i0_+ \right]^{-1}\,, \quad \text{and}
\nonumber\\
iS^R_s\left(l_0,{\bf l}\right) \!\!\!&=&\!\!\! i\left[\frac{1}{a_{nn}}
-\sqrt{-M_n l_0+\frac{{\bf l}^2}{4}-i0_+}-i0_+\right]^{-1}\,,
\label{eq:dimeron_prop1}
\end{eqnarray}
where $l\equiv(l_0,{\bf l})$ is the {\it lab}-frame four-momentum of the dimeron fields. Equivalently, in the three-body 
center-of-mass frame, the above propagators can be re-expressed in terms of the Jacobi momentum $q=|\bf q|$ and the total
three-body energy $z=E+i0_+$, after subtracting the corresponding kinetic energy of the spectator particles:
\begin{eqnarray}
iS^R_d\left(z-\frac{q^2}{2M_n},{\bf q}\right) \!\!\!&=&\!\!\! i\left[\frac{1}{a_{nD}}
-\sqrt{\frac{y}{y+1}\left(-2M_n z+\frac{y+2}{y+1}q^2-i0_+\right)}-i0_+ \right]^{-1}\,, \quad \text{and}
\nonumber\\
iS^R_s\left(z-\frac{q^2}{2yM_n},{\bf q}\right) \!\!\!&=&\!\!\! i\left[\frac{1}{a_{nn}}
-\sqrt{-M_n z+\frac{y+2}{4y}q^2-i0_+}-i0_+ \right]^{-1}\,.
\label{eq:dimeron_prop2}
\end{eqnarray}
The ${}^1S_0$ spin-projectors used in the Lagrangian are given in terms of the Pauli matrix $\sigma_2$:
\begin{eqnarray}
\hat{P}_{(nn)}^{(^1S_0)} = - i \frac{1}{2} \sigma_2\, \qquad \text{and} \qquad 
\hat{P}_{(nd)}^{(^1S_0)} = - i \frac{1}{\sqrt{2}} \sigma_2\,.
\end{eqnarray}
The couplings $g_{2s}$ and $g_{2d}$ denote the LO two-body contact interactions in the spin-singlet and spin-doublet channels, 
respectively, fixed as $g_{2s}=\sqrt{4\pi/M_n}$\cancel{,} and $g_{2d}=\sqrt{2\pi/\mu_{nD}}$. The quantities $a_{nn}$ and 
$a_{nD}$ constitute the only S-wave parameters in the LO two-body Lagrangian, where $a_{nn}=-18.63$~fm~\cite{Chen:2008zzj} 
corresponds to the phenomenologically extracted spin-singlet $n$-$n$ scattering length, and $a_{nD}= 4.141$~fm corresponds to 
the spin-doublet $n$-$D^0$ scattering length predicted in the ZCL model scenario~\cite{Raha:2017ahu,Raha:2020sse}. The latter
value was extracted from the dynamical coupled-channel model analysis with vanishing couplings to $D^0n$ sub-threshold decay 
channels. In particular, the negative signs in front of the kinetic operators for the composite dimer fields ensure that the 
corresponding two-body effective-range remains positive~\cite{Beane:2000fi,Ando:2004mm}. The three-body part of the LO 
effective Lagrangian (also referred to as the {\it three-body force} or 3BF for short) with the cutoff 
($\Lambda_{\rm reg}$)-dependent coupling, $g_3=g_3(\Lambda_{\rm reg})$, is introduced for renormalization of the two-body STM 
equations.\footnote{In Ref.~\cite{Raha:2017ahu} it was observed that the STM integral equations for the $D^0nn$ system with 
only two-body interactions ($g_{2s}$ and $g_{2d}$) become ill-defined in the UV limit. One way to remedy this malaise is to 
introduce a regulator in the form of a sharp momentum cutoff as the UV limit of the integral equations~\cite{Danilov:1961}. 
Concomitantly, scale-dependent {\it non-derivatively coupled} 3BF terms are included as a part of the LO Lagrangian as 
required by RG invariance. The calculated low-energy observables are then guaranteed to be regulator-independent.} 
The value of $g_3$ which is {\it a priori} unknown in the theory is fixed {\it via} a three-body datum, e.g., one of the 
$D^0nn$ three-body level energies $B_3$ or the corresponding S-wave scattering length $a_3$. Unfortunately, there is no 
evidence that the $D^0nn$ system is a bound state, with neither $B_3$ nor $a_3$ currently known. Therefore, the 
predictability of the theory at present hinges solely on guesstimating these quantities based on plausibility arguments.
Once $g_3$ is fixed, low-energy observables, such as form factors and other geometrical features, are readily predicted. 

With the above-mentioned ingredients of our EFT framework, it is straightforward to use Feynman diagram techniques to obtain a
set of two inhomogeneous coupled STM integral equations for the $D^0nn$ system. These represent the multiple scattering series
in the elastic ($n+d_{(nD)}\to n+d_{(nD)}$) and inelastic ($D^0+s_{(nn)}\to n+d_{(nD)}$) scattering channels (denoted by the 
subscripts $a$ and $b$), as described in terms of the half-off-shell transition amplitudes, 
$\mathcal{T}_n(p,q;E)\equiv t_a(p,q;E)$ and $\mathcal{T}_D(p,q;E)\equiv \sqrt{2}t_b(p,q;E)$, 
respectively~\cite{Raha:2017ahu,Raha:2020sse}. However, to look for Efimov trimers, the corresponding homogeneous equations must
be solved for non-trivial eigenvalues $E=-B_3$ for a given finite UV cutoff $\Lambda_{\rm reg}$. The solutions correspond to
the poles in the scattering amplitudes with residues that factor into {\it dimensionless} spectator functions of momenta $p$ and
$q$:
\begin{eqnarray}
\mathcal{T}_n(p,q;E) \stackrel{E\,\to\, -B_3}{\approx} \frac{F_n(p)F_n(q)}{E+B_3}\,, \quad \text{and}\quad 
\mathcal{T}_D(p,q;E) \stackrel{E\,\to\, -B_3}{\approx} \frac{F_D(p)F_D(q)}{E+B_3}\,.
\label{eq:T_F_relation}
\end{eqnarray}
By invoking the preceding separable ansatz, the two-variable STM equations are reduced to a single-variable coupled integral 
equation (in the Jacobi momentum $q$) described by the spectator functions $F_{n,D}$.

In order to see the equivalence between the Faddeev-type and STM integral equations, we note that in LO halo-EFT the 
effective potentials of the form, Eq.~\eqref{eq:V_eff}, simply correspond to zero-range contact interactions that vanish for 
momenta larger than the cutoff scale $\Lambda_{\rm reg}$. With the regulator functions $\chi(p)$ set as the Heaviside 
step-function, the two-body T-matrix elements are readily evaluated analytically (cf. Appendix~\ref{sec:Appendix-B}). 
Consequently, the UV limit of loop integrations is replaced by the finite cutoff $\Lambda_{\rm reg}$, as defined by the 
step-function. Now, it is easy to check that for large values of the cutoff, $\Lambda_{\rm reg}\rightarrow \infty$, both 
the inverse tangent functions in the T-matrices [see Eq.~\eqref{eq:tn_tD_S-wave}] approach $\pi/2$. As a result, the T-matrices
reduce to the standard renormalized two-body scattering amplitudes proportional to the renormalized dressed dimeron propagators:
\begin{eqnarray}
\tau_n(q;B_3) \!\!\!&=&\!\!\! \frac{y+1}{\pi y M_n} \bigg[-\frac{1}{a_{nD}}+\gamma_n(q; B_3)\bigg]^{-1} 
\equiv -\frac{y+1}{\pi y M_n} S^R_d\left(-B_3-\frac{q^2}{2M_n},{\bf q}\right)\,, \quad \text{and}
\nonumber\\
\tau_D(q;B_3) \!\!\!&=&\!\!\! \frac{2}{\pi M_n} \bigg[-\frac{1}{a_{nn}}+\gamma_D(q; B_3)\bigg]^{-1} 
\equiv -\frac{2}{\pi M_n} S^R_s\left(-B_3-\frac{q^2}{2yM_n},{\bf q}\right) \,.
\label{eq:tn_tD_S-wave_cut-off}
\end{eqnarray}
Next, we focus on the coupled Faddeev integral equations in terms of the spectator functions, Eq.~\eqref{eq:F_nD}, with
the upper integration limits replaced by the finite cutoff $\Lambda_{\rm reg}$. In this case, the only 
$x\equiv \hat{\bf q}\cdot \hat{\bf q}^\prime$ dependence arises from the free three-body Green's functions 
$\mathcal{G}^{(i)}_0$, Eq.~\eqref{eq:G0_n_D_matrix_elements}. Integration over $x$ can be carried out analytically 
yielding an expression analogous to the {\it homogeneous} part of the STM equation (i.e., without the 3BF terms) for
the $D^0nn$ system obtained within LO ${}^{\pi\!\!\!\!/}$EFT~\cite{Raha:2017ahu,Raha:2020sse}:
\begin{eqnarray}
F_n(q;\Lambda_{\rm reg}) \!\!\!&=&\!\!\! -\,\frac{1}{\pi}\int_0^{\Lambda_{\rm reg}}{\rm d}{q^\prime} q^{\prime 2} 
\left[(y+1)\mathcal{K}_{(D)}(q,q^\prime;B_3)\,S^R_d\left(-B_3-\frac{q^{\prime 2}}{2M_n},{\bf q}^{\prime}\right)\,
F_n(q^\prime;\Lambda_{\rm reg})\right.
\nonumber\\
&&\left.\hspace{3cm} +\, 2\mathcal{K}_{(n)}(q,q^\prime;B_3)\,S^R_s
\left(-B_3-\frac{q^{\prime 2}}{2yM_n},{\bf q}^{\prime}\right)\,F_D(q^\prime;\Lambda_{\rm reg})\right]\,, \quad \text{and}
\nonumber\\
F_D(q;\Lambda_{\rm reg}) \!\!\!&=&\!\!\! -\,\frac{2(y+1)}{\pi y} \int_0^{\Lambda_{\rm reg}}{\rm d}{q^\prime} q^{\prime 2}
\left[\mathcal{K}_{(n)}(q^\prime,q;B_3)\,S^R_d\left(-B_3-\frac{q^{\prime 2}}{2M_n},{\bf q}^{\prime}\right)\,
F_n(q^\prime;\Lambda_{\rm reg})\right]\,,
\label{eq:F_nD_EFT}
\end{eqnarray}
where the two kernel functions are given by
\begin{eqnarray}
\mathcal{K}_{(D)}(q,q^\prime;B_3) \!\!\!&=&\!\!\! \frac{1}{2qq^\prime} 
\ln \bigg(\frac{ay M_n B_3+q^2+q^{\prime 2}+a qq^\prime}{ay M_n B_3+q^2+q^{\prime 2}-a qq^\prime}\bigg)\,;\quad  
a=\frac{2}{y+1}\,, \quad \text{and}
\nonumber\\
\mathcal{K}_{(n)}(q,q^\prime;B_3) \!\!\!&=&\!\!\! \frac{1}{2qq^\prime} 
\ln \bigg( \frac{M_n B_3+q^2+b q^{\prime 2}+qq^\prime}{M_n B_3+q^2+b q^{\prime 2}-qq^\prime}\bigg)\,; \quad
b=\frac{y+1}{2y}\,.
\label{eq:OPE}
\end{eqnarray}
In particular, the kernel functions follow a notation distinct from the spectator labeling used in the three-particle
Green's function $\mathcal{G}^{(i)}_0$, where $i$ represents the spectator particle index. In this convention, 
$\mathcal{K}_{(\gamma)}$ represents the S-wave projected interaction kernel corresponding to the freely propagating 
exchanged particle $\gamma=n,D$ between dimers in the two different re-arrangement channels. 

As demonstrated in our previous studies~\cite{Raha:2017ahu,Raha:2020sse}, the three-body system exhibits the Eﬁmov effect.
As a straightforward consequence of the unitarity in the two-body sector, ambiguities arise in the solutions to the integral 
equations~\eqref{eq:F_nD_EFT} when the finite regulator is removed, i.e., in the limit $\Lambda_{\rm reg}\to \infty$. 
Moreover, even restoring the finite cutoff, the solutions exhibit pronounced and unnatural sensitivity to the choice of 
regulator. The root cause of this ambiguity can be traced to the breakdown of the {\it continuous scale-invariance }
(conformality) of the integral equations into a {\it discrete} scaling symmetry leading to the onset of an RG limit cycle 
behavior. This is immediately revealed {\it via} the following {\it asymptotic analysis} of the integral equations. Considering
the limit where $\Lambda_{\rm reg}\to \infty \gg q,q^\prime \gg a^{-1}_{nD},\,a^{-1}_{nn},\, B_3$, we find that the integral 
equations simplify considerably:   
\begin{eqnarray*}
F_n(q) =  \frac{(1+y)^2}{2\pi\sqrt{y(2+y)}}\int^\infty_0 \frac{{\rm d}q^\prime}{q}
\ln\left(\frac{q^2 + q^{\prime 2} + aq q^\prime}{q^2 + q^{\prime 2} - aq q^\prime}\right) F_n(q^\prime)
+ \frac{1}{\pi}\sqrt{\frac{4y}{(2+y)}}\int^\infty_0 \frac{{\rm d}q^\prime}{q}
\ln\left(\frac{q^2 + bq^{\prime 2} + q q^\prime}{q^2 + bq^{\prime 2} - q q^\prime}\right) F_D(q^\prime)\,, 
\end{eqnarray*}
and
\begin{eqnarray}
F_D(q) =  \frac{(1+y)^2}{\pi y \sqrt{y(2+y)}}\int^\infty_0 \frac{{\rm d}q^\prime}{q}\ln\left(
\frac{q^{\prime 2} + bq^2 + q q^\prime}{q^{\prime 2} + bq^2 - q q^\prime}\right) F_n(q^\prime)\,.
\end{eqnarray}
Because the above asymptotic forms are scale-free, both spectator functions must exhibit a power-law scaling behavior 
in terms of a {\it single} unknown exponent $s$, i.e., $F_i(q)\propto q^{s-1}$~\cite{Bedaque:1998km}. This follows from
the fact that the functions are proportional to one another and satisfy the same system of homogeneous equations. Next,
{\it via} a Mellin transformation~\cite{Griesshammer:2005sj}, the integral equations reduce to a single 
{\it transcendental} equation, which can then be solved numerically to determine the exponent $s$, namely 
\begin{eqnarray}
1 = \frac{(1+y)^2}{2\pi\sqrt{y(2+y)}} I_0(s;a) + \frac{2(1+y)^2}{\pi^2 y(2+y)} I_1(s;b)I_2(s;b)\,,
\label{eq:trans}
\end{eqnarray}
where the asymptotic integrals $I_{0,1,2}$ are given by
\begin{eqnarray}
I_0(s;a) \!\!\!&=&\!\!\! \int^\infty_0 \rm dx\,\, x^{s-1}\ln\left(\frac{1 + x^2 + ax}{1 + x^2 -ax}\right)
= \frac{2\pi}{s}\frac{\sin\left[ s\sin^{-1}\left(\frac12a\right)\right]}{\cos\left(\frac{\pi}{2}s\right)}\,,
\nonumber\\
I_1(s;b) \!\!\!&=&\!\!\! \int^\infty_0 \rm dx\,\, x^{s-1}\ln\left(\frac{1 + bx^2 + x}{1 + bx^2 - x}\right)
= \frac{2\pi}{s} \frac{1}{b^{s/2}}
\frac{\sin\left[s\cot^{-1}\left(\sqrt{4b-1}\right)\right]}{\cos\left(\frac{\pi}{2}s\right)}\,,  \quad \text{and}
\nonumber\\
I_2(s;b) \!\!\!&=&\!\!\! \int^\infty_0 \rm dx\,\, x^{s-1}\ln\left(\frac{b + x^2 + x}{b + x^2 - x}\right)
= \frac{2\pi}{s} b^{s/2}
\frac{\sin\left[s\cot^{-1}\left(\sqrt{4b-1}\right)\right]}{\cos\left(\frac{\pi}{2}s\right)}\,.
\end{eqnarray}
Solving the transcendental equation~\eqref{eq:trans}, yields imaginary values of the parameter $s$, i.e., 
$s=\pm is^{\infty}_0$, with $s^{\infty}_0=1.02387...$ being a transcendental number. Thus, we infer that the $D^0nn$ system 
exhibits  RG limit cycle , formally indicating its Efimovian character. It is notable 
that the transcendental number $s^{\infty}_0$ that we obtain here is slightly larger than the well-known universal 
transcendental number $s_0=1.00624...$, as obtained for S-wave systems  of three identical bosons or
nucleons (such as the triton)~\cite{Braaten:2004rn,Bedaque:1999ve,Bedaque:1998km}. The rationale for this difference lies in
the  mass ratio, $y=M_D/M_n=1.984...$, which in the present case exceeds unity, as is typical 
for systems involving identical bosons or fermions. Since it is well-known that a larger mass imbalance between the bound 
particles enhances Efimov attraction and favors three-body universality~\cite{Braaten:2004rn,Naidon:2016dpf}, it is 
conceivable that the larger value of the limit cycle parameter $s^{\infty}_0$ than what is exhibited in, e.g., triton, albeit
the small $n$-$D^0$ scattering length, might still favor the plausible existence of a {\it nearly} bound $D^0nn$ state. 

As a numerically convenient approach, employed in several of our previous EFT studies of Efimov 
universality~\cite{Ando:2015fsa,Meher:2020yjr,Meher:2020ark,Raha:2017ahu,Raha:2020sse}, we choose to renormalize the regulator
sensitivity by introducing an explicit regulator-dependent LO 3BF {\it via} the counterterm Lagrangian, Eq.~\eqref{eq:L_3BF} 
(see, e.g., Refs.~\cite{Hammer:2000nf,Epelbaum:2016ffd} for alternative schemes of renormalization, where it does not necessitate
the introduction of three-body counterterms). Moreover, since the coupled integral equations are reduced to a single integral 
equation, only one three-body coupling parameter $g_3$ is sufficient to renormalize our results. It is then straightforward to 
incorporate the 3BF terms {\it via} the modification of the single-particle exchange kernel functions, $\mathcal{K}_{(D)}$ and
$\mathcal{K}_{(n)}$, into their respective renormalized counterparts, $\mathbb{K}^R_{(D)}$ and $\mathbb{K}^R_{(n)}$ [cf. 
Fig.~\ref{fig:F_n-diagram}]:
\begin{figure}[tbp]
\centering
\includegraphics[scale=0.5]{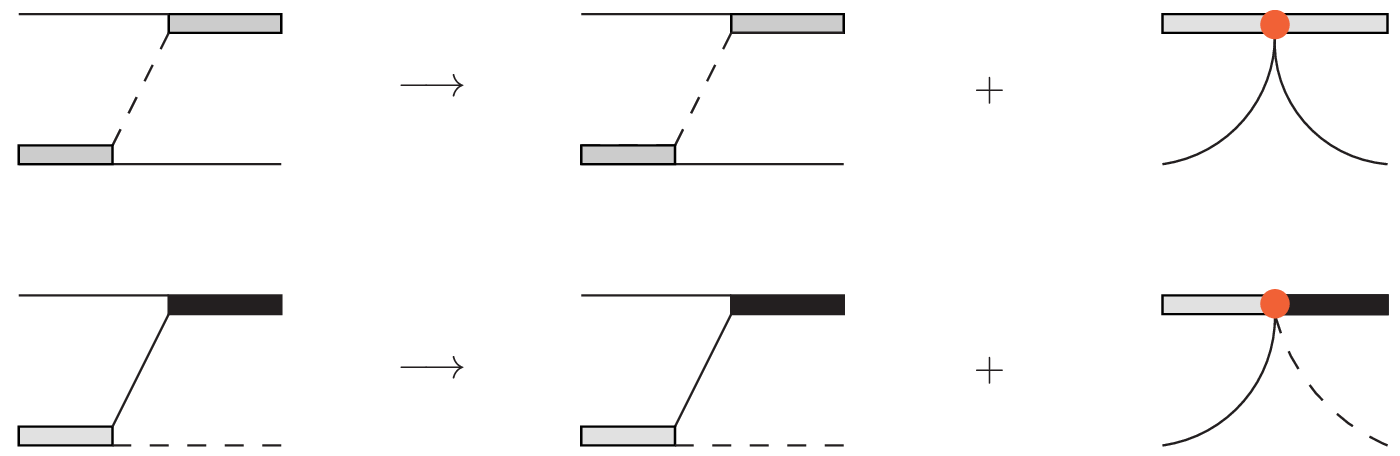}
    \caption{Feynman diagrams modifying the single $D^0$-meson and neutron exchange kernel functions, 
             $\mathcal{K}_{(D)}$ and $\mathcal{K}_{(n)}$, respectively, into their renormalized 
             counterparts, $\mathbb{K}^R_{(D)}$ and $\mathbb{K}^R_{(n)}$, contributing to the STM3 
             integral equations. The red-filled circles represent insertions of the regulator 
             ($\Lambda_{\rm reg}$) dependent three-body contact interactions with running coupling 
             $g_3=g_3(\Lambda_{\rm reg})$.}
\label{fig:F_n-diagram} 
\end{figure}
\begin{eqnarray}
\mathcal{K}_{(\gamma)}(q,q^\prime;B_3) \longrightarrow \mathbb{K}^R_{(\gamma)}(q,q^\prime,\Lambda_{\rm reg};B_3) 
\stackrel{!}{=} \mathcal{K}_{(\gamma)}(q,q^\prime;B_3)-\frac{g_3(\Lambda_{\rm reg})}{\Lambda^2_{\rm reg}}\,; \quad \gamma=n,D\,.
\label{ren_kernel}
\end{eqnarray}
Given the current paucity of available three-body data on $D^0nn$ to phenomenologically constrain the coupling $g_3$,
we  adopt a simplified strategy in which the regulator scale $\Lambda_{\rm reg}$ serves as the sole 
parameter governing the strength of the 3BF. Within this framework, we investigate the RG nature of the coupling. Such a
``minimalist'' scheme should suffice for the purposes of this exploratory EFT study, enabling us to capture 
the essential universal features of the $D^0nn$ system, without invoking more elaborate or higher-order EFT methodologies.
We then finally obtain the renormalized STM equations, the so-called {\it STM3 equations}~\cite{Braaten:2004rn} for the $D^0nn$ system:
\begin{eqnarray}
F^R_n(q) \!\!\!&=&\!\!\! -\,\frac{1}{\pi}\int_0^{\Lambda_{\rm reg}}{\rm d}{q^\prime} q^{\prime 2} 
\left[(y+1)\,\mathbb{K}^R_{(D)}(q,q^\prime,\Lambda_{\rm reg};B_3)\,
S^R_d\left(-B_3-\frac{q^{\prime 2}}{2M_n},{\bf q}^{\prime}\right)\,F^R_n(q^\prime)\right.
\nonumber\\
&&\left.\hspace{3.15cm} +\, 2\,\mathbb{K}^R_{(n)}(q,q^\prime,\Lambda_{\rm reg};B_3)\,
S^R_s\left(-B_3-\frac{q^{\prime 2}}{2yM_n},{\bf q}^{\prime}\right)\,F^R_D(q^\prime)\right]\,,\,\quad \text{and}
\nonumber\\
F^R_D(q) \!\!\!&=&\!\!\! -\,\frac{2(y+1)}{\pi y}\, \int_0^{\Lambda_{\rm reg}}\rm d{q^\prime} q^{\prime 2}
\left[\mathbb{K}^R_{(n)}(q^\prime,q,\Lambda_{\rm reg};B_3)\,
S^R_d\left(-B_3-\frac{q^{\prime 2}}{2M_n},{\bf q}^{\prime}\right)\,F^R_n(q^\prime)\right]\,.
\label{eq:F_nD_EFT_ren}
\end{eqnarray}

\begin{figure}[tbp]
\centering
\includegraphics[width=0.48\textwidth]{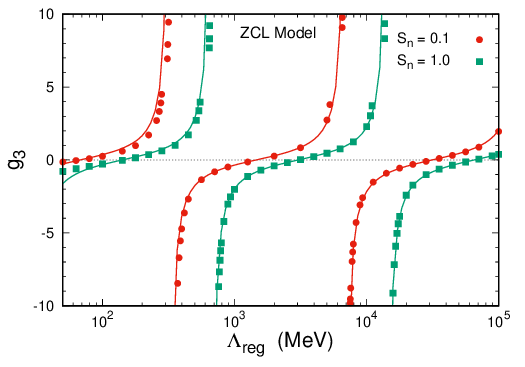} \hfill
\includegraphics[width=0.48\textwidth]{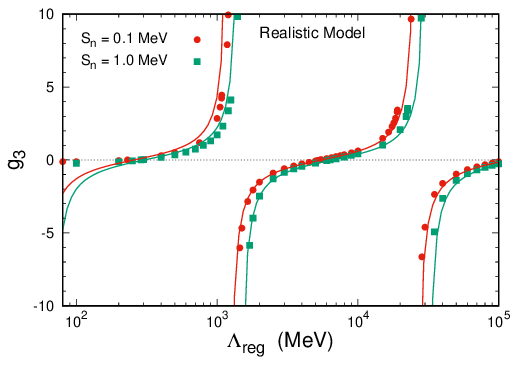}\\
\includegraphics[width=0.48\textwidth]{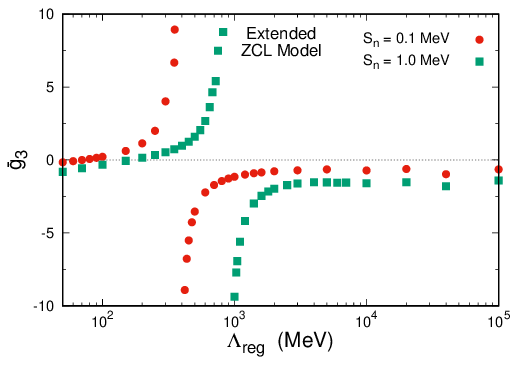}
    \caption{RG limit cycle for two choices of the most excited trimer neutron separation energy, namely 
             $S_n=0.1$~MeV and $1.0$~MeV (i.e., measured with respect to the $n+(D^0)$ particle-dimer break-up
             threshold energy $B_{nD}$). The dotted and square data points correspond to the numerical 
             solutions to the STM3 equations, while the solid lines denote their fits using the asymptotic 
             expression, Eq.~\eqref{3B_coupling_Eq}. {\bf Upper-left panel:} Results obtained in the ZCL model 
             scenario with a real-bound $D^0n$ sub-system with S-wave scattering length, $a_{nD}=4.141$~fm, and 
             $B_{nD}=1.82$~MeV. The fit parameter is obtained as $\Lambda_*=31.8$~MeV and $66.2$~MeV for 
             $S_n=0.1$ and $1.0$~MeV, respectively. {\bf Upper-right panel:} Results obtained for the 
             realistic model scenario with sufficiently shielded inelastic decay channels with a small but 
             real S-wave scattering length, ${\tilde a}_{nD}=0.764$~fm, and ${\tilde B}_{nD}=53.4$~MeV. The fit
             parameter is obtained as ${\tilde \Lambda}_*=120.0 $~MeV and $143.0 $~MeV for $S_n=0.1$~MeV and 
             $1.0$~MeV, respectively. {\bf Lower panel:} Data obtained within the extended ZCL model, 
             incorporating effective-range effects {\it via} logarithmic modifications of the STM 
             equations~\eqref{eq:F_nD_EFT}, as motivated by relativistic corrections in the analysis of
             Ref.~\cite{Epelbaum:2016ffd}. The simple modification shows that the limit cycle corresponding 
             to $\bar{g}_3$ gets driven to a well-defined UV limit. }
\label{fig:limitcycle} 
\end{figure}

The characteristic RG limit cycle behavior, arising from the residual discrete scaling symmetry of the STM3 equations, 
is manifested in the asymptotic log-periodic running of the 3BF coupling $g_3=g_3(\Lambda_{\rm reg})$, with a periodicity
set by the universal factor $\lambda^\infty_0=\exp(\pi/s^{\infty}_0)=21.5064...$. The resulting cyclic singularities are well 
described by the analytical expression~\cite{Bedaque:1998mb,Bedaque:1998kg,Bedaque:1999ve,Bedaque:1998km}
\begin{eqnarray}
g_3(\Lambda_{\rm reg}) \sim \aleph\, \frac{\sin\left[s^{\infty}_0\ln(\Lambda_{\rm reg}/\Lambda_*) 
- \arctan(1/s^{\infty}_0)\right]}{\sin\left[s^{\infty}_0\ln(\Lambda_{\rm reg}/\Lambda_*) + \arctan(1/s^{\infty}_0)\right]}\,,
\label{3B_coupling_Eq}
\end{eqnarray}
where $\Lambda_*$ is a three-body parameter determined by a single low-energy observable, such as the three-body binding energy
$B_3$. The numerical pre-factor $\aleph=0.870$ is introduced to improve the overall fit to the non-asymptotic data points. 
Interestingly, this multiplicative constant is numerically close to the corresponding value, 
$\aleph=0.879$~\cite{Braaten:2011sz,Ji:2015hha}, obtained for a system of three identical bosons, suggesting its universal 
character as a generic low-energy constant. In the absence of any binding information on the $D^0nn$ system, we may 
assume some reasonable near-threshold values of the three-body binding energy $B_3$, or equivalently, the corresponding neutron
separation energy $S_n$, for the shallowest (most excited) level state, say, $S_n=B_3-B_{nD}=0.1$~MeV and $1.0$~MeV (i.e., 
measured relative to the $n+(D^0n)$ particle-dimer break-up threshold energy $B_{nD}=1.82$ MeV). This information is sufficient
to determine the regulator dependence of three-body coupling $g_3$ by solving the STM3 integral equations in the non-asymptotic
kinematical domain. Figure~\ref{fig:limitcycle} (left upper panel) displays the typical cyclic singularities associated with the
successive formation of three-body bound states, e.g., the $D^0nn$ ground state along with its first two excited levels in the 
ZCL scenario. In addition, to obtain a rough estimate of the theoretical uncertainty associated with neglecting the 
sub-threshold $DN$ ($J=1/2,\, T=1$) decay channels in the ZCL, we emply a so-called ``realistic'' $D^0nn$ dispersive 
scattering model, whose RG limit cycle is depicted in the same figure (upper-right panel). Here, the same inputs, namely 
$S_n=0.1$~MeV and $1.0$~MeV (now measured relative to the corresponding $n+(D^0n)$ particle-dimer break-up threshold energy, 
$B_{nD}=53.4$~MeV), are used to obtain the numerical estimates. Characterized by a small but {\it real-valued} $n$-$D^0$ S-wave 
scattering length, $\tilde{a}_{nD}=0.764$~fm, this simplified scattering model is intended to emulate the essential 
dispersive effects, despite the presence of distant inelastic hadronic decay channels 
.\footnote{\label{ft:scattering_length}The scattering length in this scenario is chosen as the real 
part of the extracted value, $a^{\rm (WT)}_{nD}=0.764-i0.615$~fm, as predicted {\it via} the {\it Weinberg-Tomozawa} (WT) SU(4) 
chiral unitary model analysis, where the $D^0n$ sub-system is identified as quasi-bound~\cite{Raha:2017ahu}. It is 
noteworthy that our EFT framework in the three-body sector {\it per se} is not formulated to accommodate the dynamics 
of inelastic decay channels characterized by complex scattering lengths. Alternatively, by explicitly incorporating 
additional dynamical hadrons in coupled open channels, such as $\pi^{-}\Lambda^{+}_{c}$, $\pi^{0}\Sigma_{c}^{0}$, and 
$\pi^{-}\Sigma_{c}^{+}$, one can, in principle, recover a real-valued $n$-$D^0$ scattering length. Since such a 
treatment would significantly complicate the otherwise simple EFT framework, we instead adopt a minimal approach in which
only the real part of the scattering length, namely} $\tilde{a}_{nD}={\rm Re}\left[a^{\rm (WT)}_{nD}\right]=0.764$~fm is retained to
parametrize the dispersive effects of the decay channels in the elastic scattering region. This approximation is justified by
the fact that the open channels are sufficiently far removed from the $n$-$D^0$ threshold. A comparison between the two 
limit cycles is expected to highlight the qualitative differences between our ZCL results and those anticipated in a more
realistic scenario. 

In particular, the RG limit cycle obtained in the LO indicates the flexibility to choose the regulator scale 
$\Lambda_{\rm reg}$ in a suitable way that the 3BF terms effectively vanish~\cite{Hammer:2000nf}. In other words, we may 
prefer to simply drop the 3BF terms and work with the (unrenormalized) STM equations~\eqref{eq:F_nD_EFT} by choosing a 
finite value of the cutoff that corresponds to one of the zeros of the RG limit cycle. Figure.~\ref{fig:FD_Fn} in the 
Appendix~\ref{sec:Appendix-C} demonstrates this feature, where we obtain regulator-independent solutions $F_{n,D}$ to the
STM equations precisely at the first three zeros of the RG limit cycle, namely $\Lambda^{(0)}_{\rm reg}=67.87$~MeV, 
$\Lambda^{(1)}_{\rm reg}=1472.86$~MeV and $\Lambda^{(2)}_{\rm reg}=31467.28$~MeV, for the ZCL scenario with separation 
energy $S_n=0.1$~MeV. The lowest zero, namely $\Lambda_{\rm reg}=\Lambda^{(0)}_{\rm reg}$, is identified with the 
(inverse) interaction range, which sets the scale for the ground state or the deepest possible three-body binding energy
of the system. Beyond this scale, all deeper states get decoupled from the spectrum. 

It is important to emphasize that the result shown in Fig.~\ref{fig:limitcycle} emerges in the strict zero-range limit 
(LO) of our EFT, where the S-wave two-body scattering length $a_0$ represents the only relevant low-energy scale, while 
the effective range $r_0$ is set to zero. In this limit, the three-body renormalization group flow exhibits an ultraviolet 
limit cycle, leading to the characteristic geometrically spaced tower of Efimov states. This behavior is closely 
associated with the well-known {\it Thomas collapse}~\cite{Thomas:1935zz}, reflecting the onset of {\it exact} discrete 
scale invariance of the zero-range theory. In realistic systems, however, the two-body interaction has a finite range, or 
equivalently a non-zero effective-range $r_0$, which introduces an additional short-distance scale. As demonstrated by 
Epelbaum {\it et al.}~\cite{Epelbaum:2016ffd}, incorporating relativistic corrections in a three-boson system within the 
framework of {\it time-ordered perturbation theory} (TOPT), followed by a non-relativistic reduction using the standard 
$1/M$-expansion, effectively generates two-body ``range-like'' corrections. Importantly, this contribution is fully 
determined by the reduced masses of the interacting particles. Such a modification significantly suppresses the RG limit 
cycle beyond a characteristic cutoff scale, $\Lambda_{\rm reg}\sim 1/r_0$, leading to a well-defined ultraviolet convergent
three-boson spectrum. Since the resulting Efimov spectrum is a direct consequence of this limit cycle behavior, 
finite-range effects break the exact discrete scale invariance and restrict Efimov universality to a finite number of 
shallow states with binding momenta $\kappa_n r_0 \ll 1$, while deeper states become sensitive to the interaction range 
and decouple from the spectrum. Quantitatively, the maximum number of Efimov states supported by a finite 
interaction range can be estimated as $N_{\rm max} \simeq (s_0^\infty/\pi)\,\ln(a_0/r_0)$~\cite{HammerPlatter:2007}. Hence, 
our LO results provide only the universal zero-range description for the $D^0nn$ system. In a realistic setting, however, 
finite-range effects are expected to truncate the Efimov tower of states and modify the ultraviolet behavior of the 
spectrum. Within the halo-EFT framework employed here, the effective-range parameter enters at next-to-leading order (NLO), 
thereby allowing such finite-range corrections to be incorporated in a controlled manner. Such a systematic NLO analysis 
that includes the perturbative effect of effective-range corrections is beyond the current scope and will be pursued in 
future work.

Nevertheless, to gain  qualitative insight into the range corrections in the $D^0nn$ system, it is 
instructive-following the approach of Epelbaum {\it et al.}~\cite{Epelbaum:2016ffd}—to examine potential modifications to 
the RG limit cycle of $g_3$. Such modifications could arise from a typical $1/M$-expansion of the corresponding 
Lorentz-invariant counterpart of our unrenormalized LO STM equations~\eqref{eq:F_nD_EFT}. Without explicitly deriving the
relativistic TOPT integral equations for the $D^0nn$ system, one can instead mimic the leading effects, as suggested in 
Ref.~\cite{Epelbaum:2016ffd}, by modifying equations~\eqref{eq:F_nD_EFT}. In effect, this amounts to including a factor
$1/\left(1+\ln(1+q/M_n)\right)$ multiplying the di-neutron propagator $iS^R_s$, and a factor 
$1/\left(1+\ln(1+q/2\mu_{nD})\right)$ multiplying the $D^0n$-dimer propagator $iS^R_d$ in the STM 
equations~\eqref{eq:F_nD_EFT}, thus inducing range-like modifications in the two-body sub-systems, governed by the scales
$\rho_{\rm s}\sim 1/M_n$ and $\rho_d\sim 1/(2\mu_{nD})$.\footnote{The range-like scales $\rho_s=0.21$~fm and 
$\rho_d=0.158$~fm are, in general, distinct from the true S-wave two-body effective range parameters, $r^{(nn)}_0$ and 
$r^{(nD)}_0$. For the ${}^{1}S_0$ $nn$ (di-neutron) sub-system, the S-wave effective-range has recently been extracted 
phenomenologically as $r^{(nn)}_0 = 2.87$~fm~\cite{Malone:2022gvp}. In contrast, for the $D^0n$ ($J=1/2,, T=1$) sub-system,
the effective-range remains poorly constrained due to the scarcity of experimental data. Nevertheless, an approximate 
estimate, $r^{(nD)}_0 \approx 0.4$~fm, has been suggested in Ref.~\cite{Bayar:2012dd}.} Figure~\ref{fig:limitcycle} (lower 
panel) illustrates the modified cutoff dependence of the corresponding three-body coupling $\bar{g}_3$ within our so-called 
``extended'' ZCL model of the $D^0nn$ system, after incorporating the logarithmic modifications. In this case, we 
obtain $N_{\rm max} \simeq (s_0^\infty/\pi)\,\ln(a_{nD}/\rho_d)= 1.06$. This estimate reflects the residual universality of
the three-body hetero-nuclear system, which is governed primarily by the smallest two-body scattering length and its 
associated interaction range -- here, $a_{nD}$ and $\rho_d$, respectively. Evidently, the characteristic cyclic singularities
in $\bar{g}_3$ are suppressed beyond certain cutoff values, with the system displaying exactly one three-body bound state and
appearing to converge toward a well-defined limit. Thus, a simple modification of the integrands leads to marked improvement 
in the UV behavior of the integral equations, eliminating the need for an explicit 3BF.

\subsection{Geometrical assessment of an S-wave Efimovian $D^0nn$ system at LO}
\label{sec:2.3}
The LO halo-EFT formalism can be used to predict low-energy observables, such as the one- and two-body matter density form 
factors and their corresponding {\it root-mean-square} (rms) radii, in view of the plausible existence of a halo-bound 
$D^0nn$ system. These observables can help unravel remnant universal structural features to assess the Efimov character of the
three-body bound system. For this purpose, we need information on the full three-body wave function. Following the general
methodology outlined in Refs.~\cite{Canham:2008jd,DL-Canham:2009}, we reconstruct the $D^0nn$ three-body wave functions from 
the solutions to the spectator functions $F_n$ and $F_D$ from Eq.~\eqref{eq:Matrix_eq_Fn_FD}. These wave functions, in turn, 
are used to determine the above three-body observables. 

\begin{figure}[tbp]
\centering
\includegraphics[scale=0.45]{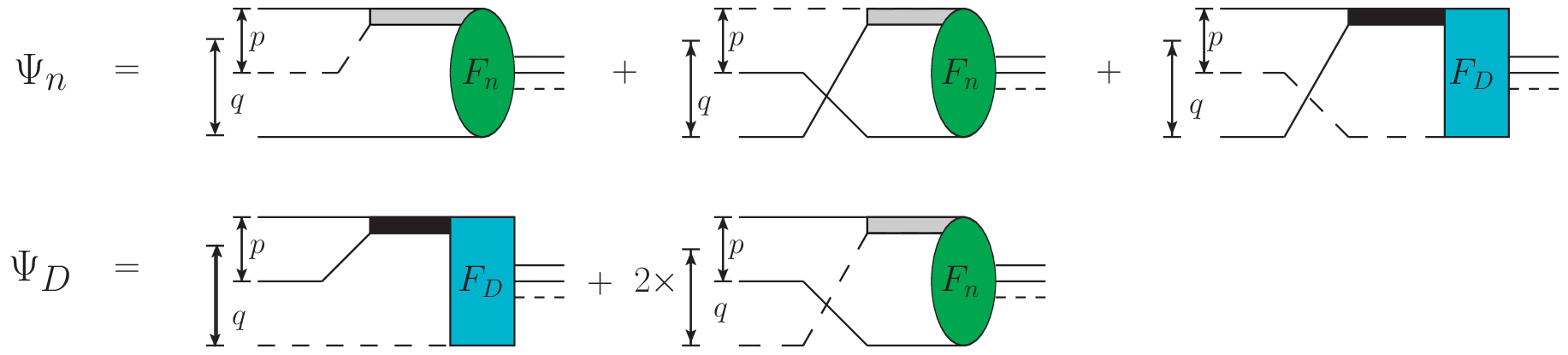}
    \caption{Reconstruction of the full three-body S-wave wave functions $\Psi_i(p,q)$  at LO with respect to the 
             $i=n,D$ fragmentation channels in Jacobi representation for the $D^0nn$ system. The solid/dashed 
             lines denote the neutron/$D^0$-meson propagators. The rectangular/oval blobs represent the three-body
             scattering kernels associated with the spectator functions $F_i(q)$. The gray/black thick-shaded 
             thick lines represent the iterated $n$-$D^0$/$n$-$n$ two-body S-wave T-matrices $\tau_i$, 
             Eq.~\eqref{eq:tn_tD_S-wave_cut-off}, which in the EFT framework are interpreted as renormalized 
             dressed auxiliary field (dimeron) propagators.}
\label{fig:wave function} 
\end{figure}

Projecting onto the appropriate S-wave spectator basis states, the two distinct three-body Jocobi wave functions (i.e., with 
respect to the two fragmentation channels labeled by $n$ or $D$) for the $D^0nn$ system (see Fig.~\ref{fig:wave function}) can
be reconstructed as\footnote{Here we recall that the binary sub-systems are assumed to be in the S-wave with vanishing
relative orbital quantum numbers, such that their total spins coincide with their intrinsic spins.}  
\begin{eqnarray}
\Psi_n(p,q) \!\!\!&=&\!\!\! \mathcal{G}^{(n)}_0(p,q;B_3)\bigg[ \chi(p)\, \tau_n(q;B_3)\, F_n(q)
 +\frac{1}{2}\int_{-1}^1 {\rm d}x \,\chi\left(\Pi_{nn}\right) \, \tau_n\left(\Pi^\prime_{nn};B_3\right)\, F_n\left(\Pi^\prime_{nn}\right)
\nonumber\\
&&\hspace{4.8cm} +\,\frac{1}{2}\int_{-1}^1 {\rm d}x \,
\chi\left(\Pi_{nD}\right)\, \tau_D\left(\Pi^\prime_{nD};B_3\right)\, F_D\left(\Pi^\prime_{nD}\right)\bigg]\,, \quad \text{and}
\nonumber\\
\Psi_D(p,q) \!\!\!&=&\!\!\! \mathcal{G}^{(D)}_0(p,q;B_3)\,\bigg[\chi(p)\, \tau_D(q;B_3)\,F_D(q) 
+ \int_{-1}^1 {\rm d}x\, \chi\left(\Pi_{Dn}\right) \, \tau_n\left(\Pi^\prime_{Dn};B_3\right) \, F_n\left(\Pi^\prime_{Dn}\right)\bigg]\,,
\label{eq:Psi_nD_pq}
\end{eqnarray}
where the following relations were used to obtain the above wave functions:
\begin{eqnarray}
\mathcal{G}^{(n)}_0(p,q;B_3)  \!\!\!&\equiv&\!\!\! \mathcal{G}^{(n)}_0(\Pi_{nn},\Pi^\prime_{nn};B_3) =
\mathcal{G}^{(D)}_0(\Pi_{nD},\Pi^\prime_{nD};B_3)\,,
\nonumber \\
 \mathcal{G}^{(D)}_0(p,q;B_3) \!\!\!&\equiv&\!\!\! \mathcal{G}^{(n)}_0(\Pi_{Dn},\Pi^\prime_{Dn};B_3)\,.
\end{eqnarray}
The expressions for the shifted three-momenta (with $x\equiv \hat{\bf p}\cdot \hat{\bf q}$) arising from re-couplings between 
the different spectator basis states are given by
\begin{eqnarray}
\Pi_{nn}(p,q;x) \!\!\!&=&\!\!\! \sqrt{\left(\frac{1}{y+1}\right)^2p^2 
+ \frac{y^2(y+2)^2}{(y+1)^4}q^2+2\frac{y(y+2)}{(y+1)^3}\,pqx}\,,
\nonumber\\
\Pi^\prime_{nn}(p,q;x) \!\!\!&=&\!\!\! \sqrt{p^2+\left(\frac{1}{y+1}\right)^2q^2-2\left(\frac{1}{y+1}\right)pqx}\,,
\nonumber\\
\Pi_{nD}(p,q;x)  \!\!\!&=&\!\!\! \sqrt{\frac{p^2}{4} + \frac{(y+2)^2}{4(y+1)^2}q^2+\frac{y+2}{2(y+1)}\,pqx}\,,
\nonumber\\
\Pi^\prime_{nD}(p,q;x)  \!\!\!&=&\!\!\! \sqrt{p^2+\left(\frac{y}{y+1}\right)^2q^2-2\left(\frac{y}{y+1}\right)\,pqx}\,,
\nonumber\\
\Pi_{Dn}(p,q;x) \!\!\!&=&\!\!\! \sqrt{\left(\frac{y}{y+1}\right)^2p^2+\frac{(y+2)^2}{4(y+1)^2}q^2+\frac{y(y+2)}{(y+1)^2}pqx}\,, 
\quad \text{and}
\nonumber\\
\Pi^\prime_{Dn}(p,q;x) \!\!\!&=&\!\!\! \sqrt{p^2+\frac{q^2}{4}-pqx}\,.
\label{eq:Pi_momenta_Dn}
\end{eqnarray}
Here, we employ the two-body T-matrices $\tau_n$ and $\tau_D$ from Eq.~\eqref{eq:tn_tD_S-wave}, together with the 
free three-body propagator functions $\mathcal{G}^{(n)}_0$ and $\mathcal{G}^{(D)}_0$ defined in 
Eq.~\eqref{eq:G0_n_D_matrix_elements}. The spectator functions are then obtained by solving the matrix equation, 
Eq.~\eqref{eq:Matrix_eq_Fn_FD}, evaluated either at the relative on-shell momentum $q$ or at a shifted off-shell momentum
$q^\prime=\Pi^\prime(p,q)$. The reconstruction of the LO wave functions $\Psi_i$ with respect to the two fragmentation channels
$i=n,D$, is illustrated using Feynman diagrams in Fig.~\ref{fig:wave function}. 

\begin{figure}[tbp]
\centering
\includegraphics[scale=0.7]{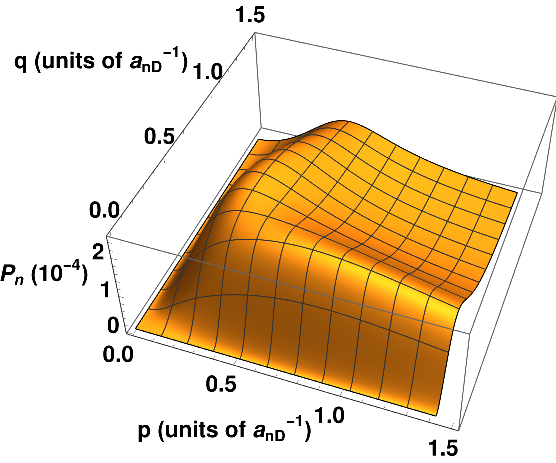} \quad \qquad
\includegraphics[scale=0.7]{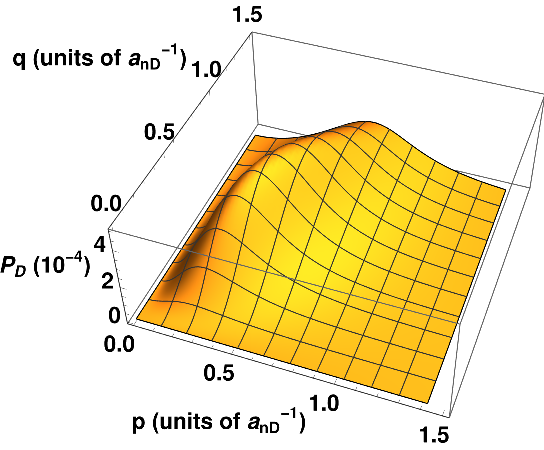}

\vspace{0.25cm}

\includegraphics[scale=0.7]{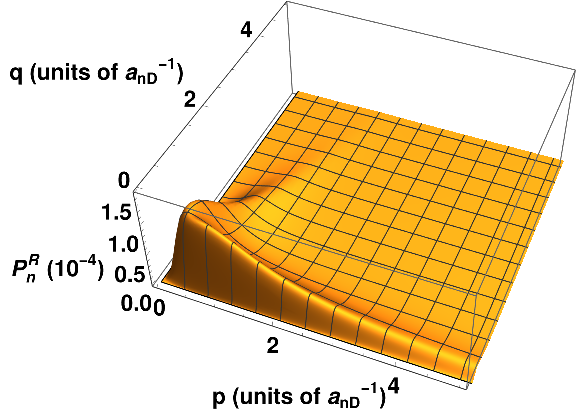} \quad \qquad
\includegraphics[scale=0.7]{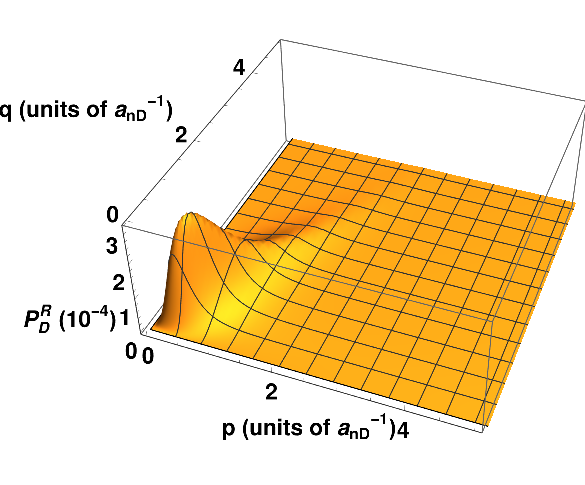}
\caption{Normalized momentum-space S-wave radial probability density, Eq.~\eqref{eq:rad_density}, for the  
         ground state wave functions $\Psi_n(p,q)$ and $\Psi_D(p,q)$ at LO in the idealized ZCL scenario 
         with input neutron separation energy $S_n=B_3-B_{nD}= 0.1$~MeV. {\bf Upper panel:} The 
         unrenormalized densities ($P_{n,D}$) at the cutoff scale $\Lambda_{\rm reg}\simeq 68$~MeV. 
         {\bf Lower panel:} The corresponding renormalized asymptotic densities ($P^R_{n,D}$) as 
         $\Lambda_{\rm reg}\to \infty$. The Jacobi momenta $p$ and $q$ are expressed in units of the 
         inverse S-wave $n$-$D^0$ scattering length or $D^0n$-dimer binding momentum, i.e., 
         $\gamma_{nD}\sim a^{-1}_{nD}=47.65$~MeV.} 
\label{fig:probability_density} 
\end{figure}

{\it Form Factors and rms Radii:} With the full three-body wave functions determined in Eq.~\eqref{eq:Psi_nD_pq}, we are now
in a position to determine the matter density form factors and their corresponding radii. In the Jacobi representation, the 
form factors are obtained by calculating the S-wave projected Fourier transforms of the corresponding matter densities 
with respect to the squares of the three-momentum transfer ${\bf k}$. They are represented by the normalized functions 
$\mathscr{F}_i(k^2=0)=1$ and $\mathscr{F}_{ni}(k^2=0)=1$, with $i=n,D$, depending on the fragmentation channel. The 
expression for the one-body matter density form factor is given by
\begin{equation}
\mathscr{F}_i (k^2) = \int {\rm d}p\, p^2\int {\rm d}q\, q^2 \int_{-1}^1 {\rm d}\xi\, 
\widetilde{\Psi}_i(p,q)\,\widetilde{\Psi}_i(p,\sqrt{q^2+k^2-2qk\xi})\,, \quad i=n,D\,,
\label{eq:one-body_FF}
\end{equation}
where $\xi\equiv \hat{\bf k}\cdot \hat{\bf q}$, and  $\widetilde{\Psi}_i(p,q)$ represents the {\it normalized} S-wave 
projected component of the full three-body wave function $\Psi_i({\bf p},{\bf q})$ in three-dimensions, namely
\begin{equation}
\widetilde{\Psi}_i(p,q) = \frac{1}{4\pi}\int {\rm d}\Omega_{\hat{p}} \int {\rm d}\Omega_{\hat{q}}\, 
\frac{\Psi_i({\bf p},{\bf q})}{\displaystyle{\sqrt{\int{\rm d}^3{\bf p} 
\int {\rm d}^3{\bf q}\, \Psi^2_i({\bf p},{\bf q})}}}\,, \quad i=n,D\,.
\end{equation}
For illustration, Fig.~\ref{fig:probability_density} displays the momentum-space radial probability densities as functions of 
Jacobi momenta $p$ and $q$, where the normalized probability densities are 
defined as 
\begin{equation}
P_i(p,q) = \frac{\displaystyle{p^2 q^2\,\Psi^2_i(p,q)}}{\displaystyle{\int{\rm d}{p}\, p^2 
\int {\rm d}{q}\, q^2 \, \Psi^2_i(p,q)}}\,, \quad i=n,D\,.
\label{eq:rad_density}
\end{equation}
Next, the expressions for the two-body matter density form factors are given by
\begin{eqnarray}
\mathscr{F}_{nD} (k^2) &=& \int {\rm d}p\, p^2\int {\rm d}q\, q^2 \int_{-1}^1 {\rm d}\zeta\, 
\widetilde{\Psi}_n(p,q)\,\widetilde{\Psi}_n(\sqrt{p^2+k^2-2pk\zeta},q)\,, \quad \text{and}
\nonumber \\
\mathscr{F}_{nn} (k^2) &=& \int {\rm d}p\, p^2\int {\rm d}q\, q^2 \int_{-1}^1 {\rm d}\zeta\, 
\widetilde{\Psi}_D(p,q)\,\widetilde{\Psi}_D(\sqrt{p^2+k^2-2pk\zeta},q)\,, 
\label{eq:two-body_FF_nD_nn}
\end{eqnarray}
where $\zeta\equiv \hat{\bf k}\cdot \hat{\bf p}$. The one- and two-body form factors provide the basis for extracting the 
universal geometrical characteristics of the S-wave $2n$-halo-bound $D^0nn$ system.

\begin{figure}[tbp]
\centering
\includegraphics[scale=0.65]{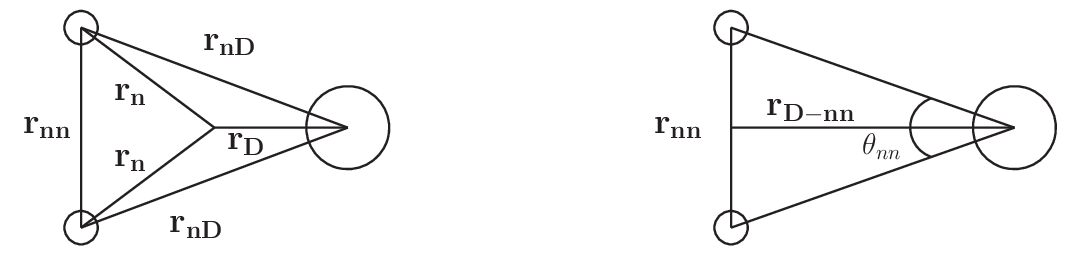}
    \caption{The matter radii defining the geometrical structure of a $2n$-halo-bound $D^0nn$ system.}
\label{fig:halo_matter_radii} 
\end{figure}

We can now extract the mean-square two-particle distances for pairs of constituent particles, namely $\langle r^2_{nn} \rangle$
and $\langle r^2_{Dn} \rangle$, as well as the mean-square distances between a spectator particle and the center-of-mass of the
corresponding binary sub-system, namely $\langle r^2_{n-nD} \rangle$ and $\langle r^2_{D-nn} \rangle$ (cf.
Fig.~\ref{fig:halo_matter_radii}). These  radii can be readily extracted from the 
slopes of the corresponding one- and two-body form factors at $k^2\equiv |{\bf k}|^2=0$, using the standard Taylor 
expansion formula: $\displaystyle{\langle r^2\rangle = -6 \left(\frac{{\rm d}\mathscr{F}(k^2)}{{\rm d}k^2}\right)_{k^2=0}>0}$. 
The slopes of $\mathscr{F}_{ni}$ give the information regarding the mean-square two-particle distances $\langle r^2_{nn}\rangle$ 
and $\langle r^2_{nD}\rangle$ for the $nn$ and $D^0n$ binary sub-systems respectively, while the slope of $\mathscr{F}_{n}$ 
($\mathscr{F}_{D}$) is related to the mean-square distance $\langle r^2_{n-nD}\rangle$ ($\langle r^2_{D-nn}\rangle$) between the
spectator neutron ($D^0$-meson) and the center-of-mass of the $D^0n$ ($nn$) sub-system. This information is also utilized to 
determine the mean-square distances $\langle r^2_i\rangle$ of each constituent particle from the $D^0nn$ 
barycenter, using the relation~\cite{Canham:2008jd}:
\begin{equation}
\langle r^2_i\rangle = -6 \left(\frac{{\rm d}\mathscr{F}_i(k^2)}{{\rm d}k^2}\right)_{k^2=0} 
\bigg[1-\frac{M_i}{(2+y)M_n}\bigg]^2\,,\; \quad i=n,D\,.
\label{eq:three-body_radius_i}
\end{equation}
Furthermore, a mean or effective geometrical matter radius can be defined for the $2n$-halo-bound $D^0nn$ system with a 
{\it point-like} core~\cite{Hammer:2017tjm}:
\begin{equation}
\langle r^2_{\rm eff}\rangle_{\rm 2n-halo} = 
\frac{2(y+1)^2}{(y+2)^3} \langle r^2_{n-nD}\rangle +\frac{4y}{(y+2)^3} \langle r^2_{D-nn} \rangle\,.
\label{eq:r_eff}
\end{equation}
Finally, through a simple geometrical analysis, one obtains the $n-D^0-n$ opening 
angle~\cite{Canham:2008jd,Acharya:2013aea}, given by
\begin{equation}
\theta_{nn}=2 \arctan\sqrt{\frac{\langle r^2_{nn} \rangle}{4\langle r^2_{D-nn} \rangle}}\,.
\label{eq:theta_nn}
\end{equation}
A numerical evaluation of the aforementioned observables is presented in the next section, allowing for the 
reconstruction of a two-dimensional geometrical representation of the halo-bound $D^0nn$ system.  

\section{ Results and Discussion}
\label{sec:results}
\begin{figure}[tbp]
\centering
\includegraphics[width=0.48\textwidth]{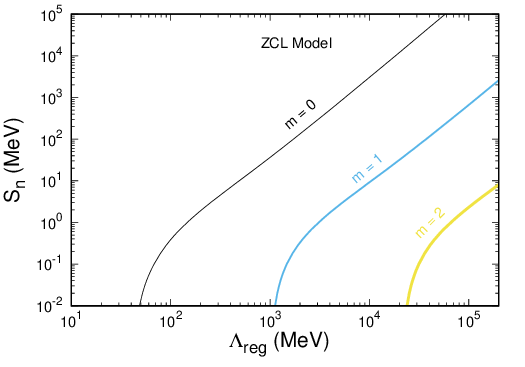} \hfill 
\includegraphics[width=0.48\textwidth]{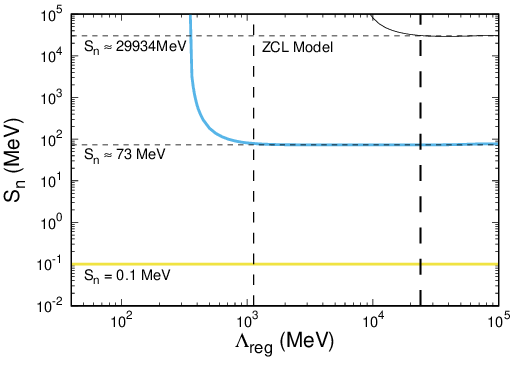}

\vspace{0.3cm}

\includegraphics[width=0.48\textwidth]{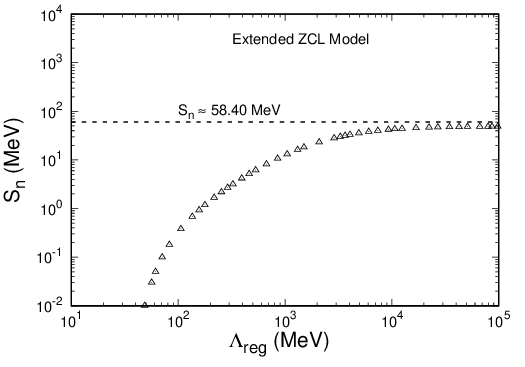} \hfill
\includegraphics[width=0.48\textwidth]{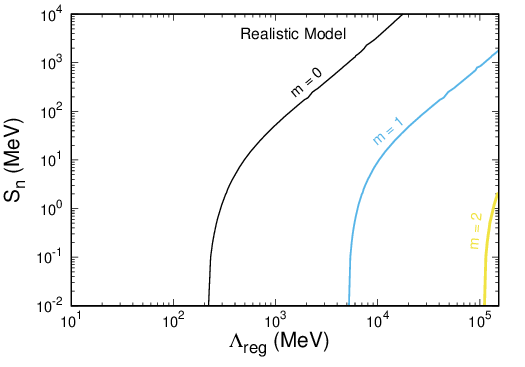}
    \caption{Cutoff regulator ($\Lambda_{\rm reg}$) dependence of the $D^0nn$ trimer binding energy $B_3$ or 
             equivalently, the corresponding neutron separation energy, $S_n=B_3-B_{nD}$, (where $B_{nD}$ is 
             $n+(D^0n)$ particle-dimer break-up threshold energy) obtained as a nontrivial solution to the STM 
             integral equations at leading order. Three sets of curves are displayed corresponding to the 
             ground ($m=0$) and the first two excited ($m=1,2$) trimer states. {\bf Upper-left panel:} Solutions 
             in the ZCL scenario with $a_{nD}=4.141$~fm and $B_{nD}=1.82$~MeV, obtained from the STM equations, 
             Eq.~\eqref{eq:F_nD_EFT}, excluding the 3BF terms. {\bf Upper-right panel:} Solutions in the ZCL 
             model scenario obtained from the normalized STM3 equations~\eqref{eq:F_nD_EFT_ren} including 
             the 3BF terms with coupling $g_3=g_3(\Lambda_{\rm reg})$ fixed {\it via} the RG limit cycle (cf. 
             Fig.~\ref{fig:limitcycle}). Assuming a threshold value of the separation energy, $S_n=0.1$~MeV, for
             the shallowest Efimov level, the regulator-independent energies, $S_n=73$~MeV and $S_n=29934$~MeV, 
             are yielded as formal solutions to the STM3 equations for the first excited and the ground Efimov 
             trimer states, respectively. The vertical lines denote the corresponding critical cutoffs, 
             $\Lambda^{(m=1)}_{\rm crit}\approx 1122$~MeV and $\Lambda^{(m=2)}_{\rm crit}\approx 23932$~MeV, at
             which the excited state appear. {\bf Lower-left panel:} Data obtained within the extended ZCL model, 
             incorporating effective-range effects {\it via} logarithmic modifications of the STM 
             equations~\eqref{eq:F_nD_EFT}, as motivated by relativistic corrections in the analysis of
             Ref.~\cite{Epelbaum:2016ffd}. {\bf Lower-right panel:} The unrenormalized neutron separation 
             energy in a plausible realistic scenario that implicitly includes inelastic effects of distant 
             sub-threshold hadronic decay channels, effectively contributing to a small but positive scattering 
             length, ${\tilde a}_{nD}=0.764$~fm. The corresponding break-up threshold energy is 
             ${\tilde B}_{nD}=53.4$~MeV.}
\label{fig:B_T_m_012} 
\end{figure}

\begin{table}[bp]
\begin{center}
\begin{tabular}{|c|c|c|}
\hline
Particle & Mass Symbol  & Numerical Value (MeV)\\
\hline\hline
$D^0$-meson   & $M_{D}$ & 1864.3  \\
Neutron ($n$) & $M_n$   & 939.565 \\
\hline\hline
\end{tabular}
\caption{PDG~\cite{Zyla:2020zbs} values of the $D^0$-meson and neutron masses used in our numerical calculations. }
\label{Tab:6.1} 
\end{center}
\end{table}
In this section, we display the LO numerical results pertaining to the S-wave $2n$-halo $D^0nn$ system. We focus mainly on the 
estimation of universal geometrical features for a plausible trimer state in the context of the idealized ZCL 
model~\cite{Raha:2017ahu,Raha:2020sse}. Here we prefer the sharp momentum cutoff regularization scheme, as incorporated {\it via}
a step-function in the T-matrices and wave function kernels. The numerical values of the masses of the particles used in our 
calculations are displayed in Table~\ref{Tab:6.1}. The sole two-body input parameters to our LO EFT are the S-wave scattering 
lengths $a_{nD}=4.141$ fm~\cite{Raha:2017ahu} and $a_{nn}=-18.63$~fm~\cite{Chen:2008zzj}. In Fig.~\ref{fig:B_T_m_012} 
(upper-left panel), we display the sequence of geometrically spaced Efimov states which progressively emerge above the $n+(D^0n)$ 
particle-dimer break-up threshold energy, $B_{nD}=1.82$~MeV, with increasing the cutoff scale $\Lambda_{\rm reg}$. The binding
energies ($B_3$) correspond to non-trivial solutions to the homogeneous STM equations~\eqref{eq:F_nD_EFT} without including the 
3BF terms. The critical cutoff values, $\Lambda_{\rm reg}=\Lambda^{(m)}_{\rm crit}$, associated with the three 
shallowest Efimov states are found to be $\Lambda^{(m=0)}_{\rm crit}\approx 48$~MeV, $\Lambda^{(m=1)}_{\rm crit}\approx 1122$~MeV,
and $\Lambda^{(m=2)}_{\rm crit}\approx 23932$~MeV. These results agree well with the LO EFT analysis of 
Ref.~\cite{Raha:2017ahu}. The small value of critical cutoff for the ground state relative to the EFT breakdown scale ($m_\pi$)
suggests a plausible Efimov-bound $D^0nn$ system, which may survive as a realistic exotic nucleus or evolve into a quasi-bound state
after relaxation of the ZCL ansatz. On the other hand, deeper levels emerging at cutoff scales substantially larger than 
$m_\pi$ are naturally excluded from the description of low-energy EFT. However, the sensitive dependence of the three-body 
binding energy on the regulator scale, as seen in Fig.~\ref{fig:B_T_m_012} (upper-left panel), can be renormalized by introducing 
the scale-dependent 3BF terms {\it via} the STM3 equations~\eqref{eq:F_nD_EFT_ren}. The neutron separation energy $S_n$ 
corresponding to the resulting renormalized three-body binding energy $B_3$ is displayed in the upper-right panel of 
Fig.~\ref{fig:B_T_m_012}. The corresponding 3BF running coupling, $g_3=g_3(\Lambda_{\rm reg})$, in the absence of any three-body 
data, is fixed using an {\it ad hoc} value of the neutron separation energy of the shallowest (most excited) Efimov level, say, 
$S_n=0.1$~MeV (i.e., $B_3=S_n+B_{nD}=1.92$~MeV) and then solving the STM3 equations [cf. Fig.~\ref{fig:limitcycle}] to produce the
RG limit cycle. Subsequently, regulator-independent values of all the deeper level energies can be readily obtained by solving the 
STM3 equations with $g_3$ fixed by the RG limit cycle. The result is illustrated in Fig.~\ref{fig:B_T_m_012} (upper-right panel), 
where the two deepest Efimov levels with regulator-independent separation energies, $S_n\approx 73$~MeV and $S_n\approx 29934$~MeV, 
appear, with the ratio of successive energies approaching the discrete scaling factor $\lambda^{\infty}_0= e^{2\pi/ s^\infty_0}$ 
as $\Lambda_{\rm reg}\to \infty$. However, these energies are far beyond the range of applicability of low-energy EFT.

To assess theoretical uncertainties in the idealized ZCL results, Fig.~\ref{fig:B_T_m_012} also displays the regulator 
dependence of the separation energy in the ``extended'' ZCL model with effective-range effects, together with the ``realistic'' 
dispersive scenario incorporating inelastic coupled-channel contributions about 200~MeV above the $DN$ threshold (cf. discussion
in Sec~\ref{sec:2.2}). In the former case (lower-left panel), the effect induced by the range-like scales, 
$\rho_{d}\sim 1/(2\mu_{nD})=0.158$~fm and $\rho_{s}\sim 1/M_n=0.21$~fm, washes out all but the ground Efimov state that emerges
from the threshold at the critical cutoff $\Lambda_{\rm crit}\approx 48$~MeV, as in the idealized scenario. The corresponding 
binding energy becomes progressively deeper and effectively cutoff independent for $\Lambda_{\rm reg}> 10^4$~MeV, asymptoting to
$S_n=58.4$~MeV, thereby obviating the need for a 3BF. Of course, such a large binding energy lies well outside the domain of 
applicability of the present low-energy EFT and, therefore, does not support a realistic interpretation of a bound state. 
Nevertheless, such a scenario highlights the sensitivity to range-like corrections -- particularly $\rho_d$ -- which could play 
a decisive role in the physical realization of such halo-bound states. For instance, treating $\rho_d$ as a free parameter and 
varying it within $\rho_d = 1.49\text{–}2.84$~fm, leads to a substantial reduction of the binding energy to the range 
$S_n\approx 0.1\text{–}1.0$~MeV. As for the latter case (lower-right plot), a small but positive value of the input $n$-$D^0$ 
S-wave scattering length is chosen, namely ${\tilde a}_{nD}=0.764$~fm, which then corresponds to the $n+(D^0n)$ particle-dimer 
break-up threshold energy, ${\tilde B}_{nD}=53.4$~MeV (cf. discussion in footnote~\ref{ft:scattering_length}). With such a 
value of the $n$-$D^0$ scattering length, the Efimov attraction in the $D^0nn$ system is substantially weakened in comparison 
to the ZCL scenario. Thus, the lowest critical cutoff corresponding to the ground Efimov state now appears around 
$\tilde{\Lambda}^{(m=0)}_{\rm crit}\gtrsim 200$~MeV. Since this value is close to the breakdown scale of the EFT, the prospect 
of a plausible Efimov-bound $D^0nn$ system becomes debatable, subject to miscellaneous realistic constraints that are not 
accounted for by our simplistic LO approach. Given that our approach is inherently universal, the observed qualitative 
differences between these model scenarios serve as a preliminary benchmark to test the consistency of more sophisticated future
model analyses or {\it ab intio} calculations.

\begin{figure}[bp]
\centering
\includegraphics[width=0.48\textwidth]{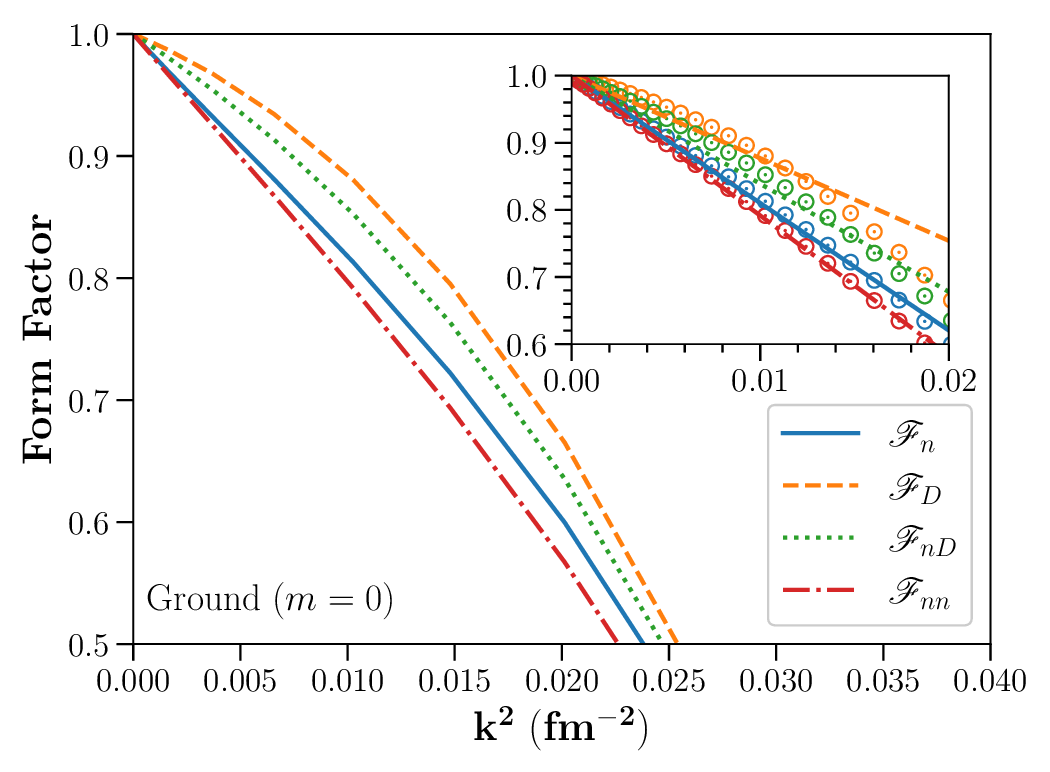}  \hfill 
\includegraphics[width=0.48\textwidth]{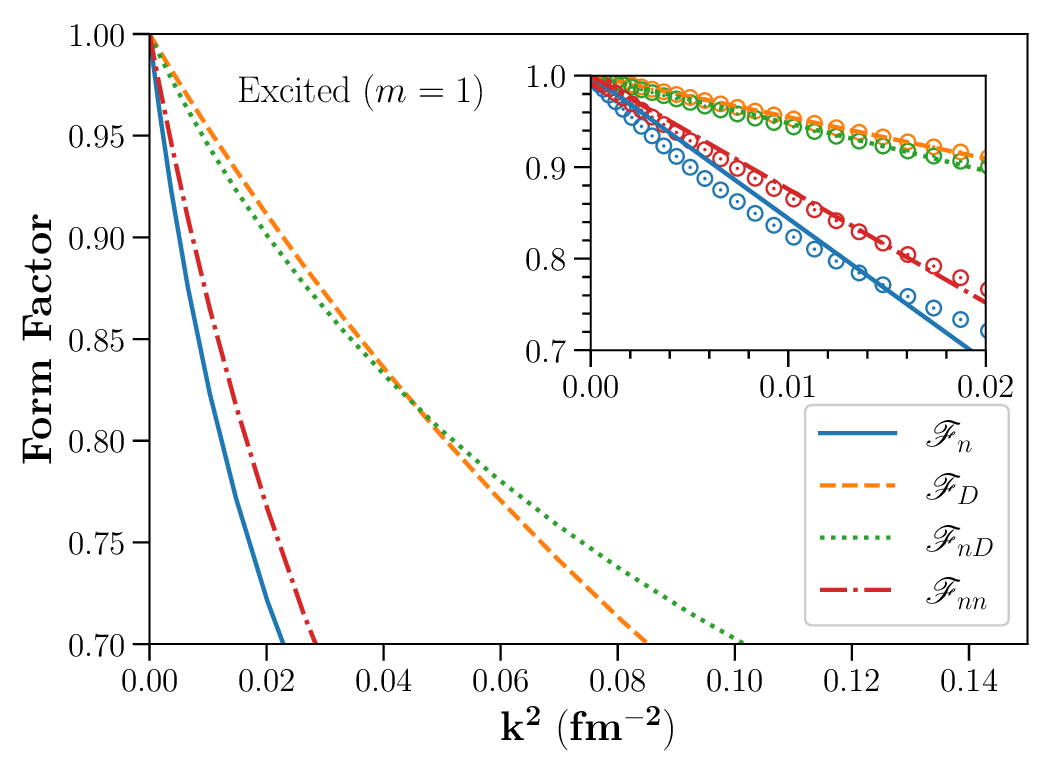} 

\vspace{0.3cm}

\includegraphics[width=0.49\textwidth]{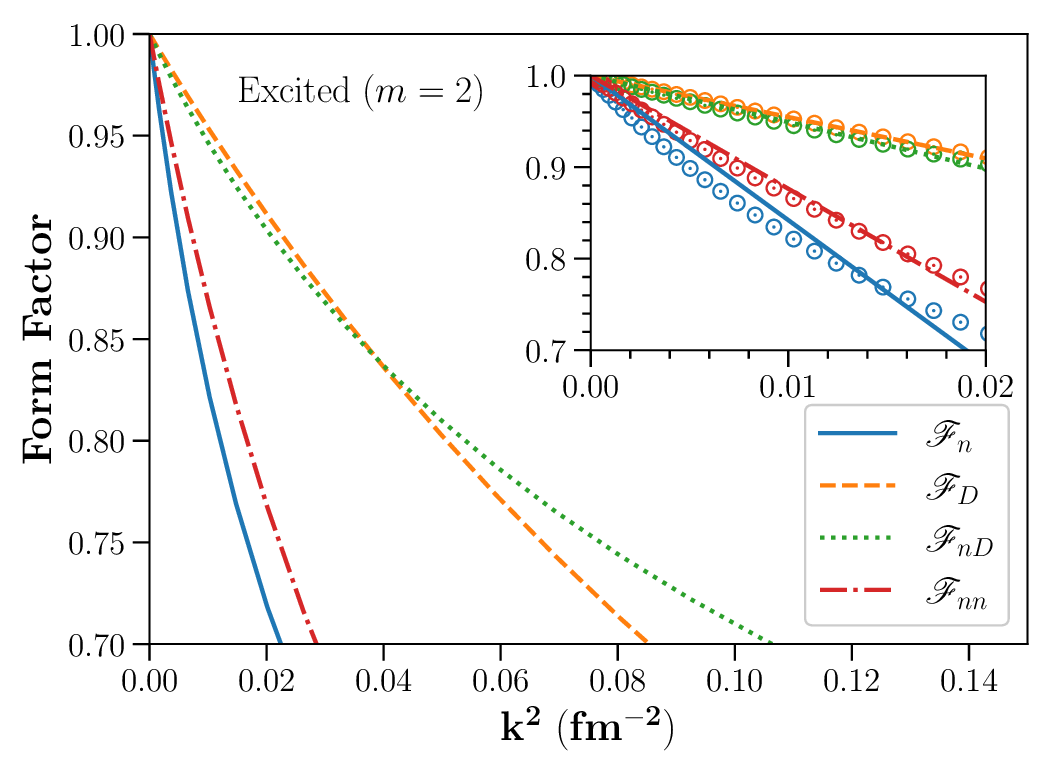}
    \caption{Leading order one- and two-body matter density form factors [see Eqs.~\eqref{eq:one-body_FF} and 
             \eqref{eq:two-body_FF_nD_nn}] as a function of the squared three-momentum transfer $k^2$ for the 
             ground state trimer ($m=0$, upper-left panel), and for the first and second excited state trimers
             ($m=1,2$, upper-right and lower panels). They are obtained using a small {\it ad hoc} input value 
             of the trimer neutron separation energy, $S_n= 0.1 $~MeV, which corresponds to the regulator values
             $\Lambda_{\rm reg} \approx 68$~MeV, $1.47$~GeV, and $31.47$~GeV for the ground ($m=0$), first 
             excited ($m=1$), and second excited ($m=2$) states, respectively. The results also correspond to 
             the two-body inputs, namely the spin-doublet $n$-$D^0$ S-wave scattering length $a_{nD}=4.141$~fm,
             extracted in the idealized ZCL model analysis of Ref.~\cite{Raha:2017ahu}, and the spin-singlet 
             $n$-$n$ S-wave scattering length $a_{nn}=-18.63$~fm, extracted phenomenologically in 
             Ref.~~\cite{Chen:2008zzj}. The form factors are normalized to unity at $k^2=0$. The inset plots 
             depict linear fits to numerical data points for very low $k^2$.}
\label{fig:form_factor_plots} 
\end{figure}

Next, we demonstrate our results for the halo-bound Samba structure~\cite{Yamashita:2004pv} of the S-wave $D^0nn$ system 
in the context of the idealized ZCL scenario only. In Fig.~\ref{fig:form_factor_plots}, we plot the various LO one- and 
two-body matter density form factors for the ground state ($m=0$) and the first two excited trimer states ($m=1,2$). 
Here, we present our numerical results by assuming a small {\it ad hoc} threshold value of the three-body binding energy,
say $B_3=1.92$~MeV (or equivalently, $S_n=0.1$~MeV) for each of the three shallowest Efimov states ($m=0,1,2$) beyond their 
respective critical cutoffs $\Lambda^{(0,1,2)}_{\rm crit}\approx 48~{\rm MeV},\, 1122~{\rm MeV}$ and $23932$~MeV). 
The excited states characterized by $S_n=0.1$~MeV are associated with regulator values 
$\Lambda_{\rm reg} \approx 1.47$~GeV ($m=1$) and $\Lambda_{\rm reg} \approx 31.47$~GeV ($m=2$). Owing to their large 
critical cutoffs, they are only marginally realizable as Efimov states (cf. upper-left panel of Fig.~\ref{fig:B_T_m_012}).
Regarding the ground state ($m=0$), the same binding energy corresponds to the regulator value 
$\Lambda_{\rm reg} \approx 68$~MeV. The non-linear behavior of the form factors is evident over a broad range of 
low-momentum {transfers, extending up to approximately} the pion mass ($k^2 \equiv |{\bf k}|^2 \lesssim m^2_\pi$). However,
in the $k\sim 0$ region, they can be approximated as linear with a constant slope, which in turn determines the 
corresponding mean squared radius. We find that the ground state form factors are significantly different from those of 
the excited states. Such differences in the behavior of the unrenormalized results (i.e., those obtained excluding the
3BF) are naturally anticipated, as only the very shallow trimer states with $S_n\sim 0$, and with regulator scales 
$\Lambda_{\rm reg}$ sufficiently exceeding $m_\pi$, fall within the domain of Efimov universality. With the binding 
momenta and regulator cutoffs of comparable size, our ground state results are likely to be plagued with poor numerical 
convergence for small regulator cutoffs. Hence, for a more robust prediction, it is desirable to compare these results with
a different regularization scheme, such as the Gaussian scheme. This would, however, require us to re-determine the 
corresponding RG limit cycle in the adopted regularization scheme, which, in principle, can exhibit qualitative differences
from our sharp cutoff scheme result displayed in Fig.~\ref{fig:limitcycle}. This comparative study shall be considered in 
future work.   

\begin{figure}[tbp]
\centering
\includegraphics[scale=0.9]{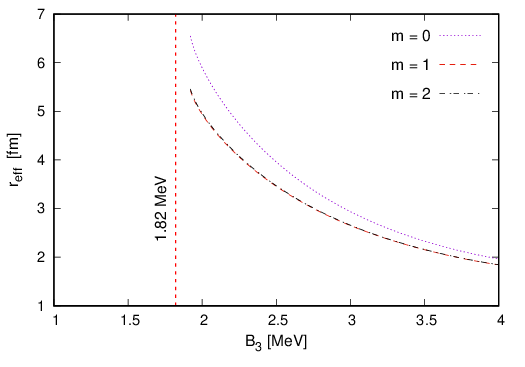}
    \caption{The variation of the unrenormalized effective or mean geometrical matter radius 
             $r_{\rm eff}\equiv \sqrt{\langle r^2_{\rm eff}\rangle_{\rm 2n-halo}}$ [see Eq.~\eqref{eq:r_eff}] 
             of the S-wave $2n$-halo-bound $D^0nn$ system for various input values of the trimer binding 
             energy $B_3$ for the three lowest Efimov states ($m=0,1,2$). The vertical dashed line denotes the 
             $n+(D^0n)$ particle-dimer break-up threshold energy $B_{nD}=1.82$~MeV, that corresponds to the 
             spin-doublet S-wave scattering length $a_{nD}=4.141$~fm extracted in the idealized ZCL model 
             analysis of Ref.~\cite{Raha:2017ahu}.}
\label{fig:radii_plot} 
\end{figure}

\begin{figure}[tbp]
\centering
\includegraphics[width=0.49\textwidth]{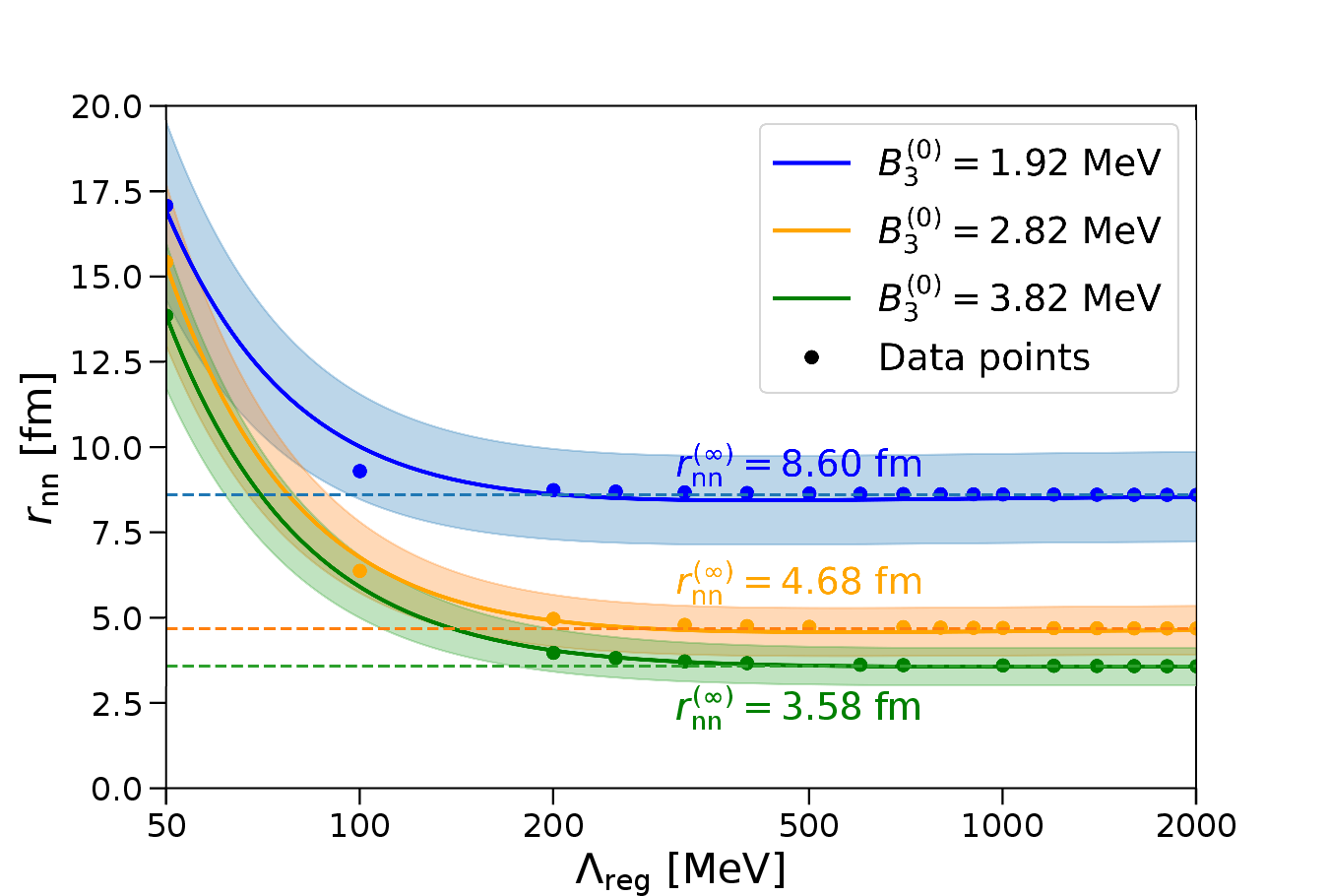} \hfill 
\includegraphics[width=0.49\textwidth]{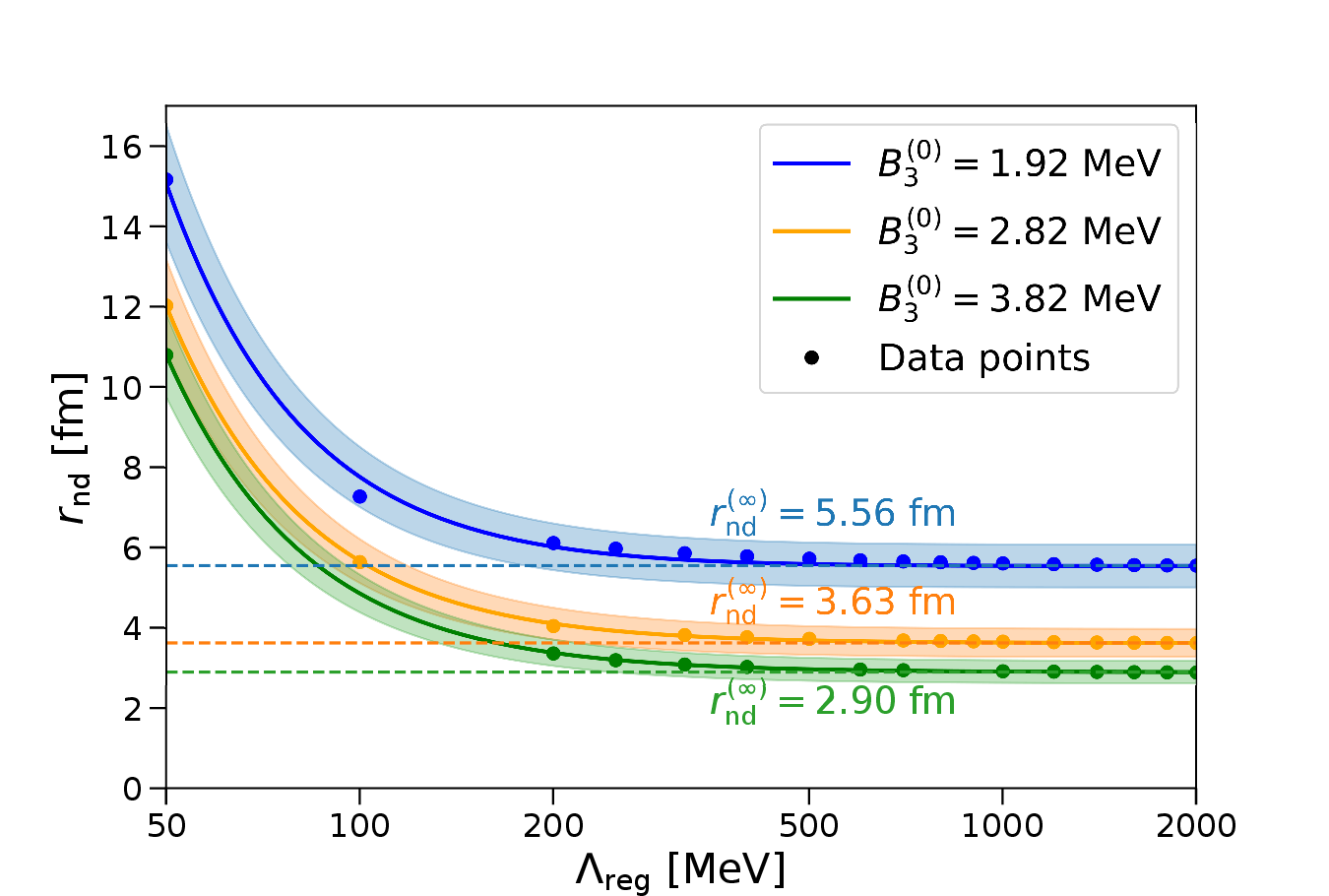}

\vspace{0.3cm}

\includegraphics[width=0.49\textwidth]{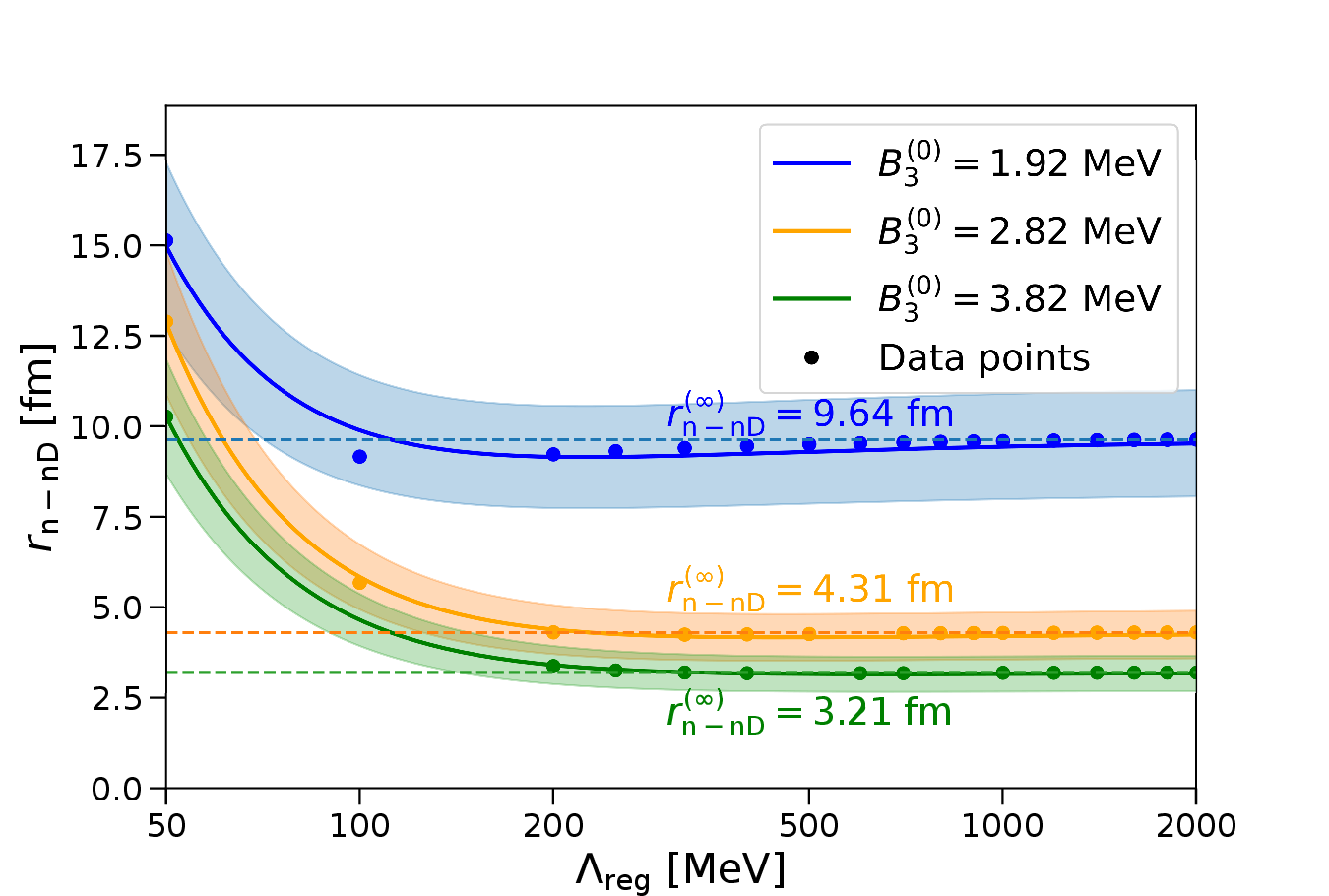} \hfill
\includegraphics[width=0.49\textwidth]{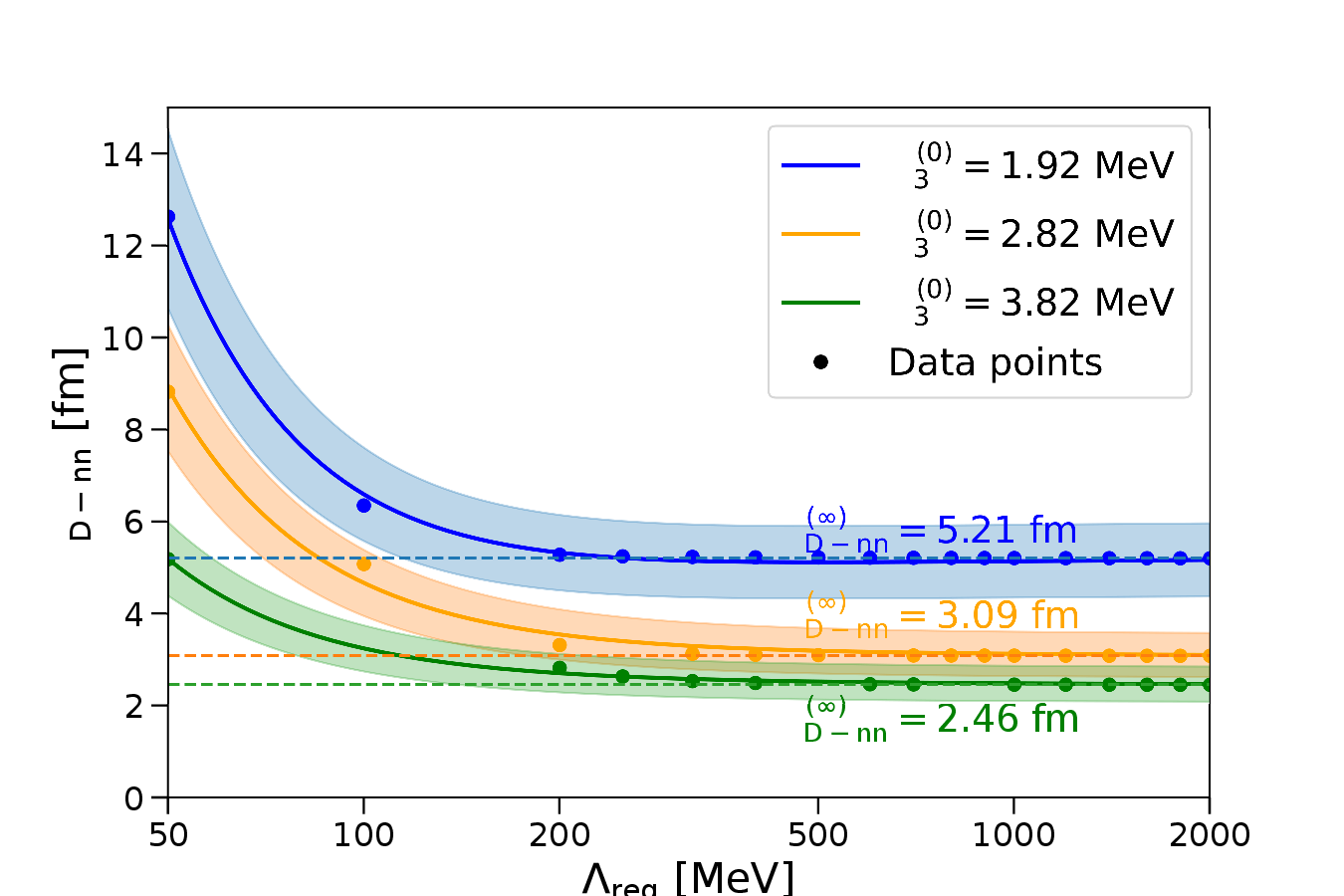}

\vspace{0.3cm}

\includegraphics[width=0.49\textwidth]{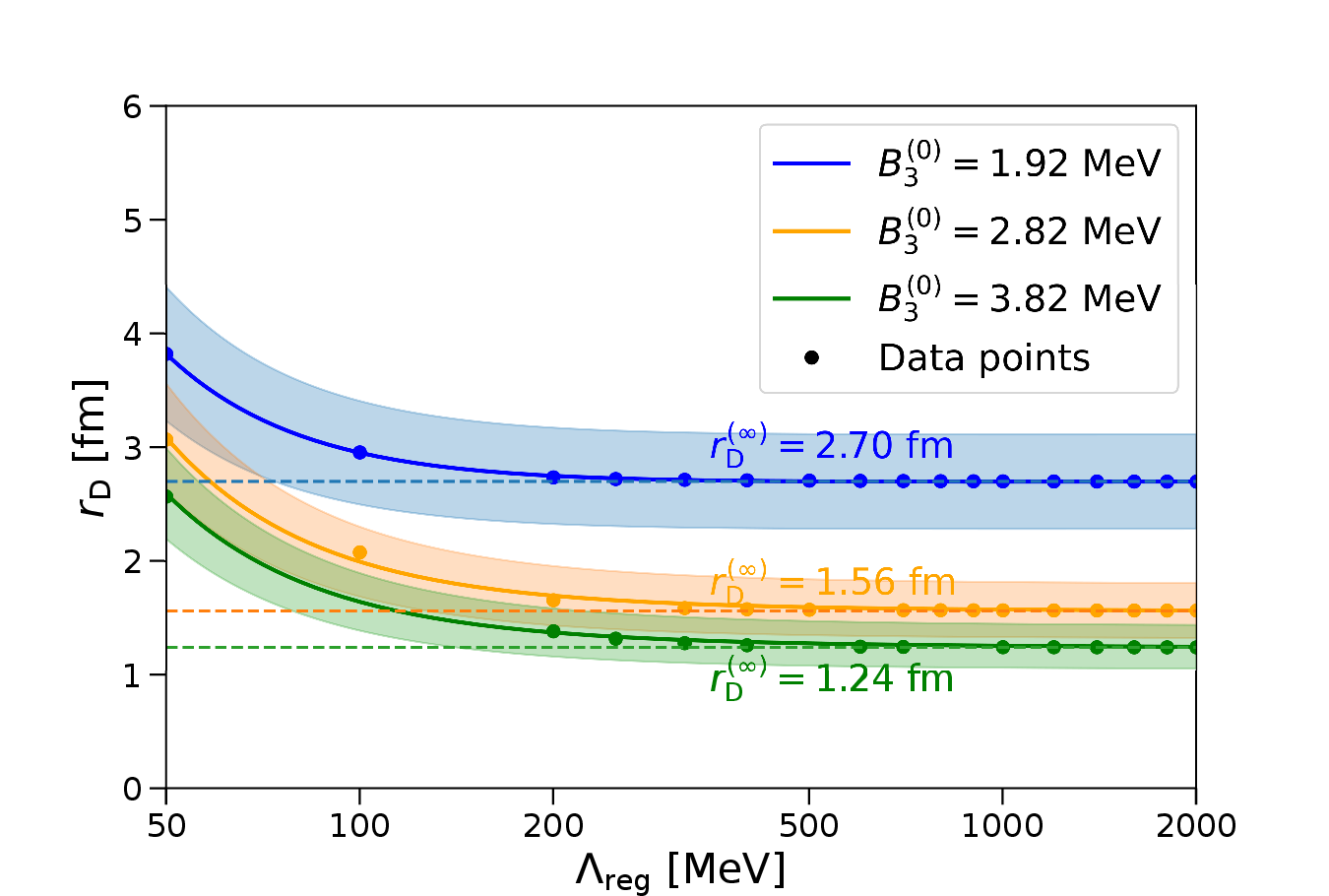} \hfill
\includegraphics[width=0.49\textwidth]{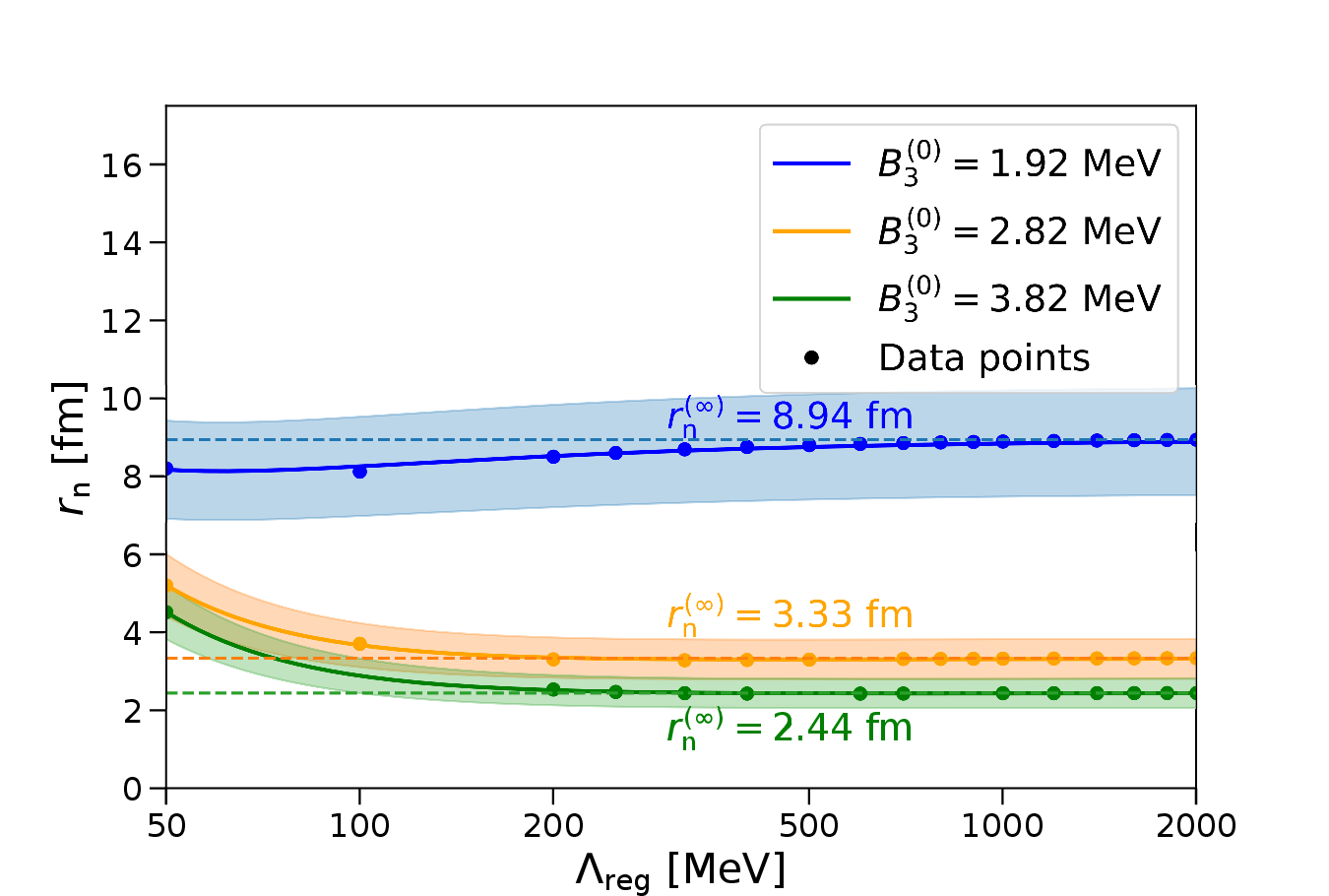}
\caption{Residual regulator $\Lambda_{\rm reg}$ dependence of the renormalized leading order root-mean-square
         radii, $r_\alpha \equiv \sqrt{\langle r^2_\alpha\rangle}$; $\alpha=1,\cdots,6$ (in fm), of the 
         S-wave $D^0nn$ ground state system for input three-body binding energies, $B^{(0)}_3=1.92,\, 2.82$, 
         and $3.82$~MeV ($S^{(0)}_n=0.1,\, 1.0$ and $2.0$~MeV). In each case, the numerical data points are 
         fitted with a smooth curve (color online) defined by Eq.~\eqref{Asymptotic_Eq}. The horizontal lines 
         denote the asymptotic values of the respective radii, $r^\infty_\alpha$ ($\alpha=1,\cdots,6$) as 
         $\Lambda_{\rm reg}\to \infty$. The results also correspond to the two-body inputs, namely the 
         spin-doublet $n$-$D^0$ S-wave scattering length $a_{nD}=4.141$~fm, extracted in the idealized ZCL 
         model analysis of Ref.~\cite{Raha:2017ahu}, and the spin-singlet $n$-$n$ S-wave scattering length 
         $a_{nn}=-18.63$~fm, extracted phenomenologically in Ref.~\cite{Chen:2008zzj}. The bands correspond 
         to LO EFT truncation errors.}
\label{fig:Asymp_radii_plots} 
\end{figure}

With the three shallowest trimer ($m=0,1,2$) state form factors evaluated at the separation energy $S_n=0.1$~MeV, 
corresponding to specific cutoffs determined by Fig.~\ref{fig:B_T_m_012} (upper-left panel), we first determine the 
regulator dependence of the rms radii defining the geometrical structure of $D^0nn$. Using 
Eq.~\eqref{eq:three-body_radius_i}, we summarize our results of our idealized ZCL scenario for rms distances in 
Table~\ref{Tab:6.2}. The inter-particle rms distance between the $(ij)$-pair is defined as 
${r}_{ij}\equiv \sqrt{\langle r^2_{ij} \rangle}$. The rms distance of the spectator particle $i$ from the center-of-mass of
the binary sub-system $jk$ is defined as $r_{i-{jk}}\equiv \sqrt{\langle r^2_{i-{jk}} \rangle}$. Finally, the rms distance of
the particle $i$ from the three-body center-of-mass is defined as $r_{i}\equiv \sqrt{\langle r^2_{i} \rangle}$. As is evident
from the table, the magnitude of all rms radii decreases with increasing $B_3$. This is naturally expected because the 
stronger the binding of a trimer level, the more compact it becomes. Especially, for small values of the trimer binding 
energy approaching the $n+(D^0n)$ particle-dimer break-up threshold, the system assumes a characteristic halo-like structure,
namely a {\it valence}-neutron orbiting around the system with a large rms radius, while the second neutron forms a bound 
$D^0n$ sub-system core. This halo-like feature is reflected in the large magnitudes of the two three-body rms radii $r_{n-nD}$
and $r_{D-nn}$, whose values could be anticipated from the naive quantitative 
estimates~\cite{Braaten:2004rn,Hildenbrand:2019sgp}:
\begin{eqnarray}
r_{i-jk} =\frac{a_{i-jk}}{\sqrt{2}}\,, \quad \text{with}\quad  a_{i-jk}=\frac{1}{\sqrt{2\mu_{i-jk}S_n}}\,.
\end{eqnarray}
With $S_n=0.1$~MeV ($B_3=1.92$~MeV), the naive estimates $r_{n-nD}\sim 12$~fm and $r_{D-nn}\sim 10$~fm agree well with the 
ground state values shown in Table~\ref{Tab:6.2}. However, these results exhibit significant regulator sensitivity in the 
small cutoff region, as seen by comparing the ground state values with those corresponding to the first excited state. 
Nevertheless, such an artifact is mostly eliminated for larger cutoffs, say beyond $\Lambda_{\rm reg}\gtrsim 200$~MeV, 
leading to converged results among the higher excited states, as expected by RG invariance. This notable feature is 
unambiguously revealed by the regulator dependence of the renormalized rms radii (shown later in 
Fig.~\ref{fig:Asymp_radii_plots}). The results for the $n-D^0-n$ opening angle $\theta_{nn}$ [see 
Eq.~\eqref{eq:theta_nn}] are also presented in the last column of the table. In this case, we observe a contrasting 
behavior, with $\theta_{nn}$ increasing with increasing binding energy for the ground state, while decreasing roughly by 
the same amount for the excited states. This indicates the strongly attractive character of the ${}^1S_0$ $nn$ (di-neutron) 
interaction in the ground state, which drives the two neutrons closer as the threshold is approached, thereby 
favoring an atypical symmetric triangular configuration over the commonly anticipated elongated, {\it obtuse} structure.
The behavior is quite the opposite for the excited states, whose structure becomes more elongated with decreasing binding 
energy, reflecting a complex interplay between the dynamics of the $nn$ and $D^0n$ sub-systems. Furthermore, our results 
shown in Fig.~\ref{fig:radii_plot}, for the variation of the effective or mean geometrical radius 
$r_{\rm eff}\equiv \sqrt{\langle r^2_{\rm eff}\rangle_{\rm 2n-halo}}$ [see Eq.~\eqref{eq:r_eff}] with trimer binding energy
also corroborates with our halo-like attribution for the ground state, provided that the small cutoff-dependent artifacts do
not affect the qualitative nature of our predictions.
\begin{table}[tbp]
\begin{center}
\scalebox{0.95}{
\begin{tabular}{|c||c|c||c|c|c|c|c|c||c|}
\hline
$m^{th}$ Efimov level & $B_3$ (MeV) & $\Lambda_{\rm reg}$ (MeV)  & $r_{nn}\equiv r_1$ & $r_{nD}\equiv r_2$ & $r_{n-nD}\equiv r_3$ 
& $r_{D-nn}\equiv r_4$ & $r_D\equiv r_5$ & $r_n\equiv r_6$ & $\theta_{nn}$ \\
\hline\hline
         & 1.92 & 67.9   & 11.20 & 9.89 & 10.70 & 9.23 & 3.35 & 8.02 & 62.47 \\
         & 2.00 & 78.6   & 9.52 & 8.27 & 8.88 & 7.60 & 2.93 & 6.47 & 64.14 \\
$m = 0$  & 2.50 & 123.8  & 6.12  & 5.14  & 5.36 & 4.44 & 2.02  & 3.96 & 69.10 \\
(Ground) & 3.00 & 157.7  & 4.94  & 4.15  & 4.26 & 3.51 & 1.68  & 3.17 & 70.23 \\
         & 3.82 & 204.1  & 3.95  & 3.35  & 3.37 & 2.8 & 1.38  & 2.52 & 70.38 \\
         & 4.00 & 213.1  & 3.80  & 3.23  & 3.24 & 2.7 & 1.33  & 2.43 & 70.33 \\
\hline   
         & 1.92 & 1473   & 8.60  & 5.57 & 9.63 & 5.20 & 2.70  & 8.92  & 79.11 \\
         & 2.00 & 1681   & 7.64  & 5.20 & 8.07 & 4.67 & 2.40  & 6.93  & 78.60 \\
$m = 1$  & 2.50 & 2608   & 5.34  & 4.02 & 5.04 & 3.44 & 1.75  & 3.94  & 75.56 \\
(First excited)&  3.00 & 3327   & 4.40  & 3.44  & 4.03 & 2.93 & 1.49 & 3.10  & 73.77 \\
         & 3.82 & 4320   & 3.57  & 2.88 & 3.21 & 2.45 &  1.24  & 2.45 & 71.99 \\
         & 4.00 & 4519   & 3.44  & 2.79  & 3.09 & 2.38 & 1.20  & 2.35 & 71.71 \\
\hline
         & 1.92 & 31467  & 8.59  & 5.51  & 9.68 & 5.20 & 2.69  & 8.99 & 79.06 \\
         & 2.00 & 35935  & 7.63  & 5.15  & 8.11 & 4.67 & 2.40  & 6.98 & 78.54 \\
$m = 2$  & 2.50 & 55878  & 5.33  & 3.99  & 5.05 & 3.44 & 1.75  & 3.96 & 75.48 \\
(Second excited)& 3.00 & 71328  & 4.39   & 3.42  & 4.04 & 2.93 & 1.48  & 3.11 & 73.69 \\
         & 3.82 & 92700  & 3.56  & 2.87  & 3.22 & 2.45 & 1.24  & 2.45 & 71.90 \\
         & 4.00 & 96961  & 3.43  & 2.78  & 3.09 & 2.38 & 1.20  & 2.36 & 71.61 \\
\hline\hline
\end{tabular}}
\caption{The leading order unrenormalized root-mean-square radii 
         $r_\alpha \equiv \sqrt{\langle r^2_\alpha\rangle}$; $\alpha=1,\cdots,6$ (in fm) and the corresponding $n-D^0-n$ 
         opening angle $\theta_{nn}$ (in degrees) for a halo-bound S-wave $D^0nn$ system. The results are displayed for
         the ground ($m=0$) and the first two excited ($m=1,2$) Efimov trimer states. The results are obtained using a
         sharp momentum cutoff regulator $\Lambda_{\rm reg}$ whose values correspond to the three-body binding energies 
         $B_3$ obtained by solving the unrenormalized STM integral equations~\eqref{eq:F_nD_EFT} at leading order in
         the idealized ZCL model scenario of Ref.~\cite{Raha:2017ahu}, as depicted in Fig.~\ref{fig:B_T_m_012} 
         (upper-left plot).}
\label{Tab:6.2}   
\end{center}
\end{table}

Finally, we turn to the issue of remnant structural universality exhibited by the $D^0nn$ halo-bound system. By approaching 
the unitary as well as the scaling limits (i.e., with $|a_{nn}|,\,a_{nD},\,\Lambda_{\rm reg}\rightarrow \infty$), 
 the square root of the ratio of the (unrenormalized) neutron separation energy 
$S_n=S_n(\Lambda_{\rm reg})$, for successive Efimov levels ($m=0,1,2,...$) is expected to converge to the universal 
transcendental factor $\lambda^{\infty}_0$, which characterizes the discrete scaling invariance associated with the 
asymptotic RG limit cycle~\cite{Braaten:2004rn,Naidon:2016dpf}:
\begin{equation}
\sqrt{\frac{B_3^{(m)}}{B^{(m+1)}_3}} \simeq \sqrt{\frac{S_n^{(m)}}{S^{(m+1)}_n}} \to 
\lambda^\infty_0= e^{\pi/s^\infty_0}=21.5064...\,,\, \text{as\, } \Lambda_{\rm reg}\to \infty\,,
\end{equation}
since, asymptotically, $S_n(\Lambda_{\rm reg}) = B_3(\Lambda_{\rm reg}) - B_{nD}\simeq B_3(\Lambda_{\rm reg}) $. Analogously, 
on dimensional grounds, the inverse ratios of the successive (unrenormalized) rms radii 
$r_\alpha(\Lambda_{\rm reg})\equiv \sqrt{\langle r^2_\alpha\rangle}$\cancel{;} ($\alpha=1,\cdots,6$), are expected to 
display the same asymptotic scaling behavior:
\begin{equation}
\frac{r^{(m+1)}_\alpha}{r^{(m)}_\alpha} \to \lambda^\infty_0 = 21.5064...\,,\, \text{as\, } \Lambda_{\rm reg}\to \infty\,.
\end{equation}
However, in the physical regime of the system (i.e., with finite $|a_{nn}|,\,a_{nD},\Lambda_{\rm reg}<\infty$), only the 
successive ratios $\sqrt{S_n^{(m)}/S_n^{(m+1)}}$ are found to approach  the asymptotic 
value $\lambda^\infty_0$ with reasonable accuracy, even at relatively small $\Lambda_{\rm reg}$, and converge 
rapidly to this limit as $\Lambda_{\rm reg}\to \infty$. In contrast, the ratios $\sqrt{B_3^{(m)}/B_3^{(m+1)}}$ deviate 
significantly from the asymptotic scaling factor $\lambda^\infty_0$ at small $\Lambda_{\rm reg}$, exhibiting 
a comparatively slow asymptotic convergence toward this limit. We find that the convergence pattern 
 of the successive ratios of the rms radii closely tracts that of the corresponding 
 inverse ratios, i.e., $1/\sqrt{B_3}$, rather than those of $1/\sqrt{S_n}$, 
thereby corroborating  similar observations reported inRefs.~\cite{Canham:2008jd,DL-Canham:2009}, i.e.,
\begin{equation}
\frac{r^{(m+1)}_\alpha}{r^{(m)}_\alpha} \simeq \sqrt{\frac{B_3^{(m)}}{B_3^{(m+1)}}}\,\,, \quad \text{but}\quad 
\frac{r^{(m+1)}_\alpha}{r^{(m)}_\alpha} \not\simeq  \sqrt{\frac{S_n^{(m)}}{S_n^{(m+1)}}}\,.
\end{equation}
The rationale behind this behavior lies in the Samba structure of the $D^0nn$ system, where the $nn$ virtual-bound sub-system
in the physical limit is still much more resonant compared to the weakly bound $D^0n$ sub-system in the ZCL scanario. In
this case, the Efimov physics is directly reflected in the spectrum of the separation energy $S_n$, rather than that of the 
trimer binding energy $B_3$. With physical $nn$ interactions already tuned sufficiently close to the unitary limit, it is 
hardly surprising that $S_n$ exhibits ``better universality'' than $B_3$. Consequently, with the rms radii proportional to 
$1/\sqrt{B_3}$ by dimensional analysis, they inherit the same slow convergence behavior as $B_3$. In contrast, 
{\it Borromean} systems (i.e., three-body clusters consisting of all virtually-bound binary sub-systems)~\cite{Yamashita:2004pv},
which are known to be much more prevalent among halo nuclei, are expected to exhibit ``good universality'' 
characterized by rapid asymptotic convergence for their binding energy ($B_3$) 
ratios~\cite{Canham:2008jd,DL-Canham:2009,Hammer:2017tjm}.    

To see this feature numerically, we revisit our results for the cutoff dependence of the neutron separation energies for
the ground ($m=0$) and the first two excited ($m=1,2$) states [cf. upper-left plot in Fig.~\ref{fig:B_T_m_012}]. For example, 
at the cutoff scale $\Lambda_{\rm reg}=1473$~MeV, when the  energy of the shallowest state, i.e., 
the first excited state ($m=1$) is fixed as $S^{(1)}_n=0.1$~MeV ($B^{(1)}_3=1.92$~MeV), the  energy of the 
ground state ($m=0$) is obtained as $S^{(0)}_n=74.81$~MeV ($B^{(0)}_3=76.63$~MeV). Thus, we can expect the ratios of the rms 
radii to approximately follow the non-asymptotic correspondence at  finite cutoff values 
$\Lambda_{\rm reg} > \Lambda^{(m=1)}_{\rm crit}$, where $\Lambda^{(m=1)}_{\rm crit}=1122$~MeV denotes the critical cutoff 
associated with the first excited state, namely
\begin{eqnarray}
\frac{r^{(1)}_\alpha}{r^{(0)}_\alpha} \simeq \sqrt{\frac{B_3^{(0)}}{B_3^{(1)}}} = 6.32\,, 
\quad \text{at } \Lambda_{\rm reg}=1473~{\rm MeV}\,. 
\end{eqnarray}
In contrast, we obtain a ratio $\sqrt{S_n^{(0)}/S_n^{(1)}}=27.35$ at the same cutoff, which rapidly approaches the 
asymptotic limit $21.5064...$ as $\Lambda_{\rm reg}$ increases. As shown in Table~\ref{Tab:6.3}, the ratios of the
rms radii are of the same order as the non-asymptotic $1/\sqrt{B_3}$ ratios and are expected to exhibit a similar 
pattern of asymptotic convergence. Quantitatively, however, the various radius ratios differ appreciably at a given 
cutoff: the ratios involving $r_{nn}$, $r_{n}$ and $r_{n-nD}$ relatively faster, whereas those associated with $r_{nD}$, 
$r_{D}$ and $r_{D-nn}$ display a slower approach to the asymptotic limit. In fact, at $\Lambda_{\rm reg}=31467$~MeV, 
when the separation energy for the second excited state ($m=2$)-- now the shallowest state -- 
is $S^{(2)}_n=0.1$~MeV ($B^{(2)}_3=1.92$~MeV), the rms radius ratios for the two deepest levels, namely 
$r^{(1)}_\alpha/r^{(0)}_\alpha$, as displayed in the table, are much closer to the asymptotic value than at 
$\Lambda_{\rm reg}=1473$~MeV. In contrast, the ratios $r^{(2)}_\alpha/r^{(1)}_\alpha$ still lag behind, exhibiting 
a noticeably slower convergence toward the asymptotic limit. Naturally, at such large cutoff values, the ground state 
becomes anomalously deep, rendering its interpretation as a physical Efimov state within the domain of a low-energy EFT 
increasingly questionable.  
\begin{table}[tbp]
\begin{center}
\scalebox{1}{
\begin{tabular}{|c||c|c|c|c|c|c|c||c|}
\hline
RMS Radius $\to$                           & $m^{\rm th}$ level                       & $r_{nn}\equiv r_1$ & $r_{nD}\equiv r_2$ & $r_{n-nD}\equiv r_3$ & $r_{D-nn}\equiv r_4$ & $r_{D}\equiv r_5$ 
& $r_{n}\equiv r_6$ & $r_{\rm eff}$ \\
\hline\hline
$\Lambda_{\rm reg}=1473$~MeV,              & 0                                        & 0.64     & 0.58     & 0.55  & 0.50       & 0.25 & 0.41       & 0.34  \\
$\sqrt{\frac{B_3^{(0)}}{B_3^{(1)}}}=6.32$  & 1                                        & 8.60     & 5.57     & 9.63  & 5.21    & 2.70 & 8.92      & 5.43  \\
\hline
$1^{\rm st}$ Ratio  $\displaystyle{\frac{r^{(1)}_\alpha}{r^{(0)}_\alpha}}$   & 1 \& 0        & 13.44    & 9.60     & 17.51 & 10.42       & 10.8 & 21.76       & 15.97  \\ 
\hline\hline
$\Lambda_{\rm reg}=31467$~MeV,             & 0                                        & 0.03     & 0.03     & 0.03  & 0.02       & 0.01 & 0.02       & 0.02  \\
$\sqrt{\frac{B_3^{(0)}}{B_3^{(1)}}}=19.88$,& 1                                        & 0.60     & 0.54     & 0.53  & 0.47       & 0.24 & 0.40       & 0.33  \\
$\sqrt{\frac{B_3^{(1)}}{B_3^{(2)}}}=6.24$  & 2                                        & 8.59     & 5.51     & 9.68  & 5.20       & 2.69 & 8.99      & 5.46  \\
\hline
$1^{\rm st}$ ratio  $\displaystyle{\frac{r^{(1)}_\alpha}{r^{(0)}_\alpha}}$   & 1 \& 0        & 20.00    & 18.00    & 17.67 & 23.50     & 24.00 & 20.00      & 16.50  \\ 
\hline
$2^{\rm nd}$ ratio  $\displaystyle{\frac{r^{(2)}_\alpha}{r^{(1)}_\alpha}}$   & 2 \& 1        & 14.32    & 10.20     & 18.26  & 11.06    & 11.21 & 22.48       & 16.55  \\ 
\hline
\end{tabular}}
\caption{The leading order unrenormalized root-mean-square radii,
         $r_\alpha \equiv \sqrt{\langle r^2_\alpha\rangle}$; $\alpha=1,\cdots,6$ (in fm), the effective 
         geometrical radius $r_{\rm eff}$ (in fm), and their successive ratios among the ground ($m=0$)
         and excited ($m=1,2$) state Efimov trimers for an S-wave halo-bound $D^0nn$ system. The 
         regulator cutoffs $\Lambda_{\rm reg}$ are chosen such that the shallowest trimer state 
         ($m=m_{\rm max}$) has a fixed binding energy, $B^{(m_{\rm max})}_3=1.92$~MeV 
         ($S^{(m_{\rm max})}_n=0.1$~MeV). The results also correspond to the two-body inputs, namely 
         the spin-doublet $n$-$D^0$ S-wave scattering length $a_{nD}=4.141$~fm, extracted in the 
         idealized ZCL model analysis of Ref.~\cite{Raha:2017ahu}, and the spin-singlet $n$-$n$ S-wave 
         scattering length $a_{nn}=-18.63$~fm, extracted phenomenologically in Ref.~\cite{Chen:2008zzj}.}
\label{Tab:6.3} 
\end{center}
\end{table}

To determine the renormalized regulator-independent radii, we introduce the LO 3BF terms and fix their couplings
$g_3=g_3(\Lambda_{\rm reg})$ from the RG limit cycle corresponding to the input trimer binding energy. To 
incorporate the 3BF, as outlined earlier, we first replace the S-wave single-particle exchange 
interaction kernels $\mathcal{K}_{(n)}$ and $\mathcal{K}_{(D)}$ with their renormalized counterparts, 
$\mathbb{K}^R_{(n)}$ and $\mathbb{K}^R_{(D)}$, respectively. We then determine the corresponding 
renormalized spectator functions, $F^R_n$ and $F^R_D$, by solving the STM3 equations [see 
Eqs.~\eqref{ren_kernel} and \eqref{eq:F_nD_EFT_ren}]. Furthermore, the three-particle free Green's functions 
$\mathcal{G}_0^{(n)}$ and $\mathcal{G}_0^{(D)}$ [see Eq.~\eqref{eq:G0_n_D_matrix_elements}], must also be 
replaced by their corresponding renormalized forms, $\mathbb{G}^{(n)R}_0$ and $\mathbb{G}^{(D)R}_0$, 
namely\footnote{Since $\mathcal{K}_{(n)}$ and $\mathcal{K}_{(D)}$ include the pre-factor 1/2 [see Eq.~\eqref{eq:OPE}], we must accordingly subtract 
$2g_3/\Lambda_{\rm reg}^2$ from $\mathcal{G}_0^{(n)}$ and $\mathcal{G}_0^{(D)}$ for consistency of 
renormalization.}
\begin{equation}
\mathcal{G}_0^{(i)}(p,q;B_3) \longrightarrow \mathbb{G}^{(i)R}_0(p,q;B_3) = 
\mathcal{G}_0^{(i)}(p,q;B_3) - 2g_3/\Lambda_{\rm reg}^2 \,,
\end{equation}
which are then used to construct the renormalized LO S-wave three-body wave functions, $\Psi^R_n$ and 
$\Psi^R_D$. The corresponding renormalized radial probability densities, $P^R_n$ and $P^R_D$ are shown in the
lower panel of Fig.~\ref{fig:probability_density}. We subsequently employ these these renormalized wave 
functions, together with the STM3 integral equations, to compute the renormalized form factors [{\it via} 
Eqs.~\eqref{eq:one-body_FF} and \eqref{eq:two-body_FF_nD_nn}], and the associated rms radii $r_\alpha$ 
($\alpha=1,\cdots,6$). The residual regulator dependence of these renormalized radii is depicted in 
Fig.~\ref{fig:Asymp_radii_plots}, where the results are displayed for three {\it ad hoc} input ground state energies, 
namely $B_3=1.92~{\rm MeV}, 2.82$~MeV and $3.82$~MeV ($S_n=0.1$~MeV $1.0$~MeV and $2.0$~MeV). The numerical data
points corresponding to each of the radii (i.e., excluding the effective radius $r_{\rm eff}$) are then fitted using 
a polynomial equation with the free parameters 
$r^\infty_\alpha,\, {\mathcal R}^{(1)}_\alpha$ and ${\mathcal R}^{(2)}_\alpha$ ($\alpha=1,\cdots,6$), having the form:
\begin{eqnarray}
    r_{\alpha}(\Lambda_{\rm reg})= r_{\alpha}^{(\infty)}\left (1+\frac{{\mathcal R}^{(1)}_\alpha}{\Lambda_{\rm reg}}
    +\frac{{\mathcal R}^{(2)}_\alpha}{\Lambda_{\rm reg}^2}\right )\,,
    \label{Asymptotic_Eq}
\end{eqnarray}
where the fitted parameters are shown in Table~\ref{Tab:6.4}. The smallness of these parameters, that is, 
${\mathcal R}^{(1)}_\alpha/m_\pi\sim {\mathcal R}^{(2)}_\alpha/m^2_\pi \ll 1$ suggests their naturalness, thus
validating the reasonable fairness of the low-energy regression. The residual regulator dependence is seen to 
rapidly converge for $\Lambda_{\rm reg}\gtrsim 150-200$~MeV, in agreement with RG invariance. The asymptotic 
values $r^\infty_\alpha$ represent our LO halo-EFT predictions for the renormalized rms radii. The error band 
around each central curve provides an approximate estimate of the theoretical uncertainty associated with higher-order 
EFT corrections. Within our LO  framework, which takes two-body observables as input, a naive 
estimate of higher-order uncertainties can be obtained by considering the ratios of the S-wave effective-range 
parameters, $r^{(nn)}_0 = 2.87$~fm~\cite{Malone:2022gvp} and $r^{(nD)}_0 \approx 0.4$~fm~\cite{Bayar:2012dd}, to the 
corresponding S-wave scattering lengths of the relevant two-body sub-systems. Thus, considering theoretical errors,
the two-body rms radii can be estimated as $\delta r_{nn}=\pm r_{nn}\left(r^{(nn)}_0/a_{nn}\right)$ with 16\% 
uncertainty, and $\delta r_{nD}=\pm r_{nD}\left(r^{(nD)}_0/a_{nD}\right)$ with 10\% uncertainty. For the rest of the 
one-body rms radii, namely $r_{n-nD}$, $r_{D-nn}$, $r_n$ and $r_D$, as well as the opening angle $\theta_{nn}$, the 
larger of the two errors, namely $16\%$, can be taken as reasonable uncertainty estimates. Here, we emphasize that 
these error estimates reflect only the theoretical uncertainties associated with our ZCL model, as 
introduced within the LO halo-EFT analysis. However, these do not capture the full uncertainties of a 
realistic prediction for all such observables, which, in principle, could be as large as $100\%$ (see the 
discussion below for a more conservative assessment). This is because the omission of decay channels introduces 
uncontrolled systematic effects that are not accounted for within our simplified framework.

{In Table.~\ref{Tab:6.5} we summarize our LO predictions of the renormalized (asymptotic) rms distances and the 
$n-D^0-n$ opening angle corresponding to a plausible $D^0nn$ Efimov ground state with input separation energies, 
$S_n^{(0)}=0.1,\, 1.0$ and $2.0$~MeV, for the three model scenarios, namely the idealized ZCL, extended ZCL, and
realistic one. As expected, we find the three-body system to shrink in size with increasing binding energy in 
each case. In particular, for both ZCL scenarios with $S^{(0)}_n=0.1$~MeV, we already observed that 
$r_{nn}^{(\infty)}$, $r_{n-nD}^{(\infty)}$ and $r_n^{(\infty)}$ supersede $r_D^{(\infty)}$ by about 
$\gtrsim 300\%$, and these differences increase progressively as the threshold approaches, i.e., 
$S^{(0)}_n\to 0$~MeV ($B^{(0)}_3\to 1.82$~MeV). This suggests the emergence of the Efimov ground state into a 
universal halo-bound structure when driven close to the $n+(D^0n)$ particle-dimer break-up threshold in the ZCL 
scenarios. Notably, a comparison between the numerical results of the extended and idealized ZCL models indicates
that range effects lead only to a slight reduction ($\lesssim 1\%$) in the asymptotic rms distances, as well as
in the $n\text{–}D^0\text{–}n$ opening angle.

Figure.~\ref{fig:ReffVSB3_plots} illustrates the same feature, showing the variation of the effective geometrical
radius $r_{\rm eff}^{(\infty)}$ as a function of the input neutron separation energy $S^{(0)}_n$ for the Efimov 
ground state. The smooth curves shown in the plot are fits to the central data points in Table.~\ref{Tab:6.5} 
using the functional form:
\begin{eqnarray}
r_{\rm eff}^{(\infty)} = \frac{1}{\sqrt{S_n^{(0)}}}\left(\gamma_1 + \frac{\gamma_2}{\sqrt{S_n^{(0)}}} 
+ \frac{\gamma_3}{S_n^{(0)}}\right)\,,
\label{eq:reff}
\end{eqnarray}
where, $\gamma_1=3.14~(2.30)~[2.73]$~MeV${}^{-1/2}$, $\gamma_2=-0.67~(-0.48)~[-0.34]$, and 
$\gamma_3=0.07~(0.06)~[0.0008]$~MeV${}^{1/2}$, are the three fitted parameters corresponding to the ZCL (realistic) 
[extended ZCL] scenario.   

\begin{table}[tbp]
\begin{center}
\scalebox{1}{
\begin{tabular}{|c||c|c|c|c|c|c|}
\hline
Parameters in Eq.~\eqref{Asymptotic_Eq} & ${\alpha}=1$ & ${\alpha}=2$ &  ${\alpha}=3$ & ${\alpha}=4$ & ${\alpha}=5$ & ${\alpha}=6$ \\
\hline
\hline
$r^{(\infty)}_{\alpha}$ (fm)             &  8.60     & 5.56     & 9.64     & 5.21     & 2.70       & 8.94  \\
\hline
$\mathcal{R}_{\alpha}^{(1)}$ (MeV)       &  -15.611  & -6.24294 & -22.3759 & -17.5094  & -2.06308  & -10.9565  \\
\hline  
$\mathcal{R}_{\alpha}^{(2)}$ (MeV${}^2$) &  3199.7   & 4587.82  & 2503.79  & 4409.3    & 1141.04   & 333.74  \\
\hline\end{tabular}}
\caption{Fit parameters $r^{(\infty)}_{\alpha}$, $\mathcal{R}_{\alpha}^{(1)}$, and $\mathcal{R}_{\alpha}^{(2)}$ 
         ($\alpha=1,\ldots,6$) of Eq.~\eqref{Asymptotic_Eq}, extracted from the central data points in 
         Fig.~\ref{fig:Asymp_radii_plots} (blue curves), for the normalized leading order root-mean-square 
         radii $r_\alpha \equiv \sqrt{\langle r^2_\alpha\rangle}$ (in fm) of the S-wave $D^0nn$ ground state 
         with input trimer binding energy $B^{(0)}_3 = 1.92$~MeV ($S^{(0)}_n = 0.1$~MeV) in the idealized ZCL
         scanario.}
\label{Tab:6.4}   
\end{center}
\end{table}
\begin{table}[h]
\begin{center}
\scalebox{0.93}{
\begin{tabular}{|c||c|c|c|c|c|c||c|c|}
\hline
$S^{(0)}_n$ (MeV) & $r_{nn}^{(\infty)}\equiv r^{\infty}_1$ & $r_{nD}^{(\infty)}\equiv r^{\infty}_2$ &  $r_{n-nD}^{(\infty)}\equiv r^{\infty}_3$ 
& $r_{D-nn}^{(\infty)}\equiv r^{\infty}_4$ & $r_D^{(\infty)}\equiv r^{\infty}_5$ & $r_n^{(\infty)}\equiv r^{\infty}_1$ & $r_{\rm eff}^{(\infty)}$ & $\theta^\infty_{nn}$\\
\hline\hline
\multicolumn{9}{|c|}{(I) Idealized ZCL Model with input $n$-$D^0$ S-wave scattering length: $a_{nD}=4.141$~fm}\\
\hline
  0.1 &  8.60 $\pm 1.32 $  & 5.56 $\pm 0.54 $ &9.64 $\pm 1.49 $ & 5.21 $\pm 0.80 $  & 2.70 $\pm 0.42 $ & 8.94 $\pm 1.38 $ &  5.44 $\pm 0.84 $ & 79.07 $\pm 12.18 $\\
  \hline
  1.0 &  4.68 $\pm 0.72 $  & 3.63 $\pm 0.35 $ &4.31 $\pm 0.66 $ & 3.09 $\pm 0.48 $ & 1.56 $\pm 0.24 $ & 3.33 $\pm 0.52 $ &   2.54 $\pm 0.39 $ & 74.27 $\pm 11.44 $\\
  \hline
  2.0 &  3.58 $\pm 0.55 $ & 2.90 $\pm 0.28 $ &3.21 $\pm 0.49 $ & 2.46 $\pm 0.38 $ & 1.24 $\pm 0.19 $ & 2.44 $\pm 0.38 $ &  1.91 $\pm 0.29 $ &  72.08 $\pm 11.10$ \\
  \hline\hline
\multicolumn{9}{|c|}{(II) Extended ZCL Model incorporating two-body effective-range corrections incorporated in the sense of Ref.~\cite{Epelbaum:2016ffd} }\\ 
\hline
  0.1 & 8.51 $\pm 1.31$ & 5.27 $\pm 0.51 $ &9.30 $\pm 1.43 $  &5.16 $\pm 0.79 $ & 2.67 $\pm 0.41$ &8.62 $\pm 1.33 $  & 5.26 $\pm 0.81$ &79.04 $\pm 12.18$  \\
  \hline
  1.0 & 4.59 $\pm 0.71$ & 3.12 $\pm  0.30$ & 4.03 $\pm 0.62 $ & 3.04 $\pm 0.47 $ & 1.54 $\pm0.24 $ &3.10 $\pm 0.48 $  &2.39 $\pm 0.37$  &74.17 $\pm 11.43$ \\
  \hline
  2.0 & 3.48 $\pm 0.54$ &2.50  $\pm 0.24$ & 2.90 $\pm 0.45$&2.40 $\pm 0.37 $  & 1.21 $\pm 0.19$ & 2.26 $\pm 0.35$ & 1.76 $\pm 0.27 $ & 71.90 $\pm 11.08$ \\
\hline\hline
\multicolumn{9}{|c|}{(III) Realistic Model with input $n$-$D^0$ S-wave scattering length: ${\tilde a}_{nD}=0.764$~fm}\\ 
\hline
  0.1 &  3.39 $\pm 0.52 $  & 1.55 $\pm 0.81 $ &8.09 $\pm 4.23 $ & 1.84 $\pm 0.96 $  & 0.95 $\pm 0.50 $ & 8.15 $\pm 4.27 $ &  4.34 $\pm 2.27 $ & 85.29 $\pm 44.65 $\\
  \hline
  1.0 &  2.24 $\pm 0.35 $  & 1.31 $\pm 0.69 $ &3.45 $\pm 1.81 $ & 1.23 $\pm 0.64 $ & 0.62 $\pm 0.32 $ & 2.71 $\pm 1.42 $ &   1.88 $\pm 0.98 $ & 84.85 $\pm 44.42 $\\
  \hline
  2.0 &  1.93 $\pm 0.30 $ & 1.22 $\pm 0.64 $ &2.55 $\pm 1.34 $ & 1.08 $\pm 0.57 $ & 0.54 $\pm 0.28 $ & 1.96 $\pm 1.03 $ &  1.41 $\pm 0.74 $ &  83.62 $\pm 43.78 $ \\
\hline
\end{tabular}}
\caption{Summary of our LO EFT predictions for the renormalized (asymptotic) rms distances, namely 
         $r^{\infty}_{\alpha}$; $\alpha=1,\cdots,6$ (in fm) and $r^{\infty}_{\rm eff}$ (in fm), as well
         as the $n-D^0-n$ opening angle $\theta^\infty_{nn}$ (in degrees), determining the 
         geometrical structure of the S-wave $D^0nn$ Efimov ground state. Three sets of results are 
         displayed for three {\it ad hoc} input choices of the trimer neutron separation energy, namely 
         $S^{(0)}_n=0.1,\,1.0$, and $2.0$~MeV, measured relative to the $n+(D^0n)$ particle-dimer 
         break-up thresholds, $B_{nD}=1.82$~MeV and $\tilde{B}_{nD}=53.4$~MeV. Set (I) corresponds to 
         the idealized ZCL scenario. Set (II) also corresponds to the extended ZCL scenario incorporating
         effective-range-like effects {\it via} logarithmic modifications of STM equation akin to the 
         treatment of Ref.~\cite{Epelbaum:2016ffd}. Finally, set (III) corresponds to the plausible 
         realistic scenario. A naive theoretical error estimate due to omission of higher-order EFT 
         corrections of the numerical results is also provided in each case. }
\label{Tab:6.5}   
\end{center}
\end{table}

Finally, we comment on the expected difference in the predicted results between those of the ZCL model scenarios and the
plausible ``realistic'' scenario. If we assume that the S-wave effective-range remains approximately the same in either 
cases, i.e, $\tilde{r}^{(nD)}_0\sim r^{(nD)}_0 \approx 0.4$~fm, then the estimated higher-order EFT error in the latter
scenario can be quantified as $\tilde {r}^{(nD)}_0/{\tilde a}_{nD}$, namely exceeding $50\%$. This is reflected in the 
results presented in Table~\ref{Tab:6.5} and Fig.~\ref{fig:ReffVSB3_plots}, where the uncertainty bands for the 
idealized and extended ZCL models are largely encompassed by those of the realistic scenario. Consequently, in the 
realistic scenario, we anticipate deviations of no less than $\sim 50\%$ in the predicted observables relative to those
obtained in the ZCL.

\begin{figure}[t]
\centering
\includegraphics[width=0.45\textwidth]{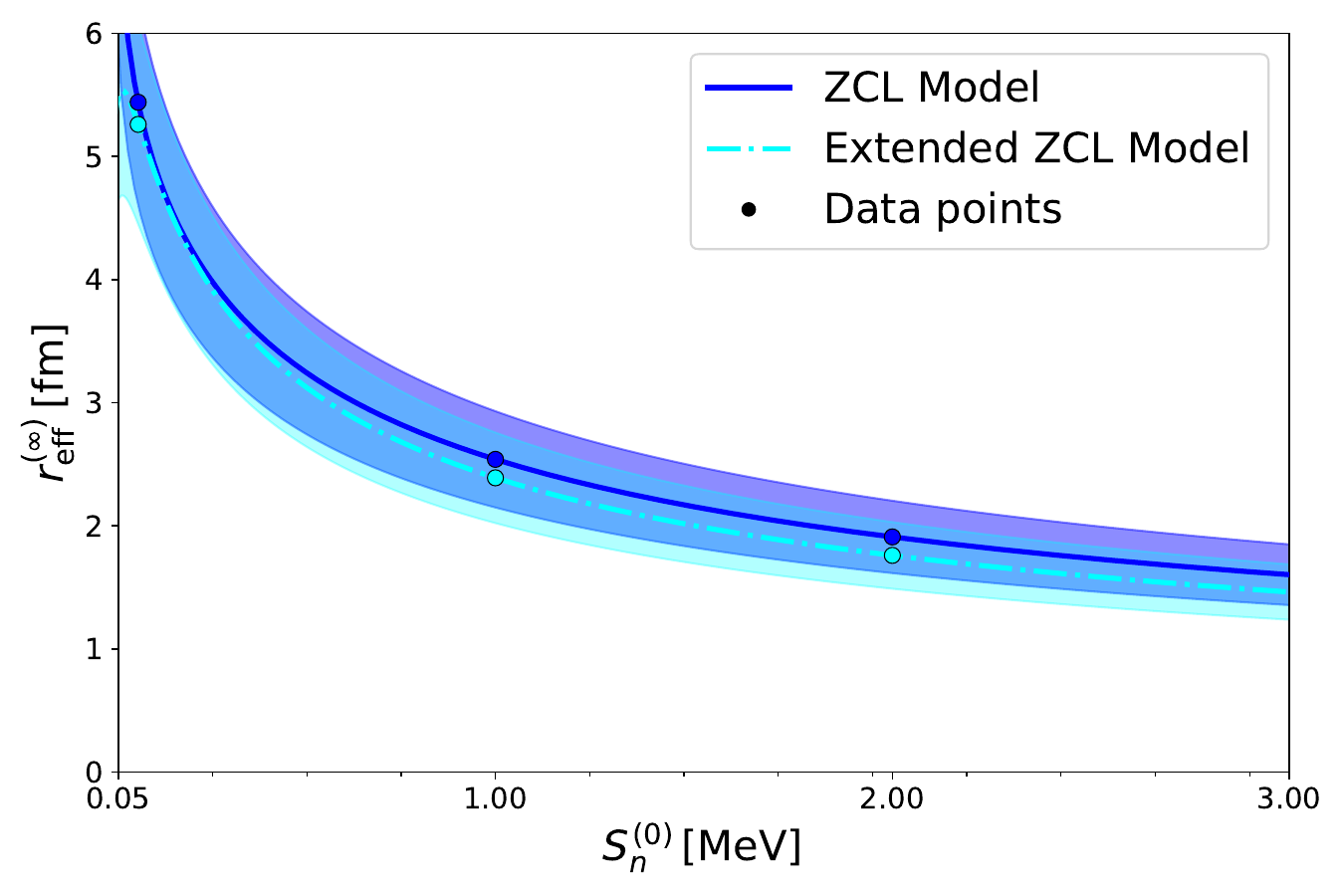} \hfill
\includegraphics[width=0.45\textwidth]{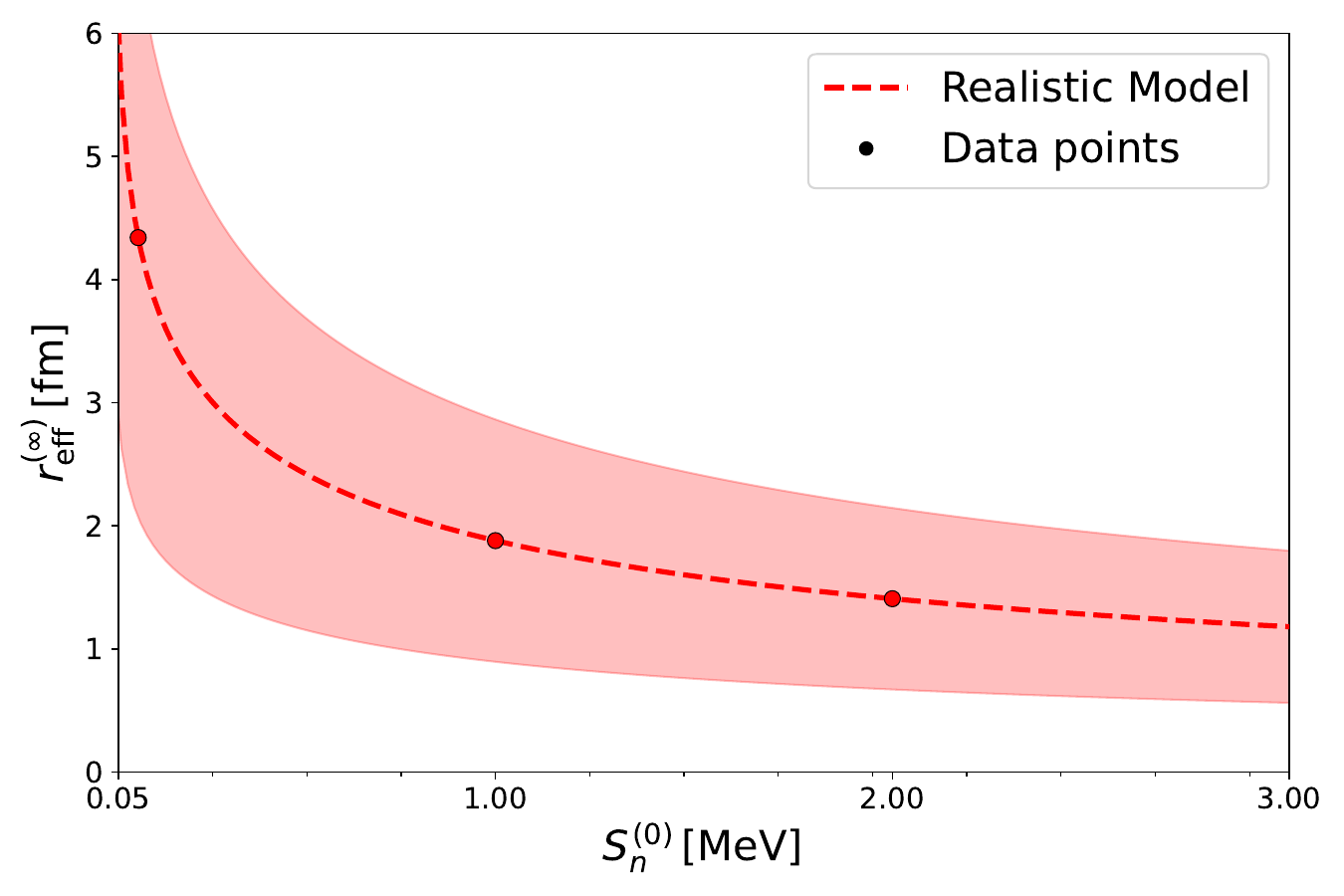} 
    \caption{Variation of the effective geometrical radius, $r_{\rm eff}^{(\infty)}$, as a function of the ground state
             trimer neutron separation energy, $S_n \equiv S_n^{(m=0)}$, for the S-wave $D^0nn$ system. The solid (blue), 
             dot-dashed (cyan), and dashed (red) curves represent fits, using Eq.~\ref{eq:reff}, to the central data points
             listed in Table~\ref{Tab:6.5}, with three free fitting parameters, $\gamma_{1,2,3}$. These correspond to the 
             idealized and extended ZCL scenarios (left panel), and the plausible realistic scenario (right panel), 
             respectively. The associated error bands reflect the $16\%$ and $50\%$ theoretical uncertainties due to the 
             omission of higher-order EFT corrections for the ZCL and realistic model scenarios, respectively. }
\label{fig:ReffVSB3_plots} 
\end{figure}

\section{Summary and Conclusion}
\label{sec:summary}
Our previous work in Refs.~\cite{Raha:2017ahu,Raha:2020sse} dealt with the novel idea of probing the remnant two- 
and three-body universalities of the S-wave $K^-nn$ and $D^0nn$ systems by extrapolating to an unphysical mass limit 
between the strange and charm sectors, where the meson-neutron interaction becomes unitary upon invoking the ZCL 
 idealization. Furthermore, feasibility studies investigating the cutoff dependence of the STM equations showed 
that the $D^0nn$ system is much more likely to become Efimov bound than the $K^-nn$ system. As a natural extension 
to those analyses, the current work deals with the problem of formulating a leading order effective quantum 
mechanical halo-EFT framework using the Faddeev technique in the momentum-space to investigate remnant 
structural universalities in the putative S-wave halo-bound $D^0nn$ system\footnote{The validity of our formalism 
inherently relies on the ``standard'' scenario of the $DN$ ($T=1$) interactions, namely the dominance of S-wave 
contribution in comparison to higher partial-waves. Interestingly, in a ``non-standard'' scenario, the analysis by
Haidenbauer {\it el al.}~\cite{Haidenbauer:2010ch} predicts a P-wave state just below the $DN$ threshold, 
 suggesting that the P-wave channel may also need to be treated at leading order. Of course, 
whether the presence of a P-wave state alters the hierarchy in the three-body sector requires explicit 
verification through calculations that include P-wave interactions. Such an analysis, however, lies beyond the 
scope of this investigation.} The theoretical framework was originally developed in 
Refs.~\cite{Platter:2004he,Platter:2004ns,Platter:2004zs} for the investigation of resonant systems
of three and four bosons, and was subsequently extended to the study of $2n$-halo nuclei, such as the 
${}^{6}$He, ${}^{20}$C, etc.~\cite{Canham:2008jd,DL-Canham:2009,Ji:2014wta,Gobel:2019jba}. In the present work,
we adopt this framework and establish its equivalence, at leading order, to the conventional pionless EFT 
description formulated through the STM integral equations~\cite{STM1,STM2}. We subsequently investigate the universal
features of the $D^0nn$ system within an idealized ZCL scenario~\cite{Raha:2017ahu,Raha:2020sse}, 
characterized by resonant S-wave interactions between the neutrons and the core $D^0$-meson. Using the Jacobi 
coordinate system, a complete set of partial-wave basis states is constructed onto which the Faddeev equations are 
projected to obtain a set of two coupled integral equations describing the multiple-scattering dynamics of the 
coupled spin and isospin channels. 
 
Investigating the asymptotic behavior of the unrenormalized STM integral equations~\eqref{eq:F_nD_EFT} reveals an 
RG limit cycle behavior (cf. Fig.~\ref{fig:limitcycle}) associated with an inherent discrete scaling symmetry, 
characterized by the universal transcendental number $\lambda_0^{\infty}=21.5064...$. Formally, this indicates the 
manifestation of Efimov states below the $n+(D^0n)$ particle-dimer break-up threshold energy, $B_{nD}=1.82$~MeV, in the
idealized ZCL scenario. Specifically, introducing a sharp momentum cutoff ($\Lambda_{\rm reg}$) in the STM equations 
and fixing the near-threshold separation energy of the shallowest Efimov level, $S_n = B_3 - B_{nD} = 0.1$~MeV, yields 
formal solutions $S_n \approx 73$~MeV and $S_n \approx 2.99\times10^{4}$~MeV for the two deepest bound states (cf. 
Fig.~\ref{fig:B_T_m_012}). In the limit $\Lambda_{\rm reg} \to \infty$, the ratios of successive levels converge to 
$\lambda_0^{\infty}$; however, such deeply bound states fall outside the domain of validity of our low-energy EFT.

Next we investigate the remnant {\it structural} universality of the halo-bound S-wave $D^0nn$ system. For 
this purpose, we reconstruct the full leading order S-wave three-body wave functions $\Psi_n(p,q)$ and $\Psi_D(p,q)$, 
Eq.~\eqref{eq:Psi_nD_pq}, in momentum-space, following the methodology outlined in 
Refs.~\cite{Canham:2008jd,DL-Canham:2009}. These wave functions are subsequently employed to determine the leading 
order S-wave one- and two-body matter density form factors for the ground and the first two excited Efimov trimer states. 
The resulting form factors are then used to extract three-body geometrical observables, such as the rms radii 
$r_{1,...,6}$, two-neutron opening angle $\theta_{nn}$, and effective matter radius $r_{\rm eff}$, which characterize 
the halo-like structure of these states. In particular, excluding 3BF, the unrenormalized results  
exhibit strong regulator dependence. Our findings in Table~\ref{Tab:6.2} reveal that the structure of the ground state has 
greater sensitivity to regulator variations than that of the excited states. Given that the effective radius $r_{\rm eff}$
grows significantly as the states approach the threshold, i.e., $S^{(0)}_n\to 0$~MeV ($B^{(0)}_3\to 1.82 $~MeV), 
the $D^0nn$ system exhibits features characteristic of a halo-like configuration. Moreover, the distinct behavior
of the opening angle $\theta_{nn}$ in the ground state, compared to the excited states, suggests that the
attractive $nn$ interaction interaction plays a more dominant role in the ground state configuration. This favors the
formation of a shallow bound state with a relatively symmetric triangular geometry, rather than the elongated, 
asymmetric obtuse structure typically expected for halo systems. Finally, in the quest for a plausible signature of remnant
structural universality, we find that the ratios of the rms distances of successive Efimov levels are commensurate with the
inverse ratios of the square root of $B_3$ (rather than $S_n$), namely 
$r^{(m+1)}_\alpha/r^{(m)}_\alpha \simeq \sqrt{B_3^{(m)}/B_3^{(m+1)}}$; $\alpha=1,\cdots,6$. However, for finite 
(non-asymptotic) cutoff values, these ratios deviate significantly from the universal asymptotic value $\lambda_0^{\infty}$ 
dictated by the RG limit cycle. Only in the limit of asymptotically large cutoffs do these quantities approach this 
universal asymptotic value. In this regime, however, the ground state becomes unphysically deeply bound, 
thereby diminishing its physical relevance and placing it beyond the domain of validity of low-energy EFT.

With 3BF terms included, the STM3 equations~\eqref{eq:F_nD_EFT_ren} are then solved to determine the renormalized 
spectator functions $F^R_{n,D}$, as well as the corresponding three-body wave functions $\Psi^R_{n,D}$. The renormalized
leading order observables obtained for the ground state trimer are displayed in 
Table~\ref{Tab:6.5}. Given the current absence of phenomenological information on the $D^0nn$ system, several 
{\it ah hoc} inputs for the three-body binding energy, namely $B^{(0)}_3=1.92,\, 2.82$, and $3.82$~MeV 
($S^{(0)}_n=0.1,\, 1.0,$ and $2.0$~MeV), are considered in order to obtain representative model predictions. As 
\st{with the} two-body input, we employ the phenomenologically extracted S-wave spin-singlet $n$-$n$ S-wave scattering
length, $a_{nn}=-18.63$~fm~\cite{Chen:2008zzj}, together with the S-wave spin-doublet $n$-$D^0$ scattering length, 
$a_{nD}=4.141$~fm, extracted from the ZCL model analysis~\cite{Raha:2017ahu}. As seen in Fig.~\ref{fig:Asymp_radii_plots}, 
the observables exhibit residual regulator sensitivity due to the lack of renormalizability at small cutoffs. Nevertheless,
asymptotic convergence is rapidly achieved beyond $\Lambda_{\rm reg}\gtrsim 150-200$~MeV. Since these results exhibit 
convergence already at momentum scales of the order of the pion mass, they suggest an encouraging possibility for 
the realization of either a very shallow Efimov-bound $D^0nn$ system or  a near-threshold quasi-bound 
resonance. This conclusion is further substantiated by the ``realistic'' dispersive scattering model results shown in 
Figs.~\ref{fig:limitcycle}, \ref{fig:B_T_m_012}, and \ref{fig:ReffVSB3_plots}, where inelastic effects from ``distant'' 
hadronic decay channels (located $\sim 200$~MeV below the $DN$ threshold) are implicitly incorporated. Given that the 
present LO results are primarily qualitative, more comprehensive treatments are required to determine the ultimate fate of 
the $D^0nn$ system. In this context, effective-range corrections may play a decisive role in establishing the existence of 
realistic $D^0nn$ bound states. Such corrections formally enter the halo-EFT framework at NLO and are beyond the scope of 
the present work. Nevertheless, guided by the relativistic TOPT formulation at LO~\cite{Epelbaum:2016ffd}, the leading 
$1/M$-expansion introduces range-like parameters $\rho_{s,d}$, which are, in general, distinct from the physical 
effective-range parameters $r^{(nn)}_0$ and $r^{(nD)}_0$, and induces logarithmic modifications of the LO STM equations. 
These modifications alter the RG limit cycle behavior, driving the system asymptotically toward a stable ultraviolet limit
(fixed point), and thereby obviating the need for a 3BF to achieve renormalization. A direct consequence is the persistence
of a single Efimov state, whose identification as a halo-bound state depends sensitively on $\rho_d$ (i.e., keeping 
$\rho_s = 1/M_n$ fixed). Within this framework, we find that a shallow bound state with $S_n=0.1\text{–}1.0$~MeV can be 
supported for $\rho_d = 1.49\text{–}2.84$~fm. A quantitatively reliable assessment of effective-range effects, however, 
requires a systematic NLO analysis, to be addressed in future work, together with the inclusion of higher partial-waves (e.g., 
P-wave channels). Progress in this direction is currently hindered by the poorly constrained $DN$ interaction, which 
precludes a reliable determination of key S-wave parameters, such as the scattering length $a_{nD}$ and the 
effective-range $r^{(nD)}_0$.

Before we finally conclude, we briefly dwell on our uncertainty estimates. A detailed error analysis is not warranted 
at this stage, due to the limited three-body phenomenology and the qualitative scope of the present study, which 
aims to probe residual universal features near the $D^0n$ unitarity limit with sub-threshold decay channels removed. 
It is therefore not unreasonable to anticipate substantial deviations -- potentially of order unity -- in the ZCL 
model predictions for more realistic estimates, even assuming that such quantities can be reliably extracted from 
three-body scattering analyses in the near future.\footnote{This possibility, however, appears rather remote even 
with current state-of-the-art facilities (e.g., ALICE at CERN~\cite{ALICE:2022enj}), where scattering experiments 
involving unstable $D^0$ mesons are challenging due to their limited production rates in femtoscopic measurements 
from $p$–$p$ and $p$–Pb collisions.} A more conservative model-based error estimate can, nevertheless, be obtained by
studying the sensitivity of our three-body observables with a realistic input for the $DN$ ($T=1$) scattering length,
namely $a^{\rm (WT)}_{Dn}=0.764-i0.615$~fm,\footnote{This estimate was obtained using a SU(4) generalization of the
chiral unitary model of Mizutani and Ramos~\cite{Mizutani:2006vq} by employing LO Weinberg-Tomozawa (WT) contact 
interactions to include the effects of $\pi^{-}\Lambda_{c},\, \pi^{0}\Sigma_{c}^{0}$, and $\pi^{-}\Sigma_{c}^{+}$ 
decay channels close to the $D^0n$ threshold. In particular, the prediction in Ref.~\cite{Raha:2017ahu} was based on
the constraint that the charmed resonance $\Sigma_c(2800)$~\cite{ParticleDataGroup:2016lqr} was reproduced as a 
sub-threshold quasi-bound state.} as predicted in Ref.~\cite{Raha:2017ahu}. Owing to the limitation of the adopted 
halo-EFT framework in treating complex scattering lengths in the three-body sector, only the real component, 
$\tilde{a}_{nD} = 0.764$~fm, is retained in generating the ``realistic'' model predictions. This leads to an error 
of no less than $\sim 50\%$ in our prediction of the effective radius $r_{\rm eff}$ [cf. Fig.~\ref{fig:ReffVSB3_plots}], 
and other related three-body observables. In addition to the overall model-based error estimate, we also assessed our 
systematic theoretical uncertainty by relying on the LO EFT truncation, i.e., the omission of higher-order 
contributions. The resulting ZCL predictions, with estimated EFT uncertainties of $\sim 16\%$, are found to be in 
reasonable agreement with those obtained in the realistic framework.

In summary, a definitive assessment of the $D^0nn$ system requires further investigation, particularly on the 
experimental side, to ascertain the existence of a bound or quasi-bound state. The scarcity of data, combined with the
challenges of systematically incorporating decay and coupled-channel dynamics, thereby limiting a universal description, 
currently precludes a reliable extraction of Efimovian signatures. However, future experiments at advanced facilities 
such as CERN, J-PARC, KEK, RIKEN, and FAIR are expected to access observables sensitive to halo or clustering 
dynamics (e.g., final-state $n$–$n$ correlations in core knock-out reactions), providing crucial input for more 
comprehensive theoretical studies.

\section*{Acknowledgments}
The authors are thankful  Daniel Phillips, Gautam Rupak, Chen Ji, Johannes Kirscher and Martin Sch\"afer for useful 
discussions. GM acknowledges financial support from the Science and Engineering Research Board (SERB) [grant number 
CRG/2022/000027] and acknowledges postdoctoral financial support from the National Science and Technology Council 
(NSTC), Taiwan. SM acknowledges financial support from the Department of Science and Technology (DST) under INSPIRE
Fellowship [grant number IF190758]. He also thanks SRM University, AP, for providing local hospitality where certain 
parts of the work were undertaken. UR acknowledges partial financial support from the Science and Engineering, 
Research Board (SERB) [grant numbers MTR/2022/000067 and CRG/2022/000027], as well as the US Department of Energy,
[award number DE-FG02-93ER-40756]. UR also gratefully acknowledges the Department of Physics and Astronomy, 
University of South Carolina, Columbia, and the Institute of Nuclear and Particle Physics, Ohio University, Athens, 
for their hospitality and financial support during his sabbatical visits, where parts of this work was carried out.

\appendix
\section{Basis states in Jacobi momentum representation }
\label{sec:Appendix-A}
The Jacobi momenta provide a convenient representation to describe the non-relativistic dynamics of a three-body 
system. By eliminating the motion of the center-of-mass of the three-body system, it expresses all the dynamics in
terms of a chosen binary sub-system's internal relative motion, as well as the motion of the binary sub-system 
relative to the corresponding spectator particle. Thus, the momentum state of an arbitrary three-body system is 
described in terms of a pair of relative three-momenta: one being the relative three-momentum ${\bf p}$ between 
two particles in the chosen two-body sub-system, and the other being the three-momentum ${\bf q}$ of the third 
spectator particle relative to the center-of-mass of the chosen binary sub-system (see 
Fig.~\ref{fig:Jacobi_momenta}). 

For concreteness, let us consider an arbitrary system of three {\it distinguishable} particles $i,j$ and $k$ with
masses $m_i$, $m_j$ and $m_k$, and laboratory/inertial frame momenta ${\bf k}_i$, ${\bf k}_j$ and ${\bf k}_k$, 
respectively. Then, the Jacobi momenta of the system, with particle $i$ chosen as the spectator, are defined 
by\footnote{The spectator notation is used throughout this work. However, this notation does not apply for defining
individual particle masses ($m_i$) and lab-frame momenta (${\bf k}_i$). Moreover, the choice of the Jacobi momenta 
is not unique. As a matter of fact, there are three equivalent descriptions of the same three-body system depending
on which particle is chosen as the spectator. The other equivalent sets of Jacobi momenta could be obtained 
{\it via} a cyclic permutation of the indices $ijk$. }
\begin{eqnarray}
{\bf p}_i = \mu_{jk}\left(\frac{{\bf k}_j}{m_j} - \frac{{\bf k}_k}{m_k}\right)\, &;& 
\qquad \mu_{jk} = \frac{m_jm_k}{m_j+m_k}\,,
\nonumber\\
{\bf q}_i = \mu_{i(jk)}\left(\frac{{\bf k}_i}{m_i}- \frac{{\bf k}_j + {\bf k}_k}{m_j+m_k}\right)\, &;& 
\qquad \mu_{i(jk)}= \frac{m_i(m_j+m_k)}{M}\,,
\end{eqnarray}
where ${\bf p}_i$ is the relative three-momentum between the particles in the center-of-mass frame of the two-body 
sub-system $j-k$, ${\bf q}_i$ is the relative three-momentum of the spectator particle $i$ with respect to the 
center-of-mass of the same binary sub-system, $M=m_i+m_j+m_k$ is the total mass of the three-body system, and 
${\bf K}= {\bf k}_i+{\bf k}_j+{\bf k}_k$ is total three-body momentum.\footnote{${\bf K}={\bf 0}$ in the 
three-body center-of-mass frame.} With the above choice of Jacobi momenta, the total kinetic energy or free 
Hamiltonian of the three-body system is given as
\begin{eqnarray}
 H_0 = \frac{{\bf K}^2}{2M}+\frac{{\bf p}^2_i}{2\mu_{jk}}+\frac{{\bf q}^2_i}{2\mu_{i(jk)}}\,.
\end{eqnarray}
\begin{figure}[tbp]
\centering
\includegraphics[scale=0.4]{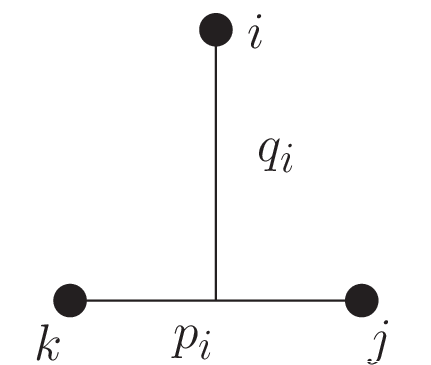}
    \caption{The Jacobi momentum ($p_i=|{\bf p}_i|,q_i=|{\bf q}_i|$) specification for an arbitrary three-body 
             system.}
\label{fig:Jacobi_momenta} 
\end{figure}

\subsection{Jacobi momentum basis in quantum mechanics}
We now briefly discuss the construction of a complete set of basis states for a quantum mechanical description of an 
arbitrary three-body system in momentum-space (see Ref.~\cite{Glockle_1983} for details). The basic objects are the
Jacobi plane-wave state vectors $|{\bf p},{\bf q}\rangle_i \equiv |{\bf p}_i,{\bf q}_i\rangle_i$, which satisfy the 
{\it orthonormality} relation:
\begin{equation}
{}_i\langle {\bf p},{\bf q} | {\bf p}^\prime,{\bf q}^\prime \rangle_i
\equiv {}_i\langle {\bf p}_i,{\bf q}_i | {\bf p}^\prime_i,{\bf q}^\prime_i \rangle_i 
= \delta^3({\bf p}_i-{\bf p}^\prime_i)\,\delta^3({\bf q}_i-{\bf q}^\prime_i)
\equiv \delta^3({\bf p}-{\bf p}^\prime)\,\delta^3({\bf q}-{\bf q}^\prime)\,,
\end{equation}
as well as the {\it completeness} relation:
\begin{equation}
\int {\rm d}^3{\bf p}\int {\rm d}^3{\bf q}\, |{\bf p},{\bf q}\rangle_i\,\, {}_i\langle {\bf p},{\bf q} |
\equiv 
\int {\rm d}^3{\bf p}_i\int {\rm d}^3{\bf q}_i\, |{\bf p}_i,{\bf q}_i\rangle_i\,\, {}_i\langle {\bf p}_i,{\bf q}_i |
= \mathbb{I}\,.
\end{equation}
Since these momentum states denote free-particle motions, they are eigenstates of the free Hamiltonian $H_0$:
\begin{equation}
H_0 |{\bf p}_i,{\bf q}_i\rangle 
= \left(\frac{{\bf p}^2_i}{2\mu_{jk}}+\frac{{\bf q}^2_i}{2\mu_{i(jk)}}\right)|{\bf p}_i,{\bf q}_i\rangle\,. 
\end{equation}

A generic partial-wave projected basis can now be constructed from the above plane wave states~\cite{Glockle_1983}: 
\begin{eqnarray}
|pq\,\mathcal{Q}_i\rangle_i \equiv
\Big|pq\big\{(l,s)j,(\lambda,\sigma)\mathcal{J}\big\}JM_J,(t,\tau)TM_T\Big\rangle_i\,,
\label{eq:jj-coupling}
\end{eqnarray}
where $\mathcal{Q}_i$ collectively specifies the set of all spin-isospin quantum numbers for a given quantum state, 
defined with respect to the spectator particle $i$. The following quantum numbers constitute the above representation:
\begin{itemize}
 \item $l$: the orbital angular momentum quantum number of the particles in binary sub-system $jk$,
 \item $s$: the intrinsic spin quantum number of the sub-system $jk$,
 \item $j=l\oplus s$: the total spin quantum number of the sub-system $jk$,\footnote{The quantum number $j$ must be 
 distinguished from the particle index $j$ throughout the text.} 
 \item $t$: the isospin quantum number of the sub-system $jk$,
 \item $\lambda\equiv l_i$: the orbital angular momentum quantum number of spectator $i$ relative to the sub-system $jk$,
 \item $\sigma\equiv s_i$: the intrinsic spin quantum number of the spectator $i$,
 \item $\mathcal{J}=\lambda \oplus \sigma$: the total spin quantum number of the spectator $i$,
 \item $\tau\equiv t_i$: the isospin quantum numbers of the spectator $i$,
 \item $J=j\oplus \mathcal{J}$: the total spin quantum number of the three-body system, 
 \item $M_J$: the quantum number for the $z$-component of total spin of the three-body system,
 \item $T=t\oplus \tau$: the total isospin quantum number of the three-body system, and
 \item $M_T$: the quantum number for the $z$-component of total isospin of the three-body system.
\end{itemize}
Note that the spectator notation also does not apply to the individual particle quantum numbers, $l_i,\, s_i,\, t_i$, etc. 
With low-energy S-wave approximation leading to vanishing orbital angular momenta, the spin-isospin quantum numbers of the
two distinct basis states of the $D^0nn$ system are specified as
\begin{eqnarray}
|pq\,{\mathcal Q}_D\rangle_D \!\!\!& \equiv &\!\!\!  
\left|pq\,\big\{(0,0)0,(0,0)0\big\}00,\left(1,\frac{1}{2}\right)\frac{3}{2}\frac{3}{2}\right\rangle_D\,, 
\quad \text{and}
\nonumber \\
|pq\,{\mathcal Q}_n\rangle_n \!\!\!& \equiv &\!\!\!  
\left|pq\,\left\{\left(0,\frac{1}{2}\right)\frac{1}{2},(0,0)0\right\}00,\left(1,\frac{1}{2}\right)\frac{3}{2}\frac{3}{2}\right\rangle_n\,.
\end{eqnarray}

Besides the aforementioned Jacobi basis representation, Eq.~\eqref{eq:jj-coupling}, in terms of the (2+1)-sub-system 
quantum numbers, it is sometimes useful to re-couple the orbital angular momentum, intrinsic spin, and isospin quantum
numbers of the full three-body system. This is effected {\it via} a unitary transformation of the basis 
$|pq\,\mathcal{Q}\rangle_i$ employing the {\it Wigner's 9j} symbol: 
\begin{eqnarray}
|pq\,\mathcal{Q}_i\rangle_i = \sum_{LS} \sqrt{\hat{j}\,\hat{\mathcal{J}}\,\hat{L}\,\hat{S}}
\begin{Bmatrix} l       & s      & j\\
                \lambda & \sigma & \mathcal{J}\\
                 L      &   S    & J         \\
\end{Bmatrix}
\Big|pq\big\{(l,\lambda)L,(s,\sigma)S\big\}JM_J,(t,\tau)TM_T\Big\rangle_i\,, 
\end{eqnarray}
where
\begin{itemize}
\item $L=l\oplus \lambda$: the total orbital quantum number of three-body system, and
\item $S=s\oplus \sigma$: the total intrinsic spin quantum number of three-body system.
\end{itemize}
It is also customary to introduce the hatted quantum number notation, e.g., 
$\hat{j}\equiv 2j+1$, etc. 

These partial-wave-projected basis state vectors likewise satisfy the following {\it orthonormality} relation:
\begin{equation}
 {}_i\langle pq\,\mathcal{Q}_i|p^\prime q^\prime\,\mathcal{Q}_i\rangle_i 
 = \frac{\delta(p-p^\prime)}{pp^\prime}\frac{\delta(q-q^\prime)}{qq^\prime}\,,
\end{equation}
as well as the {\it completeness} relation:
\begin{equation}
\sum_{\mathcal{Q}_i} \int_0^{\infty} {\rm d}p\,\, p^2 \int_0^{\infty} {\rm d}q \,\,q^2 \,\, 
|pq\,\mathcal{Q}_i\rangle_i\,\,{}_i\langle pq\,\mathcal{Q}_i|=\mathbb{I}\,,
\label{eq:complete}
\end{equation}
such that the Faddeev components $|\psi_i\rangle$ could be expanded in the Jacobi basis as follows:
\begin{equation}
|\psi_i\rangle = \sum_{\mathcal{Q}_i} \int_0^{\infty} {\rm d}p\,\, p^2 \int_0^{\infty} {\rm d}q \,\,q^2 \,\, 
{}_i\langle pq\,\mathcal{Q}_i|\psi_i\rangle\,\,|pq\,\mathcal{Q}_i\rangle_i\,.
\end{equation}
The derivation of the Faddeev coupled integral equations~\eqref{eq:F_nD} also require us to re-couple two
Jacobi basis states corresponding to different spectator representations. This leads to the overlap matrix 
element of the form:
\begin{eqnarray}
{}_i\langle pq\,\mathcal{Q}_i|p^\prime q^\prime\,\mathcal{Q}_j\rangle_j 
\!\!\!&=&\!\!\! \sum_{LS} \sum_{L^\prime, S^\prime}
\sqrt{\hat{j}\hat{\mathcal{J}}\hat{L}\hat{S}}\sqrt{\hat{j^\prime}\hat{\mathcal{J}^\prime}\hat{L^\prime}\hat{S^\prime}}
\begin{Bmatrix} l       & s      & j\\
                \lambda & \sigma & \mathcal{J}\\
                 L      &   S    & J         \\
\end{Bmatrix}
\begin{Bmatrix} l^\prime       & s^\prime      & j^\prime\\
                \lambda^\prime & \sigma^\prime & \mathcal{J}^\prime\\
                 L^\prime      &   S^\prime    & J^\prime         \\
\end{Bmatrix}
\nonumber \\
&&\hspace{1cm} \times\,\, {}_i\langle pq(l,\lambda)L|p^\prime q^\prime(l^\prime,\lambda^\prime)L^\prime\rangle_j  \,\,\,
{}_i\langle (s,\sigma)S | (s^\prime,\sigma^\prime)S^\prime \rangle_j \,\,\, 
{}_i\langle (t,\tau)T | (t^\prime,\tau^\prime)T^\prime\rangle_j \,,\quad\,
\end{eqnarray}
where the re-coupling of the spins and isospin states is obtained employing the {\it Wigner's 6j} symbol: 
\begin{eqnarray}
{}_i\langle (s,\sigma)S | (s^\prime,\sigma^\prime)S^\prime \rangle_j \!\!\!&=&\!\!\! 
\delta_{SS^\prime} \sqrt{\hat{s}\hat{s}^\prime}(-1)^{s_i+s_j+s_k+S}
\begin{Bmatrix} s_j   & s_k  & s       \\
                s_i   &  S   & s^\prime\\
\end{Bmatrix}\,, \quad \text{and}
\nonumber\\
{}_i\langle (t,\tau)T | (t^\prime,\tau^\prime)T^\prime \rangle_j \!\!\!&=&\!\!\! 
\delta_{TT^\prime} \sqrt{\hat{t}\hat{t}^\prime}(-1)^{t_i+t_j+t_k+T}
\begin{Bmatrix} t_j   & t_k  & t       \\
                t_i   &  T   & t^\prime\\
\end{Bmatrix}\,.
\end{eqnarray}
The overlap matrix elements ${}_i\langle pq(l,\lambda)L|p^\prime q^\prime(l^\prime,\lambda^\prime)L^\prime\rangle_j$ 
involve re-couplings between momenta and orbital angular momenta belonging to different spectator representations. 
The general methodology of evaluating such matrix elements is quite elaborate but well-known in the 
literature~\cite{Canham:2008jd,DL-Canham:2009,Platter:2004he,Platter:2004ns,Platter:2004zs,Glockle_1983}. Thus, we 
refrain from repeating their evaluations in this work, and only quote the generic result, 
Eq.~\eqref{eq:ij_overlap_matrix}, for the matrix elements of the $D^0nn$ system in the main text. 

\subsection{Jacobi momenta for the $D^0nn$ system}
The three different re-arrangement channels for the $D^0nn$ system are displayed in Fig.~\ref{fig:channels}. For the 
simplicity of analytical derivations, we find it convenient to express the mass of the $D^0$-meson in terms of the 
nucleon (neutron) mass, namely $M_D=yM_n$. Also, to distinguish the momenta of the two identical neutrons, we find 
it convenient to tag them as $n_1$-neutron and $n_2$-neutron. Quantum mechanically, however, they are 
indistinguishable, demanding anti-symmetry under their permutations to preserve the parity of the component wave 
functions [see Eq.~\eqref{eq:antisymmetry}]. Depending on the choice of the spectator, the following {\it equivalent} 
kinematical description of the $D^0nn$ system can be given:  
\begin{itemize}
\item {\bf Channel-1:}  $D^0$-meson is the spectator, and the two neutrons form the binary sub-system. In this case, 
the choice of the Jacobi momenta is
\begin{equation}
{\bf p}_D = \frac{1}{2} ({\bf k}_{n_1}-{\bf k}_{n_2})\,, \qquad \text{and} \qquad
{\bf q}_D = \frac{2 {\bf k}_D-y({\bf k}_{n_1}+{\bf k}_{n_2})}{y+2}\,,
\end{equation}
where ${\bf k}_{n_1}$, ${\bf k}_{n_2}$, and ${\bf k}_D$ are individual momenta of $n_1$-neutron, $n_2$-neutron and 
the $D^0$-meson, respectively. The corresponding free Hamiltonian of the system in this representation is given by
\begin{equation}
H_0^{(D)} = \frac{{\bf p}_D^2}{2\mu_{nn}} + \frac{{\bf q}_D^2}{2\mu_{D(nn)}} 
= \frac{1}{M_n}p_D^2 + \left(\frac{y+2}{4yM_n}\right)q_D^2\,,
\end{equation}
where the reduced masses are defined as
\begin{equation}
 \mu_{nn} = \frac{M_n}{2} \quad \text{and} \quad \mu_{D(nn)} = \frac{2M_n M_D}{M_D+2M_n} = \frac{2y}{y+2}M_n\,.
\end{equation}
\item {\bf Channels-2, -3:} Either $n_1$-neutron or $n_2$-neutron is the spectator, while the other non-spectator 
neutron forms the binary sub-system with the $D^0$-meson. The two re-arrangement channels yield identical 
configurations since the two neutrons are identical. As noted earlier, using the permutation operator $\mathcal P$, 
the Faddeev component $|\psi_{n2}\rangle$ can be expressed in term of the component $|\psi_{n1}\rangle$. In this 
case, the choice of the Jacobi momenta is
\begin{eqnarray}
{\bf p}_{n_1} = \frac{1}{y+1} \left(y{\bf k}_{n_2}-{\bf k}_D\right)\, &,&  
{\bf p}_{n_2} = \frac{1}{y+1} \left({\bf k}_D-y{\bf k}_{n_1}\right)\,, \qquad \text{and}
\nonumber\\
{\bf q}_{n_1} = \frac{(y+1){\bf k}_{n_1}-({\bf k}_{n_2}+{\bf k}_D)}{y+2}\, &,& 
{\bf q}_{n_2} = \frac{(y+1){\bf k}_{n_2}-({\bf k}_{n_1}+{\bf k}_D)}{y+2}\,.
\end{eqnarray}
The corresponding free Hamiltonian of the system in this representation is given by
\begin{equation}
H_0^{(n)} = \frac{{\bf p}_n^2}{2\mu_{nD}}+\frac{{\bf q}_n^2}{2\mu_{n(nD)}} 
= \frac{y+1}{2yM_n}p_n^2+\left(\frac{y+2}{2M_n(y+1)}\right)q_n^2\,,
\end{equation}
where the reduced masses are defined as
\begin{equation}
 \mu_{nD} = \frac{M_n M_D}{M_D+M_n} = \frac{y}{y+1}M_n, \quad \text{and} \quad \mu_{n(nD)} 
 = \frac{M_n (M_D+M_n)}{M_D+2M_n} =\frac{y+1}{y+2} M_n\,.
\end{equation}
\end{itemize}
\begin{figure}[tbp]
\centering
\includegraphics[scale=0.6]{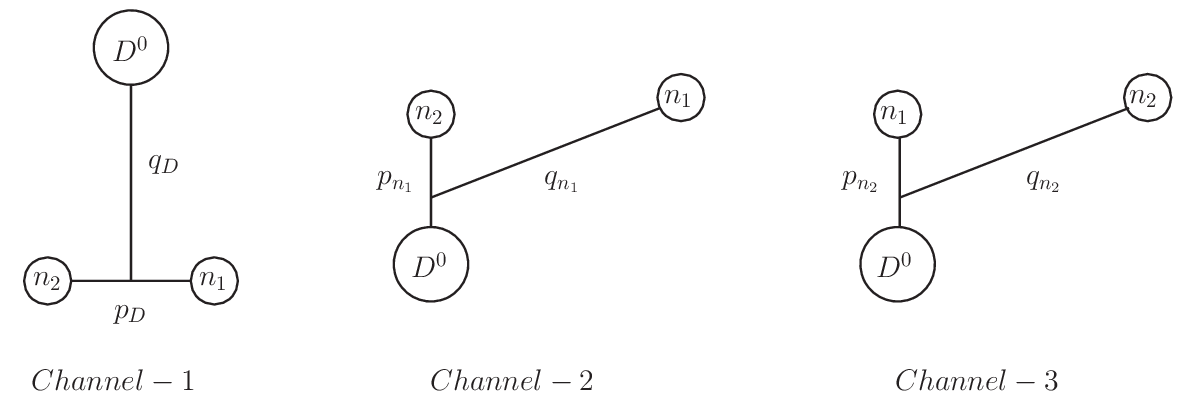}
    \caption{The re-arrangement channels and Jacobi momenta for the $D^0nn$ system.}
\label{fig:channels} 
\end{figure}

\section{Two-body T-matrix with a separable potential }
\label{sec:Appendix-B}
Here, we review the quantum mechanical framework, developed originally in the
Refs.~\cite{DL-Canham:2009,Platter:2004he,Platter:2004ns,Platter:2004zs}, and subsequently used in the study of 
two-body S-wave bound states for the choice of a single-channel (rank 1) attractive local separable potential. In 
the momentum-space, such a model potential is represented as
\begin{equation}
\frac{1}{(2\pi)^3}\langle {\bf k}^\prime|V^S|{\bf k}\rangle=V({\bf k},{\bf k^\prime}) = C_0 \chi(k)\chi(k^\prime)+\cdots\,,
\label{Eq6.39}
\end{equation}
where $C_0<0$ is a two-body coupling constant, and the ellipses denote contributions from higher partial-waves which
do not concern us. The potential can also be expressed as a projection operator 
$V^S \equiv C_0 |\chi\rangle\langle \chi |$, where  $\chi(p)=\langle {\bf p}|\chi\rangle$ and 
$\chi(p^\prime)=\langle {\bf p^\prime}|\chi\rangle$ are the so-called {\it form factor} functions. Here, for 
instance, one can use simple functional forms for the $l^{\rm th}$ partial-wave contribution, such as the Heaviside 
step-function, $\chi_l(k,\beta_l) = k^l\Theta(\beta_l-k)$, where the pre-factor $k^l$ constrains the T-matrices 
to satisfy the well-known {\it Wigner threshold law}~\cite{Wigner:1948zz}. Such regulator functions are employed to 
suppress dynamical contributions arising from the high-momentum modes, i.e., for $k,k^\prime \geq \beta_l$, where 
$\beta_l$ are cutoff parameters defining the (inverse) effective-range of the potential. In other words, taking 
$\beta_l \to \infty$ is equivalent to considering a zero-range potential (i.e., the scaling limit). In our treatment, 
however, we only consider S-wave contribution, namely $\chi_0\equiv \chi$, with $\beta_0\equiv \Lambda_{\rm reg}$ 
being the regulator scale.

Now, it is well-known that for a separable potential $V^S$, the two-body T-matrix $t(z)$ is itself separable:  
\begin{eqnarray}
t(z) = V^S+V^Sg_0(z)t(z) \!\!\!&=&\!\!\!  C_0|\chi \rangle \langle \chi| + C_0 |\chi \rangle \langle \chi | g_0(z) t(z)
\nonumber\\
\!\!\!&\equiv &\!\!\! |\chi\rangle\tau(z)\langle\chi|\,, \qquad \text{where} \qquad 
\tau(z) = \frac{1}{C^{-1}_0-\langle\chi|g_0(z)|\chi\rangle}\,,
\end{eqnarray}
and $g_0$ is the free two-body Green's function. Consequently, for the S-wave T-matrix, 
$T({\bf k},{\bf k^\prime};{\mathcal E})\equiv - \langle {\bf k}^\prime|t(z)|{\bf k}\rangle/(2\pi)^3$, we obtain 
{\it via} the Lippmann-Schwinger equation, the following separable form in momentum representation:
\begin{eqnarray}
T({\bf k},{\bf k^\prime};{\mathcal E})  &=&  V({\bf k},{\bf k^\prime}) + 
\int {\rm d}^3{\bf p}\, \frac{V({\bf k},{\bf p})}{{\mathcal E}-\frac{p^2}{2\mu}+i\varepsilon}T({\bf p},{\bf k^\prime};{\mathcal E})
\nonumber\\
&=&  C_0 \chi(k,\Lambda_{\rm reg})\,\chi(k^\prime,\Lambda_{\rm reg}) + C_0\chi(k,\Lambda_{\rm reg})\int {\rm d}^3{\bf p}\, 
\frac{\chi(p,\Lambda_{\rm reg})}{{\mathcal E}-\frac{p^2}{2\mu}+i\varepsilon}T({\bf p},{\bf k^\prime};{\mathcal E})
\nonumber\\
&\equiv& \chi(k,\Lambda_{\rm reg})\,\chi(k^\prime,\Lambda_{\rm reg}) \frac{1}{C^{-1}_0-{\mathcal I}({\mathcal E},\Lambda_{\rm reg})} \,,
\end{eqnarray}
where
\begin{eqnarray}
{\mathcal I}({\mathcal E},\Lambda_{\rm reg}) &=& 
\int{\rm d}^3{\bf p}\, \frac{|\chi(p,\Lambda_{\rm reg})|^2}{{\mathcal E}-\frac{p^2}{2\mu}+i\varepsilon} 
= 8\pi\mu \int_0^{\Lambda_{\rm reg}} {\rm d}p\, p^2 \frac{1}{2\mu{\mathcal E}-p^2+i\varepsilon}\,.
\end{eqnarray}
Since we are concerned here with real- or virtual-bound two-body sub-systems where the total energy is negative 
(${\mathcal E}<0$), the $i\varepsilon$ prescription in the denominator of integral $\mathcal I$ becomes redundant. A
straightforward evaluation of the above integration leads to the following expression:
\begin{equation}
{\mathcal I}({\mathcal E},\Lambda_{\rm reg})= 
-4\pi^2\mu \bigg[\frac{2}{\pi}\Lambda_{\rm reg}
-\sqrt{-2\mu {\mathcal E}}\,\frac{2}{\pi}\arctan\bigg(\frac{\Lambda_{\rm reg}}{\sqrt{-2\mu {\mathcal E}}}\bigg)\bigg]\,.
\label{Eq6.44}
\end{equation}
Taking the zero-range limit, i.e., $\Lambda_{\rm reg}\rightarrow \infty$, the first term in the above expression yields 
an ultraviolet divergence. Since low-energy observables are expected to be regulator-independent, such terms are 
eliminated {\it via} a renormalization prescription. For a binary S-wave system with large positive (negative) scattering
length $a_0$, a two-body shallow bound (virtual) state with energy ${\mathcal E}=-B_2=-1/(2\mu a^2_0)$ is produced at 
leading order. This is indicated by the pole of the two-body T-matrix, represented by the renormalization condition,
$C^{-1}_0 = {\mathcal I}({\mathcal E}=-B_2, \Lambda_{\rm reg})$. Utilizing this pole position, we fix the running of the 
two-body coupling $C_0\to C_0(\Lambda_{\rm reg})$ in order to reproduce the scattering length $a_0$. Especially, with 
$a_0\Lambda_{\rm reg}>>1$,  yields a simple expression for the two-body coupling:
\begin{equation}
C_0(\Lambda_{\rm reg}) = \frac{a_0}{4\pi^2\mu}\bigg[1-\frac{2a_0\Lambda_{\rm reg}}{{\pi}}\bigg]^{-1}\,.
\end{equation}
Hence, the leading order renormalized two-body T-matrix for negative energies (${\mathcal E}<0$) in the vicinity of an 
S-wave bound/anti-bound state is given by the separable form:
\begin{eqnarray}
T({\bf k},{\bf k^\prime};{\mathcal E}) \!\!\!&=&\!\!\! \chi(k,\Lambda_{\rm reg})\,
\chi(k^\prime,\Lambda_{\rm reg})\,\frac{1}{4\pi^2 \mu}\bigg[\frac{1}{a_0}\,\frac{2}{\pi}\arctan\big(|a_0|{\Lambda_{\rm reg}}\big)
-\sqrt{-2\mu {\mathcal E}}\,\frac{2}{\pi}\arctan\bigg(\frac{\Lambda_{\rm reg}}{\sqrt{-2\mu {\mathcal E}}}\bigg)\bigg]^{-1}.\quad\,
\label{eq:two-body_T-matrix}
\end{eqnarray}

\section{Numerical implementation of Faddeev equations}
\label{sec:Appendix-C}
Finally, we elaborate on our numerical procedure for solving the Faddeev integral equations for the $D^0nn$ system, 
Eq.~\eqref{eq:Matrix_eq_Fn_FD}, to determine the spectator functions $F_{n,D}$, as eigenvectors of the matrix 
equation:\footnote{The numerical implementation that we demonstrate here for the unrenormalized equations is 
straightforwardly extended to the renormalized counterparts, including 3BF terms. }
\begin{equation}
\lambda \begin{bmatrix} F_n\\F_D \end{bmatrix} = \begin{bmatrix}\mathcal{K}_{nn}&\mathcal{K}_{nD}\\
\mathcal{K}_{Dn}&0\end{bmatrix} \begin{bmatrix} F_n\\F_D \end{bmatrix}\,,
\end{equation}
where of all possible solutions, the physical solutions correspond to the eigenvalue $\lambda=1$.

Since we perform the numerical integration over $N$ Gaussian quadratures, the integration measure ${\rm d}q^{\prime}$ 
is replaced by a summation over the $N$ Gaussian weights $W_i$, with $q=q_i$ and $q^\prime=q^\prime_j$ being the 
discrete Gaussian points lying between 0 and the UV cutoff momentum $\Lambda_{\rm reg}$. Thus, the spectator functions
$F_n$ and $F_D$ represent $N$-dimensional arrays while the kernel functions $\mathcal{K}_{nn}$, $\mathcal{K}_{Dn}$, 
and $\mathcal{K}_{nD}$ represent $N\times N$-dimensional matrices defined over the mesh points $(q_i,q^{\prime}_j)$. 
For example, the kernel function $\mathcal{K}_{Dn}$ can be written in the discretized form
\begin{eqnarray}
\mathcal{K}_{Dn}(q_i,q^{\prime}_j) = -
\frac{2(y+1)}{\pi y}W_j q^{\prime \,2}_j\,
\mathcal{K}_{(n)}(q_j^\prime,q_i;B_3)\,S^R_d\left(-B_3-\frac{q^{\prime 2}_j}{2M_n},{\bf q}^{\prime}_j\right)\,; \quad j=1,...,N\,,
\end{eqnarray}
where $W_j=W(q^\prime_j)$ are {\it weight factors} associated with the quadrature points $q^{\prime}_j$. Here, 
${\mathcal K}_{(n)}$ denotes the S-wave projected interaction kernel with a freely propagating neutron exchanged 
between the two spin-doublet $D^0n$-dimer fields [see Eqs.~\eqref{eq:OPE}] with renormalized dressed propagator 
$S^R_d$ [see Eqs.~\eqref{eq:dimeron_prop1} and \eqref{eq:dimeron_prop2}]. It is then a straightforward numerical 
task to determine the $2N$ dimensional eigenvector $\begin{bmatrix} F_n(q_j)\\F_D(q_j) \end{bmatrix}$ corresponding 
to the eigenvalue $\lambda=1$. As depicted in Fig.~\ref{fig:FD_Fn}, the solutions to the spectator functions are
shown for selected cutoff values, $\Lambda_{\rm reg}=\Lambda^{(n)}_{\rm reg}$, corresponding to vanishing three-body
coupling $g_3(\Lambda^{(n)}_{\rm reg})=0$ (i.e., at the $n^{th}$ zero of the RG limit cycle), with an input trimer 
neutron separation energy, $S_n=0.1$~MeV. In particular, to obtain unique solutions of the linear homogeneous 
equations, we impose the normalization condition $F_D(q=0)=1$. As expected, the spectator functions display the 
characteristic oscillatory behavior modulated by a $1/q$ falloff. Furthermore, the results show negligible regulator
dependence, indicating that the solutions are effectively renormalized at the zeros of the RG limit cycle. 

In addition, Fig.~\ref{fig:FD_Fn} presents a comparison with the results obtained in our so-called ``extended''
ZCL scenario. For this purpose, we choose the largest of the above cutoff values, for which the differences between
the two scenarios can be most clearly resolved. In this case, the renormalized di-neutron and $D^0n$-dimer 
propagators, $iS^R_{n,d}$, are effectively modified by logarithmic contributions governed by the range-like 
parameters $\rho_s \sim 1/M_n$ and $\rho_d \sim 1/(2\mu_{nn})$ (see the discussion in the main text, 
Sec.~\ref{sec:2.2}). Evidently, such leading order range-like corrections produce hardly any noticeable change in 
the spectator functions, with little or no sensitivity to the range parameters.

\begin{figure}[h]
\centering
\includegraphics[width=0.49\textwidth]{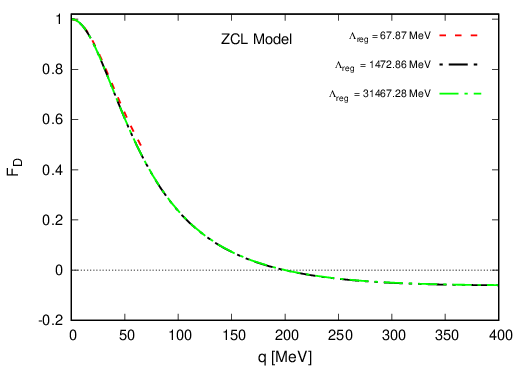} \hfill
\includegraphics[width=0.49\textwidth]{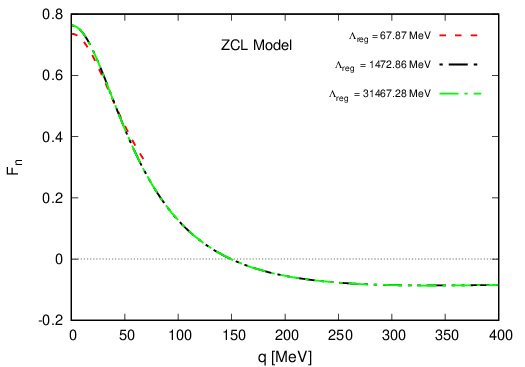} 

\vspace{0.3cm}

\includegraphics[width=0.49\textwidth]{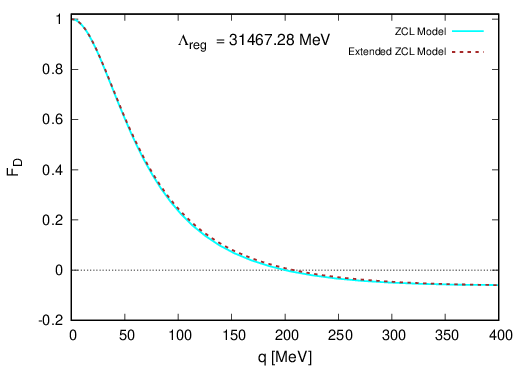} \hfill
\includegraphics[width=0.49\textwidth]{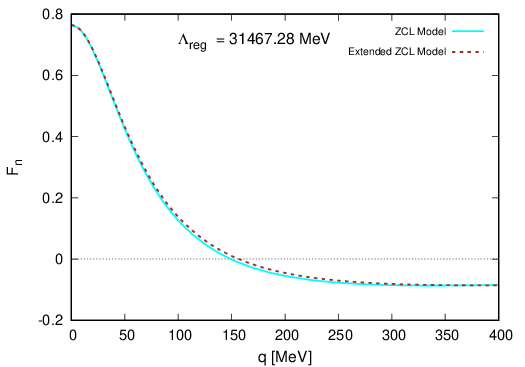}
    \caption{The two Faddeev-component spectator functions, $F_D$ and $F_n$, shown as functions of the
             spectator-particle momentum transfer $q$, for three selected values of the sharp cutoff 
             regulator $\Lambda_{\rm reg}$ corresponding to vanishing three-body coupling ($g_3=0$). 
             The solutions are obtained for an input trimer binding energy of $B_3=1.92$~MeV, subject
             to the normalization condition $F_D(0)=1$. {\bf Upper panel:} Results in the idealized ZCL
             model scenario. {\bf Lower panel:} Comparison between the idealized and extended ZCL 
             results for the specific cutoff value $\Lambda_{\rm reg}=31467.28$~MeV. }
\label{fig:FD_Fn} 
\end{figure}

Furthermore, for the purpose of evaluating the three-body wave functions $\Psi_{n,D}$, Eq.~\eqref{eq:Psi_nD_pq}, it 
is required to evaluate the spectator functions at shifted values of the off-shell relative three-momentum 
$q^\prime=\Pi^\prime$ [see Eq.~\eqref{eq:Pi_momenta_Dn}]. To this end, we use interpolated values of these functions
as obtained {\it via} a {\it cubic spline} interpolating algorithm of the form:
\begin{equation}
F_i(\Pi^\prime) \approx \sum_j {\mathcal S}_j(\Pi^\prime)F_i(q^\prime_j)\,; \quad i=n,D\,,
\end{equation}
where ${\mathcal S}_j$ are the spline elements evaluated at the desired momentum, $q^\prime=\Pi^\prime$. Likewise, 
the spline interpolator is needed for evaluating the matter density from factors, Eqs.~\eqref{eq:one-body_FF} and 
\eqref{eq:two-body_FF_nD_nn}, at the shifted values of the momentum transfer $p \to |{\bf p} - {\bf k}|$ and 
$q\to |{\bf q} - {\bf k}|$.

\bibliographystyle{elsarticle-num}


\end{document}